\documentclass[12pt]{extarticle}

\usepackage{url}

\usepackage{graphicx}
\usepackage{soul}
\usepackage{xcolor}
\usepackage{bm} 
\usepackage{amsmath}
\usepackage{cancel}
\usepackage{enumitem}
\usepackage{float}

\usepackage{subfigure}
\usepackage{multirow}
\usepackage{amsmath}
\usepackage{amsfonts}
\usepackage{bm}
\usepackage{bbding}
\usepackage{hyperref}
\newtheorem{Remark}{Remark}

\def\bx{{\bf x}}

\def\by{{\bf y}}

\def\y{{\bf  y}}
\def\x{{\bf  x}}
\newcommand{\post}{\bar{\phi}}

\def\btheta{\boldsymbol \theta}
\def\bTheta{\boldsymbol \Theta}

\def\mR{\mathbb{R}}

\def\cY{{\cal Y}}

\def\mR{\mathbb{R}}

\def\E{\mathbb{E}}

\UseRawInputEncoding

\title{
A unifying view of contrastive learning, importance sampling, and bridge sampling for energy-based models
}
\author{Luca Martino, \footnote{The author have not conflicts of interest.  The code and data used in this work will be made publicly available upon publication. } \\
 University of Catania, Italy.\\Ê
 {\small Corresponding author: luca.martino@unict.it}
}

\begin{document}
%

\maketitle 

%






\begin{abstract}
In the last decades, energy-based models (EBMs) have become an important class of probabilistic models in which a component of the likelihood is intractable and therefore cannot be evaluated explicitly. Consequently, parameter estimation in EBMs is challenging for conventional inference methods. This work is a review paper that brings together these different methodological contributions within a unified framework, highlighting the connections among  noise contrastive estimation (NCE), reverse logistic regression (RLR), Bennett's acceptance-ratio (BAR)  method, multiple importance sampling (MIS), and bridge sampling within the context of EBMs. Some of them  are equivalent under specific conditions. This unified perspective clarifies relationships among existing methods and enables the development of new estimators, with the potential to improve statistical and computational efficiency. Furthermore, this study helps elucidate the success of NCE in terms of its flexibility and robustness, while also identifying scenarios in which its performance can be further improved. Our aim is to provide a self-contained and accessible treatment, with clear and straightforward derivations, that brings together these contributions within a common framework and makes the main results and their underlying ideas understandable to a broader readership. 
 The contribution is twofold. First, this work serves as a comprehensive tutorial for researchers entering the field, introducing the fundamental concepts and derivations. Second, it offers a thorough review for experienced practitioners. The code used in the numerical experiments is available to support the reproducibility of the results.
\newline
\newline
{\bf Keyword:} Contrastive learning; bridge sampling; reverse logistic regression;  Bennett's acceptance-ratio; multiple importance sampling; binary classification.
\end{abstract}

\section{Introduction}\label{intro}

\begin{figure}[htbp]
   \centering
   \centerline{
   \includegraphics[width=17cm]{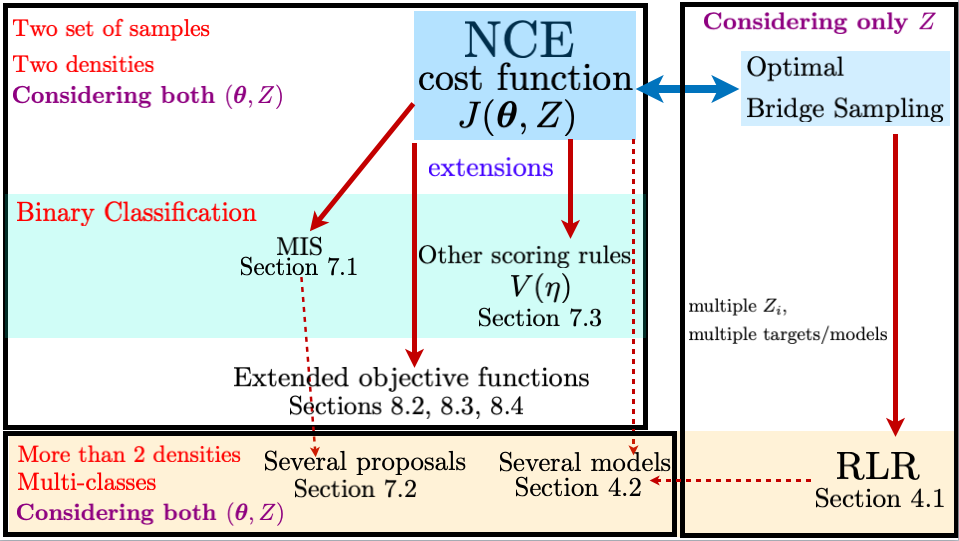} 
    }
   \caption{{\small Graphical summary of the connections and extensions {provided in the literature}, and described in this work. The noise contrastive estimation (NCE) method provides estimators of  $\btheta_\texttt{tr}$ and $Z_\texttt{tr}=Z(\btheta_\texttt{tr})$ designing a  binary classification problem, {and treating the variable $Z$ as a free parameter during the optimization.} We show that NCE operates as an optimal bridge estimator in the $Z$-domain. The reverse logistic regression (RLR) coincides with NCE  in the $Z$-domain and as an extension of bridge sampling, when several models/targets are considered. Several other generalizations  (even for the estimation of $\btheta$) are described: different scoring rules $V(\eta)$,  multiple importance sampling (MIS) procedures \cite{OwenZhou2000,MIS2019} {and extended objective functions  (see Sections \ref{NovelSect} and \ref{BClassSect}).}}   }
   \label{fig_super}
\end{figure}

Energy-based models (EBMs), denoted as $\post(\by|\btheta) =\frac{\phi(\by|\btheta)}{Z(\btheta)}$,  provide a flexible and powerful framework for probabilistic modeling. Here, $Z(\btheta)$ is an intractable partition function, and $\btheta \in \bTheta \subseteq \mR^{d_\theta}$ is the object of interest for inference  \cite{du2019implicit,dawid2024introduction,LeCun2006Energy,Wainwright2008Graphical,llorenteREV2}.  Despite their flexibility and expressive capability, inference and learning in EBMs are inherently challenging due to the intractability of the normalizing constant $Z(\btheta) \in \mathbb{R}$, which is typically unknown. As a result, EBMs are often referred to as unnormalized models, since the numerator is $\phi(\by|\btheta)$ can be evaluated pointwise, whereas
 $Z(\btheta)$ cannot.   In a Bayesian framework, such likelihood functions give rise to so-called doubly intractable posteriors \cite{caimo2015efficient,liang2010double,murray2012mcmc,park2018bayesian}. The intractability of the partition function $Z(\btheta)$, especially in high-dimensional settings, severely hinders likelihood-based inference, complicating model comparison and parameter estimation.
\newline
\newline
Several strategies have been proposed to enable practical inference in these models{:  likelihood approximation-based approaches \cite{Geyer1991MarkovChain,Geyer1994Convergence},  score matching methods \cite{Hyvarinen2005ScoreMatching,Hyvarinen2007SomeExtensions},
  pseudo-likelihood techniques \cite{Besag1974Spatial,Hyvarinen2006Consistency}, and the contrastive learning schemes  \cite{Gutmann2010NCE,Gutmann2022Contrastive}. }
 In this work, we focus on the last family of methods, and in particular on noise-contrastive estimation (NCE), which recasts parameter estimation as a classification problem between {$N$ observed data $\y_1,...,\y_N$, and $M$ artificially generated samples $\x_1,....,\x_M$ \cite{Gutmann2010NCE,Gutmann2022Contrastive,merda2025}}.
 NCE builds a cost function $J(\btheta,Z)$ over the augmented parameter space $\bTheta\times \mathbb{R}$. Minimizing $J(\btheta,Z)$ yields estimates of both the model parameters $\btheta_{\texttt{tr}}$ and the corresponding normalizing constant $Z_{\texttt{tr}}=Z(\btheta_{\texttt{tr}})$. {Here, $\btheta_{\texttt{tr}}$ denotes the parameter vector associated with the observed data $\by_n$, i.e., assumed to be generated according to the normalized model $\bar{\phi}(\y|\btheta_{\texttt{tr}})$, hence  $\by_n \sim \bar{\phi}(\y|\btheta_{\texttt{tr}})$.} 
 Owing to its effectiveness and flexibility, NCE has been widely studied and applied in a variety of settings{, such as statistical physics, machine learning, natural language processing, computer vision, speech recognition, and generative modelling, to name a few (see, e.g., \cite{Gutmann2010NCE,Gutmann2022Contrastive,RiouDurand2019NCE,LeKhac2020,li2021contrastive}).} Recently in \cite{MartinoScaffidi26}, the authors study the NCE performance focusing mainly on the estimation in the $\btheta$-space, { and the relationship with the Geyer's approach for likelihood approximation \cite{Geyer1991MarkovChain,Geyer1994Convergence}. Several combinations of these two techniques, NCE and Geyer's idea,  have been proposed in \cite{MartinoScaffidi26}.}
\newline
\newline
 {This work provides a review of a broad range of techniques, ideas, and approaches related to NCE and the estimation of partition functions, drawing on contributions from different areas of the literature. We highlight connections among these methods, many of which have already been pointed out in previous works (e.g., \cite{Chehab2023OptimizingNoise,Pihlaja2010,Matsuda2019Estimation,GutmannHirayama2011}), and provide clear and detailed derivations of the various schemes and their relationships. Our aim is to offer a self-contained and accessible treatment that brings together these contributions within a common framework and makes the main results understandable to a broader readership. Perhaps some of the derived estimators (contained in some example) and/or the detailed derivations may also contain elements of novelty, although the primary aim of the paper is to review, clarify, and unify existing contributions.} Unlike in \cite{MartinoScaffidi26}, we mainly focus on the estimation of the normalizing constant $Z_{\texttt{tr}}=Z(\btheta_{\texttt{tr}})$ (or more generally $Z=Z(\btheta)$ given a fixed $\btheta$) by NCE-type approaches{, and on the relationship between these estimators and classification cost functions. We also present several ways of extending the NCE cost function, leading to generalized objective functions that yield consistent estimators of both \(\btheta_{\texttt{tr}}\) and \(Z_{\texttt{tr}}\), while treating \(Z\) as a free parameter in the optimization problem, despite the fact that, by definition, \(Z(\btheta)\) is a function of \(\btheta\) \cite{Chehab2023OptimizingNoise,Pihlaja2010}.} More specifically, we provide a unifying view that connects NCE,  { reverse logistic regression (RLR),} multiple importance sampling (MIS), and bridge sampling within a common framework for EBMs, {and we show and recall their equivalence under some specific conditions \cite{Chehab2023OptimizingNoise,Pihlaja2010,Matsuda2019Estimation,GutmannHirayama2011}. Figure \ref{fig_super} provides a summary of the main relationships investigated in this study.} Although these methods originate from different communities and are often presented from distinct perspectives, they clearly share a common underlying structure: all rely on comparing samples drawn from the model of interest $\post(\y|\btheta_{\texttt{tr}})$,  with samples generated from an auxiliary proposal/reference distribution, denoted as $q(\y)$.  In particular, contrastive learning methods frame the problem as a classification task between data and noise, while importance sampling and bridge sampling construct estimators of normalizing constants through weighted combinations of samples from multiple distributions.
 \newline
 This unified view  not only clarifies the relationships among existing methods, but also enables the design of new estimators that interpolate between NCE, { reverse logistic regression (RLR) \cite{geyer1994estimating,chen1997monte} and  Bennett's acceptance-ratio (BAR) Ê\cite{BENNETT1976245} (see also \cite[Section 3.2]{Chehab2023OptimizingNoise} and \cite{Matsuda2019Estimation}), }  multiple importance sampling \cite{OwenZhou2000,MIS2019} and bridge sampling \cite{meng1996simulating,llorenteREV}, potentially offering improved statistical and computational properties: {regarding these connections and NCE extensions, see specifically the seminal and relevant contributions \cite{Pihlaja2010,GutmannHirayama2011,Chehab2023OptimizingNoise,Matsuda2019Estimation} }  (see also Figure \ref{fig_super}). 
 These generalizations extend the standard frameworks to encompass a broader class of importance sampling schemes, which jointly exploit samples from the model's data distribution and artificial samples generated from a proposal/reference density. { In this sense, this work can also be viewed as a focused survey of methods for estimating normalizing constants based on two densities: the target model and a proposal (or reference) distribution.} Moreover, the proposed unified formulation naturally enables the development of new estimation schemes for $\btheta$, which are also introduced and empirically evaluated.
\newline %
Thus, in line with other works in the literature {that pursue a similar goal \cite{llorenteREV2,Gelman98,Storvik11,Martino2013}, which review different methodological schemes and elucidate their theoretical and practical relationships}, the connections established in this work offer a twofold contribution. {First of all,} they provide a unifying perspective on existing methods and a principled framework for designing novel estimation schemes. Furthermore, this study helps to elucidate the success of the NCE method in terms of its flexibility and robustness, while also highlighting scenarios in which its performance may be further improved. 
Additionally, some of the described schemes may admit a more tractable theoretical analysis, which in turn can simplify the characterization of the \emph{optimal} proposal/reference density,  an aspect that is not straightforward in standard NCE \cite{Chehab2022OptimalNoise,Chehab2023OptimizingNoise}.
{ The reduction of the optimization space based on the results presented in this work and in previous contributions \cite{Chehab2023OptimizingNoise,Pihlaja2010,Matsuda2019Estimation,GutmannHirayama2011} is discussed in Section \ref{SuperGenOfPreviousPaper}. This strategy also extends and generalizes the idea introduced in \cite{MartinoScaffidi26}.}
\newline
{ In the remainder of this work, Sections \ref{SectNCE_RLR}, \ref{NCE_BridgeSect}, and \ref{IS_sect} are devoted to estimating a specific value of the partition function, denoted by $Z$, for a generic fixed value of $\btheta$ (one can consider $\btheta=\btheta_{\texttt{tr}}$). Accordingly, in some cases, we omit the dependence on $\btheta$ in the notation for simplicity. By contrast, Sections \ref{NCEsect}, \ref{NCEmixModelsSect}, \ref{NovelSect}, and \ref{BClassSect} address the joint estimation of $\btheta_{\texttt{tr}}$ and $Z_{\texttt{tr}}=Z(\btheta_{\texttt{tr}})$. More generally, these sections consider optimization over the extended parameter space $(\btheta,Z)$, treating $Z$ as a free optimization parameter rather than explicitly constraining it to be equal to $Z(\btheta)$ during the optimization procedure. }
 The Matlab code related to the experiments is also provided.\footnote{The code is publicly available at \url{http://www.lucamartino.altervista.org/PUBLIC_CODE_NCE_BRIDGE.zip}.}



%


\section{Preliminaries}\label{sec:EBMs0}

{
\subsection{Problem statement and main notation}
}

In this work, we mainly focus on the so-called energy-based models (EBMs).
Let us define $\phi(\by|\btheta)\geq0$  a function parametrized by a vector of parameters $\btheta$ taking values in $\bTheta \subseteq \mR^{d_\theta}$,
and $\y \in \mathcal{Y} \subseteq \mR^{d_y}$. We assume that $\phi(\by|\btheta)$ is analytically known and we can evaluate it.
An energy-based model is represented by the probability density function (pdf), 
\begin{equation}
\bar{\phi}(\y|\btheta) =\frac{\phi(\by|\btheta)}{Z(\btheta)} \propto \phi(\by|\btheta), 
\label{px|theta-Z}
\end{equation}
parametrized  by the vector $\btheta$. In many applications,  the following integral cannot be evaluated analytically:
\begin{equation}
Z(\btheta)=\int_{\cY} \phi(\by|\btheta) d \by\label{Z(theta)}.
\end{equation}
Namely, $Z(\btheta): \bTheta \to \mathbb{R}^+$ is  positive function that is unknown since the integral above cannot be solved analytically in closed form, i.e.,  is intractable.\footnote{We assume that $\by$ is a continuous vector, although several considerations are also valid  for the discrete case.} Hence, the normalizing constant $Z(\btheta)$, often called {\em partition function}, cannot be evaluated point-wise. For this reason, sometimes they are also known as {\it non-normalized models}. This represents a challenge for making inference on $\btheta$. Note that fixing $\btheta$,  $Z(\btheta)$ is a positive (unknown) normalizing constant. 
{We also assume that the model is identifiable, meaning that the parameterization $(\btheta,Z(\btheta))\mapsto\bar{\phi}(\y|\btheta)=\frac{\phi(\y|\btheta)}{Z(\btheta)}$ is injective. Namely, distinct parameter values must correspond to distinct normalized probability distributions.}
\newline
\newline
{\bf Observed data.} Let us assume that we have an observed dataset  $\y_{1:N}=\{\by_1, \ldots, \by_N \}\in \cY^N$, that contains i.i.d. realizations distributed as the  the EBM in Eq. \eqref{px|theta-Z} for a specific unknown vector of parameters $\btheta_{\texttt{tr}}$ (true vector of parameters), i.e., 
\begin{align}
\by_n \sim p_{\texttt{tr}}(\y)= \bar{\phi}(\y|\btheta_{\texttt{tr}})=\frac{\phi(\by|\btheta_{\texttt{tr}})}{Z(\btheta_{\texttt{tr}})}, \qquad n=1,...,N.
\end{align}
Note that $Z(\btheta_{\texttt{tr}})$ is a scalar normalizing constant, i.e., the true partition function evaluated at $\btheta_{\texttt{tr}}$. 
\newline
\newline
{\bf Goal.} Given the observed data $\y_{1:N}$,  the goal  is to infer the parameter vector $\boldsymbol{\theta}_{\texttt{tr}}$ and the scalar value $Z_{\texttt{tr}}=Z(\btheta_{\texttt{tr}})$, or related to other generic $\btheta$, i.e., $Z=Z(\btheta)$.
For this reason,  in many sections, we will simplify the notation as 
\begin{align}
\bar{\phi}(\y)=\bar{\phi}(\y|\btheta_{\texttt{tr}}), \quad \phi(\by)=\phi(\by|\btheta_{\texttt{tr}}), \quad Z_{\texttt{tr}}=Z(\btheta_{\texttt{tr}}).
\end{align}

{
\subsection{The importance of estimating the partition function $Z(\btheta)$}\label{IMP_Z_sect}

The estimation of the partition function \(Z(\btheta)\), i.e., obtaining an approximation $\widehat{Z}(\btheta)$ for each $\btheta$, is important for several reasons, both from a theoretical and a practical perspective. For instance, in statistical physics, the partition function is directly related to the free energy and other thermodynamic quantities.  Below, we list some statistical reasons:
\begin{itemize}
\item {\bf Frequentist inference.} Once an approximation $\widehat{Z}(\btheta)$ of the partition function is available, we can approximate the likelihood function of an EBM and, consequently, perform standard likelihood-based frequentist inference. This includes parameter estimation $\widehat{\btheta}_{\texttt{tr}}$ (e.g., maximum likelihood estimation), hypothesis testing, confidence interval construction, and likelihood-based model comparison (which involves the values $Z_{\texttt{tr}}$ of each model). Furthermore, if the same EBM is used in different inference tasks involving different datasets (i.e. different observed data), a reusable estimate of $Z(\btheta)$ can facilitate likelihood evaluation and reduce the computational cost of subsequent analyses (avoiding the estimation of the partition function from scratch for each new dataset, since $Z({\bm \theta})$ does {\em not} depend on the observed data).
\item {\bf Bayesian inference.}  An approximation $\widehat{Z}(\btheta)$ allows the possible application of a Bayesian analysis. Indeed, in this case, we can approximate the complete likelihood is given by 
\begin{equation}\label{likeEq_otraVEzBaysian}
 p(\by_1,\by_2,\ldots, \by_N|\btheta) \approx \widehat{L}(\y_{1:N}|\btheta)=
\frac{1}{\widehat{Z}(\btheta)^N} \prod_{n=1}^N \phi(\by_n|\btheta).
\end{equation}
 This is particularly useful for allowing Bayesian inference  (even considering different sets of observed data $\y_{1:N}$) where, assuming a prior distribution $p(\btheta)$, a posterior density is analyzed:  
\begin{align}
p(\btheta|\y_{1:N}) &\approx\frac{1}{\widehat{p}(\y_{1:N})}\widehat{L}(\y_{1:N}|\btheta) p(\btheta) \propto \widehat{L}(\y_{1:N}|\btheta) p(\btheta).
\end{align}
{\bf Marginal likelihood and double intractability.} A second normalizing constant arises when considering Bayesian inference based on an EBM likelihood. In particular,  the approximated {\it marginal likelihood} (also  a.k.a. {\it Bayesian evidence}) is given by:
\[
\overline{Z}
= \widehat{p}(\y_{1:N}) =\int_{\Theta} \widehat{L}(\y_{1:N}\mid\btheta)p(\btheta)\,d\btheta,
\]
 This second normalizing constant can also be estimated by slightly  adapting the methods discussed in this work. The key idea is to consider the parameter space as the sampling space, use the posterior distribution \(p(\btheta|\y_{1:N})\) as the target density, and generate samples of \(\btheta\) (rather than artificial data points) from suitable proposal distributions.
\newline
The marginal likelihood is particularly important for Bayesian model selection, since it provides the basis for computing Bayes factors and comparing competing statistical models. However, when the likelihood itself contains an unknown partition function and the marginal likelihood must also be evaluated, two levels of normalization are involved. More precisely, the likelihood requires the evaluation of \(Z(\btheta)\), whereas the posterior distribution requires the evaluation of the marginal likelihood \(\overline{Z}\). This nested structure gives rise to the so-called {\it doubly intractable} posterior distributions \cite{caimo2015efficient,liang2010double,llorenteREV2,Sonobe2026Contrastive}.
\end{itemize}
 The accuracy of the resulting inferential procedures will, of course, depend on the accuracy and uncertainty of the partition-function estimate $\widehat{Z}(\btheta)$.
}


\section{Noise contrastive estimation (NCE) }\label{NCEsect}

In this section, we present one of the most prominent methods for performing inference in EBMs, i.e., the noise-contrastive estimation (NCE).
NCE is a contrastive learning (CL)  approach applied in EBMs. The inference is driven by comparing samples from the observed data distribution against samples from a reference/noise distribution.
More specifically, the idea in NCE is to learn ${\bm \theta}$, and a pointwise estimation of $Z(\btheta)$, by designing a suitable binary classification problem.
Let us define a  generic input vector ${\bf u}\in\mathbb{R}^{d}$ and a binary label $a\in \{0,1\}$, more specifically, $\y_n \sim p({\bf u}|a=1)$ and ${\bf x}_m \sim p({\bf u}|a=0)$, 
where ${\bf x}_m \in \mathcal{Y} \subseteq \mR^{d_y}$, i.e., each $\y_n$ and ${\bf x}_m$ live in the same space.
This framework  can be rewritten as
$$
\y_n \sim \post({\bf u}|{\bm \theta}_{\texttt{tr}})=\frac{\phi({\bf u}, {\bm \theta}_{\texttt{tr}})}{Z({\bm \theta}_{\texttt{tr}})}, \qquad n=1,...,N,
$$
and
$$
{\bf x}_m \sim q({\bf u}), \qquad m=1,...,M,
$$
i.e., $p({\bf u}|a=1)=\post({\bf u}|{\bm \theta}_{\texttt{tr}})$  and again $ p({\bf u}|a=0)=q({\bf u})$ is a density chosen by the user.\footnote{We assume that $q$ is normalized (i.e., $\int_{\mathcal{Y}}q(\y) d\y=1$)} Thus, we have  $M+N$ labelled inputs ${\bf u}_i$, i.e.,  $\{{\bf u}_i,a_i\}_{i=1}^{M+N}$, set as
\begin{align}\label{EqU}
\underbrace{{\bf u}_1={\bf y}_1, \ldots, {\bf u}_N={\bf y}_N}_{a=1},\underbrace{ {\bf u}_{N+1}={\bf x}_1, \ldots, {\bf u}_{N+M}={\bf x}_{M},}_{a=0}. 
\end{align}
Namely, the first $N$ inputs are labelled with $a=1$, and the rest $M$ inputs are labelled with $a=0$. In the CL context, the samples ${\bf x}_1,...,{\bf x}_M$ are usually called reference/noise data and $q$ is often {referred to} as {\it reference density}. In this work, we will call it  {\it proposal density}, to clarify the link with the importance sampling framework. 
 \newline
\newline
 Thus, we can consider a binary classification problem with the entire dataset  $\{{\bf u}_i,a_i\}_{i=1}^{M+N}$, formed by the union of the two sets of vectors of $\y$'s and ${\bf x}$'s.  Then, we can apply a binary classifier in order to  estimate the unknown variables ${\bm \theta}_{\texttt{tr}}$ and $Z({\bm \theta}_{\texttt{tr}})$, comparing the two sets of data. The marginal (prior) probabilities of the labels can be approximated as $p(a=1)\approx \alpha_1=\frac{N}{M+N}$, $p(a=0)\approx\alpha_2=\frac{M}{M+N}$.  Setting  $\nu=\frac{p(a=0)}{p(a=1)}\approx \frac{M}{N} $ and ${\bm \xi}=[{\bm \theta},Z]$, the posterior probabilities are
 \begin{align}
p(a=1|{\bf u})=p(a=1|{\bf u},{\bm \xi})=\eta({\bf u},{\bm \xi})&=\eta({\bf u},{\bm \theta},Z), \nonumber\\
&=\frac{p({\bf u}|a=1)p(a=1)}{p({\bf u}|a=1)p(a=1)+p({\bf u}|a=0)p(a=0)}, \nonumber\\
&=\frac{\post({\bf u}|{\bm \theta})}{\post({\bf u}|{\bm \theta})+\nu q({\bf u})},\\
&=\frac{\phi({\bf u}, {\bm \theta})}{\phi({\bf u}, {\bm \theta})+\nu Z({\bm \theta}) q({\bf u})},
 \end{align}
 Clearly, we also have $p(a=0|{\bf u})=1-\eta({\bf u},{\bm \theta},Z)$. Note that $\eta$ depends on the analytic form of $\phi$ and $q$ and on the unknown values of ${\bm \theta}$ and $Z({\bm \theta})$, i.e., the parameter vector  ${\bm \xi}=[{\bm \theta},Z]$.
 Note that here we are considering a generic vector ${\bm \theta}$ and a generic function $Z({\bm \theta})$.
\newline
\newline
Moreover, a Bernoulli model can be considered with parameter $p(a=1|{\bf u})=\eta({\bf u},{\bm \theta},Z)$ and build a likelihood function $L({\bm \xi})=L({\bf y}_{1:N},{\bf x}_{1:M}|{\bm \xi})$ (according to the data) exactly as in a logistic regression {, i.e.,
\begin{align}
L({\bm \xi})=L({\bm \theta},Z)& = \prod_{n=1}^N p(a=1|{\bf y}_n)   \prod_{m=1}^M p(a=0|{\bf x}_m), \nonumber \\
&=\prod_{n=1}^N \eta\left({\bf y}_n, {\bm \theta},Z\right) \prod_{m=1}^M (1- \eta\left({\bf x}_m, {\bm \theta},Z\right)).
\end{align}}
Thus, the corresponding negative log-likelihood function $J_{\texttt{NCE}}\big({\bm \xi}\big)=-\log L({\bm \xi})$ is:
\begin{gather}\label{Eq_Neg_log_BERNOULLI}
\left\{
\begin{split}
&J_{\texttt{NCE}}\big({\bm \xi}\big)=-\sum_{n=1}^{N} \log\left(\eta\left({\bf y}_n, {\bm \theta},Z\right)\right) -\sum_{m=1}^{M}\log\left(1- \eta\left({\bf x}_m, {\bm \theta},Z\right)\right), \quad \mbox{ with } \\
&\eta({\bf u},{\bm \theta},Z)=\frac{\post({\bf u}|{\bm \theta})}{\post({\bf u}|{\bm \theta})+\nu q({\bf u})}, \qquad 1-\eta({\bf u},{\bm \theta},Z)= \frac{\nu q({\bf u})}{\post({\bf u}|{\bm \theta})+\nu q({\bf u})}.
\end{split}
\right.
\end{gather}
Recalling $\nu=\dfrac{M}{N}$, the final cost function to minimize is
\begin{align}
J_{\texttt{NCE}}({\bm \theta},Z)&=-\sum_{n=1}^{N} \log\left[\frac{\post({\bf y}_n|{\bm \theta})}{\post({\bf y}_n|{\bm \theta})+\nu q({\bf y}_n)}\right] -\sum_{m=1}^{M}\log \left[\frac{\nu q({\bf x}_m)}{\post({\bf x}_m|{\bm \theta})+\nu q({\bf x}_m)}\right],  \label{CLeqFin0}\\
&=-\sum_{n=1}^{N} \log\left[\frac{\phi({\bf y}_n, {\bm \theta})}{\phi({\bf y}_n, {\bm \theta})+\nu Z({\bm \theta}) q({\bf y}_n)} \right] -\sum_{m=1}^{M}\log \left[\frac{\nu Z({\bm \theta}) q({\bf x}_m)}{\phi({\bf x}_m,{\bm \theta})+\nu Z({\bm \theta}) q({\bf x}_m)}\right]. \label{CLeqFin}
\end{align}
We can minimize $J_{\texttt{NCE}}({\bm \theta},Z)$ with respect to ${\bm \theta}$ and $Z$, i.e.,
\begin{align}
[\widehat{{\bm \theta}}_{\texttt{NCE}},\widehat{Z}_{\texttt{NCE}}]=\arg\min J_{\texttt{NCE}}({\bm \theta},Z),
\end{align}
where $\widehat{{\bm \theta}}_{\texttt{NCE}}\longrightarrow{\bm \theta}_{\texttt{tr}}$  and
\begin{align}
\widehat{Z}_{\texttt{NCE}} \longrightarrow  Z_{\texttt{tr}}=Z({\bm \theta}_{\texttt{tr}}),
\end{align}
is a scalar value, that is  the approximation of function $Z({\bm \theta})$ in one specific point, ${\bm \theta}_{\texttt{tr}}$.  
For considerations about the optimality of proposal/reference density in NCE see \cite{Chehab2022OptimalNoise,Chehab2023OptimizingNoise}.  
{Recall that we have also assumed that the model is identifiable.}

{\Remark
Note that, within the objective function $J_{\texttt{NCE}}({\bm \theta},Z)$, the variable $Z$ is treated as an auxiliary additional optimization variable independent on ${\bm \theta}$. Namely, $Z$ represents an additional dimension in the optimization problem, independent to ${\bm \theta}$. Consequently, NCE avoids the direct evaluation and differentiation of the partition function $Z({\bm \theta})$, which is one of its main computational advantages.
}

\section{{  Handling multiple unnormalized models}} 

{ 
\subsection{From NCE to reverse logistic regression}\label{SectNCE_RLR}

As first noted in \cite[Section 3.2]{Chehab2023OptimizingNoise}, the NCE objective function $J_{\mathrm{NCE}}(\boldsymbol{\theta},Z)$ can be connected to the so-called {\it reverse} logistic regression (RLR) method introduced by \cite{geyer1994estimating}. This connection arises under the following conditions:
\begin{itemize}
\item The objective function $J_{\mathrm{NCE}}(\boldsymbol{\theta},Z)$ is considered as a function of $Z$ alone, with $\boldsymbol{\theta}$ held fixed (e.g., at $\boldsymbol{\theta}=\boldsymbol{\theta}_{\texttt{tr}}$).
\item We consider a collection of $K$ unnormalized models, $\phi_k$, for $k=1,\ldots,K$, each associated with its own normalizing constant. This can seen as an extension of the standard NCE framework.
\item  RLR considers observations drawn from the target distributions $\phi_k$. No additional artificial noise data are required (as in NCE).
\end{itemize}
{\bf Joint estimation of several normalized constants.} Under these conditions, the resulting objective function can be interpreted as the RLR criterion. We emphasize that the name reverse logistic regression may be somewhat misleading: RLR is primarily a method for estimating multiple normalizing constants (or, more generally, ratios of normalizing constants), rather than a conventional logistic regression technique. The method uses a criterion with a logistic-regression structure, adapted to the estimation of these normalizing constants (this is the reason of the term ``reverse''). In what follows, we provide a detailed derivation to clarify this connection.}
We can rewrite Eq. \eqref{CLeqFin} as
$$
\begin{aligned}
J_{\mathrm{NCE}}(\boldsymbol{\theta},Z) & =-\sum_{n=1}^N \log \left[\frac{N \post\left(\mathbf{y}_n |  \boldsymbol{\theta}\right)}{N \post\left(\mathbf{y}_n |  \boldsymbol{\theta}\right)+M q\left(\mathbf{y}_n\right)}\right]-\sum_{m=1}^M \log \left[\frac{M q\left(\mathbf{x}_m\right)}{N \post\left(\mathbf{x}_m |  \boldsymbol{\theta}\right)+M q\left(\mathbf{x}_m\right)}\right],\\
& =-\sum_{n=1}^N \log \left[\frac{\alpha_1 \post\left(\mathbf{y}_n |  \boldsymbol{\theta}\right)}{\alpha_1 \post\left(\mathbf{y}_n |  \boldsymbol{\theta}\right)+\alpha_2 q\left(\mathbf{y}_n\right)}\right]-\sum_{m=1}^M \log \left[\frac{\alpha_2 q\left({\bf x}_m\right)}{\alpha_1 \post\left({\bf x}_m |  \boldsymbol{\theta}\right)+\alpha_2 q\left({\bf x}_m\right)}\right], 
\end{aligned}
$$
where we have multiplied numerators and denominators of the fractions (inside the log) by $\frac{1}{M+N}$, and we have also defined 
$$
\alpha_1=\frac{N}{M+N} \quad \mbox{ and } \quad \alpha_2=\frac{M}{M+N}.
$$
Note that $\alpha_1+\alpha_2=1$. Furthermore, using the property $\log (a b)=\log (a)+\log (b)$, we obtain:
$$
\small
\begin{aligned}
J_{\mathrm{NCE}}(\boldsymbol{\theta},Z) & =-\sum_{n=1}^N \log \left[\frac{\post\left(\mathbf{y}_n |  \boldsymbol{\theta}\right)}{\alpha_1 \post\left(\mathbf{y}_n |  \boldsymbol{\theta}\right)+\alpha_2 q\left(\mathbf{y}_n\right)}\right]-\sum_{m=1}^M \log \left[\frac{q\left(\mathbf{x}_m\right)}{\alpha_1 \post\left(\mathbf{x}_m |  \boldsymbol{\theta}\right)+\alpha_2 q\left(\mathbf{x}_m\right)}\right]-N \log \alpha_1-M \log \alpha_2, \\
& =-\sum_{n=1}^N \log \left[\frac{\post\left(\mathbf{y}_n |  \boldsymbol{\theta}\right)}{\alpha_1 \post\left(\mathbf{y}_n |  \boldsymbol{\theta}\right)+\alpha_2 q\left(\mathbf{y}_n\right)}\right]-\sum_{m=1}^M \log \left[\frac{q\left(\mathbf{x}_m\right)}{\alpha_1 \post\left(\mathbf{x}_m |  \boldsymbol{\theta}\right)+\alpha_2 q\left(\mathbf{x}_m\right)}\right]+C_0.
\end{aligned}
$$ 
Taking the minus-expectation of the last expression above,  we finally have: 
\begin{align}
\exp\left(-J_{\mathrm{NCE}}(\boldsymbol{\theta},Z)\right)& \propto\prod_{n=1}^N \frac{\post\left(\mathbf{y}_n |  \boldsymbol{\theta}\right)}{\alpha_1 \post\left(\mathbf{y}_n |  \boldsymbol{\theta}\right)+\alpha_2 q\left(\mathbf{y}_n\right)} \prod_{m=1}^M \frac{q\left(\mathbf{x}_m\right)}{\alpha_1 \post\left(\mathbf{x}_m |  \boldsymbol{\theta}\right)+\alpha_2 q\left(\mathbf{x}_m\right)}, \nonumber \\
 &\propto\prod_{n=1}^N  \frac{\frac{\phi\left(\mathbf{y}_n |  \boldsymbol{\theta}\right)}{Z({\bm \theta})}}{\alpha_1 \frac{\phi\left(\mathbf{y}_n |  \boldsymbol{\theta}\right)}{Z({\bm \theta})} +\alpha_2 q\left(\mathbf{y}_n\right)} \prod_{m=1}^M  \frac{q\left(\mathbf{x}_m\right)}{\alpha_1 \frac{\phi\left(\mathbf{x}_m |  \boldsymbol{\theta}\right)}{Z({\bm \theta})}+\alpha_2 q\left(\mathbf{x}_m\right)}.
\end{align}
We now fix $\boldsymbol{\theta}=\boldsymbol{\theta}_{\mathrm{tr}}$ and focus on the computation of $Z=Z\left(\boldsymbol{\theta}_{\mathrm{tr}}\right)$. The resulting (pseudo-)likelihood can be written as
\begin{align}
L(Z)&=\exp \left(-J_{\mathrm{NCE}}(Z)\right), \nonumber Ê\\
&\propto \prod_{n=1}^N \frac{\frac{\phi\left(\mathbf{y}_n\right)}{ Z}}{\alpha_1 \frac{\phi\left(\mathbf{y}_n\right)}{ Z}+\alpha_2 q\left(\mathbf{y}_n\right)} \prod_{m=1}^M \frac{q\left(\mathbf{x}_m\right)}{\alpha_1 \frac{\phi\left(\mathbf{x}_m\right)}{Z}+\alpha_2 q\left(\mathbf{x}_m\right)} .
\end{align}
Here, $L(Z)=L\left(\mathbf{y}_{1: N}, \mathbf{x}_{1: M} | Z\right)$ denotes a (pseudo-)likelihood function used to obtain an estimate $\widehat{Z}$ of $Z$ by maximization. This likelihood can be obtained knowing that $\mathbf{y}_{1: N} \sim \post({\bf y})$, $\mathbf{x}_{1: M} \sim q({\bf y})$  are data generated from, respectively, a first and second component of the mixture, 
\begin{align}
q_{\texttt{mix}}(\y)&=\alpha_1 \bar{\phi}(\y)+\alpha_2 q(\y)=\alpha_1 \bar{\phi}(\y)+\alpha_2 q(\y),
\end{align}
that is the denominator of the ratios above. {This approach can be also interpreted as the application of the {\it reverse logistic regression} (RLR) method with $K=2$ models $\bar{\phi}_1(\y)=\bar{\phi}(\y)$ and $\bar{\phi}_2(\y)=q(\y)$\cite{geyer1994estimating,chen1997monte,cameron2014recursive}.} This is equivalent to the NCE {considering only the estimation of $Z_1=Z$ (since the second model $\phi_2({\bf y})=q({\bf y})$ is normalized chosen by the user, hence $Z_2=1$). Above, we have considered only one unnormalized model $\phi$}, 
but the RLR scheme was proposed in a more generic scenario with more than one normalizing constant to estimate {(and without using artificial data):} let $\left\{\post_k(\y)\right\}_{k=1}^K$ be a collection of nonnegative functions on a common space $\mathcal{Y}$, and define the corresponding normalized densities
$$
\post_k({\bf y})=\frac{\phi_k({\bf y})}{Z_k}, \quad Z_k=\int_{\mathcal{Y}} \phi_k(\y) d \y.
$$
where the normalizing constants $Z_k$ are unknown.
Assuming that,  we have access to different sets of samples 
\begin{align}\label{WeknowLabelEq}
\fbox{$\displaystyle \y_{k, 1}, \ldots, \y_{k, N_k} \sim \post_k({\bf y}),$}
\end{align}
 for each $k=1, \ldots, K$, the objective of RLR is to estimate $\log Z_k$ or $Z_k$ values up to an additive constant. RLR considers the conditional probability of $a=k$ given $\y$ as
\begin{align}
\fbox{$\displaystyle p(a=k |\y_{k,n},Z_{1:K})=\frac{N_k \phi_k(\y_{k,n})/Z_k}{\sum_{j=1}^K N_j \phi_j(\y_{k,n})/Z_j}=\frac{\beta_k \phi_k(\y_{k,n})/Z_k}{\sum_{j=1}^K \beta_j \phi_j(\y_{k,n})/Z_j}$,}
\end{align}
{ where $\beta_k=\frac{N_k}{\sum_{j=1}^K N_j}$. These conditional probabilities for different $k$ form a multinomial logistic regression model (i.e., a classification problem with $K$ classes),} where the parameters $\left\{Z_k\right\}$ (that can expressed as $Z_k=e^{\lambda_k}$, if desired) play the role of regression coefficients. The parameters  $Z_k$ (or $\lambda_k$)
are estimated by maximizing the (pseudo-)likelihood function:\footnote{{It is a pseudo-likelihood rather than an ordinary likelihood because, in RLR, the labels are already known by construction: each sample is drawn from its respective density, corresponding to one of the components of the mixture. The samples are then pooled together, as if placed in the same urn, and reinterpreted as if their labels had been generated by the hypothetical mixture-classification mechanism, as in standard NCE.}}
{\begin{align}
\fbox{$\displaystyle
L(Z_{1:K})=\prod_{k=1}^K  \prod_{n=1}^{N_k} p(a=k |\y_{k,n},Z_{1:K})
.$}
\end{align}}
{Identifiability is ensured by fixing one parameter, typically, e.g., one $Z_k=1$ for some $k$, otherwise only ratios of normalizing constants can be estimated. }
{
{\Remark RLR could  not require artificial data generation $\x_1,...,\x_M \sim q({\bf y})$ (as derived above),  because generally RLR estimates ratios of normalizing constants, e.g., quantities such as $\frac{Z_k}{Z_1}$, with $k=2, \ldots, K$, not their absolute values, since the pseudo-likelihood is invariant to a common multiplicative scaling.}
}

{\Remark
Hence, when focusing exclusively on the estimation of $Z$ and setting $K=2$, with $\bar{\phi}_1(\mathbf{y})=\bar{\phi}(\mathbf{y})$, $\bar{\phi}_2(\mathbf{y})=q(\mathbf{y})$, and $Z_2=1$, we can conclude that the two methods, NCE and RLR, coincide in this setting.}
\newline
\newline
{ More generally, again focusing solely on the estimation of normalizing constants, RLR can be interpreted as an extension of the JCE approach to the case of multiple unnormalized models $\phi_k$, with $k=1,\ldots,K$, whose corresponding normalizing constants $Z_k$ are estimated simultaneously. }

{ 
\subsection{ NCE for a mixture of unnormalized models}\label{NCEmixModelsSect}

In \cite{Matsuda2019Estimation}, the authors extend  NCE for making inference in a mixture of unnormalized models (mixture-NCE).  They suppose to observe $N$ samples ${\bf y}_1, \cdots, {\bf y}_N$ from a finite mixture distribution of $K $ components,
$$
\post({\bf y} | \btheta_{1:K}, \beta_{1:K})=\sum_{k=1}^K \beta_k \ {\bar \phi}_k\left({\bf y} | \btheta_k\right) = \sum_{k=1}^K \beta_k \frac{\phi_k({\bf y}|\btheta_k)}{Z_k(\btheta_k)},
$$
where $\beta_k>0, and \sum_{k=1}^K \beta_k=1$.
We desire to learn jointly $\btheta_{1:K}=\left[\btheta_1, \cdots, \btheta_K\right]$, $\beta_{1:K}=\left[\beta_1, \cdots, \beta_K\right]$  and the normalization constants $Z_k\left(\btheta_k\right)$ of each component $\post_k\left({\bf y} | \btheta_k\right)$ (which are intractable). Additionally, we consider $M$ artificial data $\x_{1:M} \sim q(\y)$, hence overall we have $K+1$ classes. Recalling that $\alpha_1=\frac{N}{M+N}$ and  $\alpha_2=\frac{M}{M+N}$, to extend NCE to this framework, we have to write the  posterior $p(a=k | {\bf y},\btheta_{1:K}, \beta_{1:K})$, i.e.,  
$$
 p(a=k |  {\bf y}_n)= p(a=k |  {\bf y}_n,\btheta_{1:K}, \beta_{1:K},Z_{1:K})=\frac{\alpha_1 \beta_k \post_k\left({\bf y}_n | \btheta_k\right)}{\alpha_1\sum_{j=1}^K \beta_j \post_j\left(\y_n | \btheta_j\right)+\alpha_2 q(\y_n)},
$$
for $k=1, \cdots, K$ and
$$
{\small
 p(a=K+1|{\bf x}_m)= p(a=K+1 |  {\bf x}_m,\btheta_{1:K}, \beta_{1:K},Z_{1:K})=\frac{\alpha_2 q(\x_m)}{\alpha_1\sum_{j=1}^K \beta_j \post_j\left(\x_m | \btheta_j\right)+\alpha_2 q(\x_m)},}
$$ 
then we can maximize the likelihood function:
\begin{align}
\fbox{$\displaystyle
L(\btheta_{1:K}, \beta_{1:K},Z_{1:K})=\prod_{k=1}^K \left[ \prod_{n=1}^{N} p(a=k |  {\bf y}_n) \right] \prod_{m=1}^{M} p(a=K+1 |  {\bf x}_m).$}
\end{align}
With respect to  RLR, the authors in \cite{Matsuda2019Estimation} generalizes the RLR method in different important directions:
\begin{itemize}
\item {\bf Estimating also ${\bm \theta}_{1:K}$.} First, it jointly estimates the model parameters $\boldsymbol{\theta}_k$ and the normalizing constants $Z_k$, rather than focusing exclusively on the estimation of the latter. 
\item {\bf Unknown sample labels (estimation of $\beta_{1:K}$).} Second, it addresses finite mixture models in which the component labels associated with the observed samples are unknown, unlike in Eq. \eqref{WeknowLabelEq}, hence  requiring the simultaneous estimation of coefficients of the mixture $\beta_k$ as well (as in clustering problem). 
\item Another difference: the presence of artificial noise samples is another distinction with respect to RLR. Indeed,  RLR uses samples from the component distributions with known source labels, whereas mixture-NCE in \cite{Matsuda2019Estimation} uses also noise/artificial  samples for discrimination as in standard NCE.
\end{itemize}
Hence, the approach in \cite{Matsuda2019Estimation} extends the estimation setting considered in standard RLR, although the authors formulate their method explicitly as an extension of NCE.

}

\section{From NCE and RLR to bridge sampling}\label{NCE_BridgeSect}

In the next sections, we fix $\boldsymbol{\theta}=\boldsymbol{\theta}_{\mathrm{tr}}$ and use the simplified notation $
\bar{\phi}(\y)=\bar{\phi}(\y|\btheta_{\texttt{tr}})$, $\phi(\by)=\phi(\by|\btheta_{\texttt{tr}})$, and  $Z=Z(\btheta_{\texttt{tr}})$.
We first show how the optimal bridge sampling formula can be obtained by deriving the NCE cost function (or, equivalently, the negative log-likelihood of reverse logistic regression){, as firstly highlighted in \cite{Pihlaja2010,GutmannHirayama2011,Chehab2023OptimizingNoise}.} We then recall the standard derivation of bridge sampling.

\subsection{Equivalence to optimal bridge sampling}\label{ArriveToBridge}

Let consider the  log-likelihood, $
\log L(Z)=-J_{\mathrm{NCE}}(Z)$ or Eq. \eqref{CLeqFin}, i.e.,
\begin{align}
-J_{\mathrm{NCE}}(Z)
&= \sum_{n=1}^N \log  \frac{\phi(\y_n)}{\alpha_1 \phi(\y_n)+\alpha_2 Zq(\y_n)}+\sum_{m=1}^M \log  \frac{Z q({\bf x}_m)}{\alpha_1 \phi({\bf x}_m)+\alpha_2 Zq({\bf x}_m)}.
	\end{align}
	For minimizing $J_{\mathrm{NCE}}(Z)$, we can take the derivative with respect to $Z$ and equaling to zero. Using the following rules and properties, 
\begin{align*}
	\frac{d\log(\frac{c}{a+bZ})}{dZ}=-\frac{b}{a+bZ}, \qquad \qquad  
	\log\left(\frac{cZ}{a+bZ}\right)&= \log(cZ)-\log(a+bZ),
\end{align*}
and hence
$$
\frac{d\log(\frac{cZ}{a+bZ})}{dZ}	= \frac{1}{Z}-\frac{b}{a+bZ},
$$	
we can write:
\begin{align*}
\fbox{$\displaystyle\frac{d J_{\mathrm{NCE}}}{d Z}=+\sum_{n=1}^N \frac{\alpha_2 q(\y_n)}{\alpha_1 \phi(\y_n)-\alpha_2 Zq(\y_n)}-\sum_{m=1}^M\frac{1}{Z}+\sum_{m=1}^M\frac{\alpha_2 q({\bf x}_m)}{\alpha_1 \phi({\bf x}_m)+\alpha_2 q({\bf x}_m)Z} =0. $} 
\end{align*}
With some additional algebra, we obtain
\begin{align*}
\frac{d J_{\mathrm{NCE}}}{d Z}&= + \sum_{n=1}^N\frac{\alpha_2 q(\y_n)}{\alpha_1 \phi(\y_n)+\alpha_2 Zq(\y_n)}- \sum_{m=1}^M\frac{\alpha_1 \phi({\bf x}_m)+\cancel{\alpha_2 Zq({\bf x}_m)}-\cancel{\alpha_2Z q({\bf x}_m)}}{Z\left(\alpha_1 \phi({\bf x}_m)+\alpha_2 q({\bf x}_m)Z\right)} =0.
\end{align*}
so finally we get 
\begin{align}
&\sum_{n=1}^N\frac{\alpha_2 q(\y_n)}{\alpha_1 \phi(\y_n)+\alpha_2 Zq(\y_n)}-\sum_{m=1}^M\frac{\alpha_1 \phi({\bf x}_m)}{ Z\left(\alpha_1 \phi({\bf x}_m)+\alpha_2  Zq({\bf x}_m)\right)} =0, \\
&Z\sum_{n=1}^N\frac{\alpha_2 q(\y_n)}{\alpha_1 \phi(\y_n)+\alpha_2 Z q(\y_n)} = \sum_{m=1}^M\frac{\alpha_1 \phi({\bf x}_m)}{ \alpha_1 \phi({\bf x}_m)+\alpha_2  Zq({\bf x}_m)} .
\end{align}
The expression above can be rewritten as fixed-point equation:
\begin{align}\label{EqOBs}
\displaystyle Z=\frac{\alpha_1\sum\limits_{m=1}^M\dfrac{ \phi({\bf x}_m)}{ \alpha_1 \phi({\bf x}_m)+\alpha_2  Zq({\bf x}_m)} }{\alpha_2\sum\limits_{n=1}^N\dfrac{ q(\y_n)}{\alpha_1 \phi(\y_n)+\alpha_2 Zq(\y_n)}}, \quad  \y_{1:N} \sim \bar{\phi}(\y), \quad {\bf x}_{1:M}\sim q(\y),
	\end{align}
where $Z$ appears in the two sides of the equation.	 Recall that $\bar{\phi}(\y)=\frac{\phi(\y)}{Z}$ and  
$$
\frac{\alpha_1}{\alpha_2}=\frac{N}{M}=\frac{\frac{1}{M}}{\frac{1}{N}}.
$$
 However, this is a a fixed point equation on $Z$.
Thus, the expression above suggests the iterative procedure (with iteration index $t\in \mathbb{N}$) for obtaining an estimator $\widehat{Z}$:
\begin{align}\label{optimal_bridge_eq}
\fbox{$\displaystyle \widehat{Z}_{t+1}=\frac{\dfrac{1}{M}\sum\limits_{m=1}^M\dfrac{ \phi({\bf x}_m)}{ \alpha_1 \phi({\bf x}_m)+\alpha_2  \widehat{Z}_t q({\bf x}_m)} }{\dfrac{1}{N}\sum\limits_{n=1}^N\dfrac{ q(\y_n)}{\alpha_1 \phi(\y_n)+\alpha_2 \widehat{Z}_t q(\y_n)}}, \quad  \y_{n} \sim \bar{\phi}(\y), \quad {\bf x}_{m}\sim q(\y),$}
	\end{align}
that coincides exactly with iteration procedure of the {\bf optimal bridge sampling} \cite{meng1996simulating,Gelman98,llorenteREV,BENNETT1976245}.
 Considering the asymptotic case, i.e., $M\rightarrow \infty$, $N \rightarrow \infty$, the expression above represents  the Eq. \eqref{EqAsyBridge} below. {This connection has been noted in \cite{Chehab2023OptimizingNoise}, based on several NCE extensions provided in  \cite{Pihlaja2010,GutmannHirayama2011}. See also Section \ref{BClassSect}.} 	

{\Remark With respect to the estimation of the normalizing constant $Z$ (with $\theta=\theta_{\mathrm{tr}}$ fixed), the three methodologies, (a) NCE, (b) reverse logistic regression, and (c) optimal bridge sampling, coincide.}

{\Remark\label{Rem3} Note that, in this work, we are not assuming to be able to draw samples from the model $\post(\y)$. The $N$ samples 
$$
\y_1,...,\y_N \sim \post(\y),
$$
are the observed data.  Moreover, the posterior density $\post(\y)$ cannot be completely evaluated because the normalizing constant $Z$ is unknown. This difficulty is usually addressed by employing recursive procedures in most of the estimators discussed above. }
\newline
\newline
The considerations in Remark \ref{Rem3} are also relevant for the estimators described in Section \ref{IS_sect}.


\subsection{Classical derivation of bridge sampling}
Let us define with $b(\y)>0$ an arbitrary, positive, generic function defined on the support of $\bar{\phi}(\y)$ i.e., $\mathcal{Y}$. Moreover, $b(\y)$ must be such that $b(\y) q(\y)$ and $b(\y) \post(\y)$ are both integrable. Bridge sampling can be derived from the following identity \cite{meng1996simulating,llorenteREV}: 
\begin{align}
\frac{\int_\mathcal{Y} b(\y) \bar{\phi}(\y) q(\y) d\y}{\int_\mathcal{Y} b(\y) \bar{\phi}(\y) q(\y) d\y} =1,
\end{align}
that is true since numerator and denominator are the exactly the same integral. This integral can be expressed as expectation with respect to $q$, i.e., $\E_q[b(\y) \bar{\phi}(\y)]$, or as expectation with respect to $\bar{\phi}$, i.e. $\E_{\bar{\phi}}[b(\y) q(\y)]$, hence
\begin{align}\label{BridgeFirstId}
\frac{\E_q[b(\y) \bar{\phi}(\y)]}{\E_{\bar{\phi}}[b(\y) q(\y)]}&=\frac{\dfrac{1}{Z}\E_q[b(\y) \phi(\y)]}{\E_{\bar{\phi}}[b(\y) q(\y)]}=1. 
\end{align}
Then, we arrive to the main bridge sampling identity:
\begin{align}\label{MainBridgeIdEq}
\fbox{$\displaystyle\frac{\E_q[b(\y) \phi(\y)]}{\E_{\bar{\phi}}[b(\y) q(\y)]}=Z$}
\end{align}
It is possible to show that the choice
\begin{align}
b(\y) =\frac{1}{\alpha_1 {\bar {\phi}}(\y)+\alpha_2 q(\y)}=\frac{1}{\alpha_1 \frac{1}{Z} {\phi}(\y)+\alpha_2 q(\y)},
\end{align}
is optimal \cite{meng1996simulating,Gelman98,llorenteREV,BENNETT1976245}. It yields the optimal bridge sampling scheme,
\begin{align}\label{EqAsyBridge}
\frac{\E_q\left[ \frac{\phi(\y)}{\alpha_1 {\bar {\phi}}(\y)+\alpha_2 q(\y)}\right]}{\E_{\bar{\phi}}\left[ \frac{q(\y)}{\alpha_1 {\bar {\phi}}(\y)+\alpha_2 q(\y)}\right]}=Z,
\end{align}
by replacing the expectations above with empirical estimators as in Eq. \eqref{EqOBs}.  { An interesting alternative formulation of bridge sampling can be also found in literature and can be derived as following. Considering a positive function $\gamma(\y)$, let start now from the identity and express the integrals as different expectations:
\begin{align*}
\frac{\int_\mathcal{Y} \gamma(\y) \bar{\phi}(\y) d\y}{\int_\mathcal{Y} \gamma(\y) \bar{\phi}(\y) d\y} &=1,\Longrightarrow \frac{\frac{1}{Z} \mathbb{E}_q\left[\gamma(\y) \frac{\phi(\y)}{q(\y)}\right]}{\mathbb{E}_{\bar{\phi}}\left[ \gamma(\y)\right] } =1, \\
\frac{ \mathbb{E}_q\left[\gamma(\y) \frac{\phi(\y)}{q(\y)}\right]}{\mathbb{E}_{\bar{\phi}}\left[ \gamma(\y)\right] } &=Z.
\end{align*}
The relationship between $ \gamma(\y)$, $b(\y)$ is $b(\y)=\dfrac{\gamma(\y)}{q(\y)}$ and, hence,  $\gamma(\y)=b(\y)q(\y)$.
See also Section \ref{GenPihSect} and \cite{Chehab2023OptimizingNoise} for further details. 
}

{
\subsection{Bennett's acceptance-ratio (BAR) method}

 A related scheme is {\it Bennett's acceptance-ratio} (BAR) method Ê\cite{BENNETT1976245}: it was originally introduced in statistical mechanics to estimate free-energy differences between two thermodynamic states from samples generated under the corresponding distributions, without requiring the individual partition functions to be known Ê\cite{BENNETT1976245}. The BAR scheme can be interpreted as a particular instance of bridge sampling, employing an optimal bridge function to estimate the ratio of the two normalizing constants. The connection between the two methods has been established explicitly in \cite{Gelman98}.
\newline
{\bf Metropolis-Hastings (MH) interpretation.} An interesting question is about the reason of its name and where BAR involves an ``acceptance-ratio''. The name comes from an interpretation involving a MH algorithm on an augmented state space $[\btheta, t]$, where $t \in\{0,1\}$ indicates which of the two distributions is active \cite{MHluca,robert1999monte}. This interpretation is useful because it explains another way to obtain the bridge identities, and why they use samples from two distributions. The key idea is to construct a Markov chain that switches between two distributions \cite{Gelman98}. Let us consider two unnormalized densities, $
\phi_0(\btheta)$,  $\phi_1(\btheta)$
with unknown normalizing constants $Z_0$ and $Z_1$. The corresponding normalized distributions are:
$$
\post_0(\btheta)=\frac{\phi_0(\btheta)}{Z_0}, \quad \post_1(\btheta)=\frac{\phi_1(\btheta)}{Z_1} .
$$
Assume that we desire to estimate the ratio $r=Z_1 / Z_0$.
Let us introduce an additional discrete variable $t \in\{0,1\}$, which identifies the active distribution. Then we can define a joint target distribution on the augmented space $\Theta \times\{0,1\}$, namely:
$$
\bar{p}(\btheta, t)=\frac{\phi_t(\btheta)}{Z_0+Z_1},
$$
where $\phi_t(\btheta)=\phi_0(\btheta)$ if $t=0$ and $\phi_t(\btheta)=\phi_1(\btheta)$ if $t=1$. Note that $\bar{p}(\btheta, t)$ is normalized, indeed, $\sum_{t=0}^1 \int \bar{p}(\btheta, t) d \btheta=1$.
Its marginal probabilities for the discrete variable $t$ are:
$$
\operatorname{Pr}(t=0)=\frac{Z_0}{Z_0+Z_1}, \quad \operatorname{Pr}(t=1)=\frac{Z_1}{Z_0+Z_1}.
$$
Therefore, we also have:
$$
\frac{\operatorname{Pr}(t=1)}{\operatorname{Pr}(t=0)}=\frac{Z_1}{Z_0}=r .
$$
This is the key observation: estimating the ratio of normalizing constants is equivalent to estimating the ratio of the marginal probabilities $\frac{\operatorname{Pr}(t=1)}{\operatorname{Pr}(t=0)}$ of the two states in this augmented target distribution. The idea would be to run an MH chain  $(\btheta^{(k)},t^{(k)})$, where $k$ denotes the iteration index, and estimate the ratio  $\frac{\operatorname{Pr}(t=1)}{\operatorname{Pr}(t=0)}$ from the empirical frequencies of the two states. This is possible even if the normalizing constants are unknown, since the MH acceptance probabilities can be expressed in terms of the unnormalized densities, as shown below.
\newline
Let us consider to propose a switch between the two states (fixing $\btheta$, for simplicity). Suppose that, at the current state $(\btheta, t)$, we propose to switch the label $t$ while keeping $\btheta$ fixed: $(\btheta, 0) \longleftrightarrow(\btheta, 1)$.  Assume for simplicity that the use of symmetric proposal densities within MH (i.e., probabilities for the two directions are equal. Fixing $\btheta$, the MH acceptance probability \cite{MHluca,robert1999monte} for a proposed switch from state 0 to state 1 is:
$$
\alpha_{01}(\btheta)=\min \left\{1, \frac{\bar{p}(\btheta, 1)}{\bar{p}(\btheta, 0)}\right\} = \min \left\{1, \frac{\phi_1(\btheta)}{\phi_0(\btheta)}\right\},
$$
while the reverse switch has acceptance probability
$$
\alpha_{10}(\btheta)=\min \left\{1, \frac{\bar{p}(\btheta, 0)}{\bar{p}(\btheta, 1)}\right\}=\min \left\{1, \frac{\phi_0(\btheta)}{\phi_1(\btheta)}\right\} .
$$
Note that the unknown normalizing constants do not appear in these expressions. They cancel in the ratio of the augmented target densities.

{\Remark  After the burn-in period, once the chain has reached stationarity, the marginal probability of occupying each state is proportional to the corresponding normalizing constant.
}
\newline
Now, we can connect the acceptance probabilities to the bridge sampling identities. Recall that $a_{01}(\btheta)$ and $a_{10}(\btheta)$ denote the probabilities of accepting a switch in one of the two possible directions. For all $\btheta$, supposing that the switch proposals are symmetric, the acceptance rule above satisfies detailed balance condition \cite{MHluca,Martino2013,Storvik11}, so that (considering only  the probability of ``jumps'', to simplify the derivation):
$$
\phi_0(\btheta) a_{01}(\btheta)=\phi_1(\btheta) a_{10}(\btheta).
$$
Integrating both sides with respect to $\btheta$, i.e.,
$
\int \phi_0(\btheta) a_{01}(\btheta) d \btheta=\int \phi_1(\btheta) a_{10}(\btheta) d\btheta.
$
and writing the integrals as expectations under the normalized distributions, this yields:
$$
Z_0 \mathbb{E}_{\post_0}\left[a_{01}(\btheta)\right]=Z_1 \mathbb{E}_{\post_1}\left[a_{10}(\btheta)\right].
$$
Hence,
\begin{align}\label{MainBridgeIdEq2}
\fbox{$\displaystyle\frac{Z_1}{Z_0}=\frac{\mathbb{E}_{\post_0}\left[a_{01}(\btheta)\right]}{\mathbb{E}_{\post_1}\left[a_{10}(\btheta)\right]},$}
\end{align}
that recalls Eq. \eqref{MainBridgeIdEq}. This is the key connection: the ratio of normalizing constants equals a ratio of expected switching-acceptance probabilities \cite{Gelman98}.

{\Remark The expectations in Eq. \eqref{MainBridgeIdEq2} can be estimated using samples already drawn from $\post_0$ and $\post_1$: we do not need to run the augmented MH chain itself.}
 \newline
\newline 
Another suitable acceptance probability within MH, that still satisfies the balance conditions (i.e., $
\phi_0(\btheta) a_{01}(\btheta)=\phi_1(\btheta) a_{10}(\btheta)$)  is the well-known {\it Barker acceptance rule} \cite{MHluca,robert1999monte,Martino2013,Storvik11}. Using the Barker acceptance rule, we have:
$$
a_{01}(\btheta)=\frac{\phi_1(\btheta)}{\phi_0(\btheta)+\phi_1(\btheta)}, \quad a_{10}(\btheta)=\frac{\phi_0(\btheta)}{\phi_0(\btheta)+\phi_1(\btheta)} .
$$
Replacing them into Eq. \eqref{MainBridgeIdEq2}, we obtain:
\begin{align}\label{MainBridgeIdEq3}
\fbox{$\displaystyle
\frac{Z_1}{Z_0}=\frac{\mathbb{E}_{\post_0}\left[\frac{\phi_1(\btheta)}{\phi_0(\btheta)+\phi_1(\btheta)}\right]}{\mathbb{E}_{\post_1}\left[\frac{\phi_0(\btheta)}{\phi_0(\btheta)+\phi_1(\btheta)}\right]}.$}
\end{align}
Considering one normalized density as $\post_0(\btheta)$, i.e., 
$\post_0(\btheta)=\phi_0(\btheta)=q(\btheta)$ (chosen by the user) hence $Z_0=1$, and changing the notation of  $\post_1(\btheta)=\post(\btheta)$, $\phi_1(\btheta)=\phi(\btheta)$, $Z_1=Z$, then Eq. \eqref{MainBridgeIdEq3} above becomes exactly the other bridge identity in Eq. \eqref{EqAsyBridge}  (setting also $\alpha_1=\alpha_2$).

 }

\section{Related importance sampling (IS) estimators}\label{IS_sect}

\subsection{Samples from two densities}

In this section, we introduce other schemes for  estimating of $Z=Z({\bm \theta}_{\texttt{tr}})$ where  $\bar{q}(\y)$ {\it and} $\post(\y)$ are employed separately or jointly. We begin by describing estimators that leverage both densities jointly. In this setting, the model  $\post(\y)$  is also used as a proposal distribution.
Note that drawing $N$ samples from $\post(\y)$ and $M$ samples from $q(\y)$ is equivalent to sampling by a {\it deterministic} approach from the mixture \cite{OwenZhou2000,MIS2019},
 \begin{align*}
 q_{\texttt{mix}}(\y)&= \alpha_1 \post(\y)+\alpha_2 q(\y),  \\
 &= \alpha_1 \frac{1}{Z} \phi(\y)+\alpha_2 q(\y),  
 \end{align*}
 i.e., a single density defined as mixture of the two densities \cite{MIS2019,llorenteREV}. 
The first estimator is based on the following classical equality: 
\begin{align}
Z&=\int \phi(\y) d\y=\E_{q_{\texttt{mix}}}\left[ \frac{\phi(\y)}{q_{\texttt{mix}}(\y)}\right]=\int \frac{\phi(\y)}{q_{\texttt{mix}}(\y)} q_{\texttt{mix}}(\y) d\y. 
\end{align}
Hence, applying a deterministic mixture sampling approach from $q_{\texttt{mix}}(\y)$,
\begin{align}\label{EqU0}
\y_1,...,\y_N \sim \bar{\phi}(\y),  \quad  \bx_1,...,\bx_M \sim q(\y),
\end{align}
and denoting
\begin{align}\label{EqU2}
{\bf u}_1={\bf y}_1, \ldots, {\bf u}_N={\bf y}_N, \quad {\bf u}_{N+1}=\x_1, \ldots, {\bf u}_{N+M}=\x_{M}, 
\end{align}
we can consider ${\bf u}_i \sim q_{\texttt{mix}}({\bf u}_i)$ \cite{OwenZhou2000,MIS2019},
we have the IS estimator 
\begin{align}\label{MIX-1_0}
\displaystyle\widehat{Z}=\frac{1}{N+M} \sum_{i=1}^{N+M} \frac{\phi({\bf u}_i)}{q_{\texttt{mix}}({\bf u}_i)},
\end{align}
that can be rewritten expressed with a recursive procedure as in the bridge sampling:
\begin{align}\label{MIX-1}
\fbox{$\displaystyle\widehat{Z}_{t+1}=\frac{1}{N+M} \sum_{i=1}^{N+M} \frac{\widehat{Z}_t \phi({\bf u}_i)}{ \alpha_1 \phi({\bf u}_i)+\alpha_2 \widehat{Z}_t q({\bf u}_i)},\qquad \{{\bf u}_i\}=\left\{\y_n\right\} \cup  \left\{\x_m\right\}.$}
\end{align}
 We call it as {\bf MIS} estimator. Note that this estimator can be rewritten as
 \begin{align}\label{MIX-1_otro}
\widehat{Z}_{t+1}=\alpha_1\frac{1}{N} \sum_{n=1}^{N} \frac{\widehat{Z}_t \phi({\bf y}_n)}{ \alpha_1 \phi({\bf y}_n)+\alpha_2 \widehat{Z}_t q({\bf y}_n)}+\alpha_2\frac{1}{M} \sum_{m=1}^{M} \frac{\widehat{Z}_t \phi(\x_m)}{ \alpha_1 \phi(\x_m)+\alpha_2 \widehat{Z}_t q(\x_m)},
\end{align}
where the expression separates into two components (i.e., just divided in two disjoint parts), one involving ${\y_n}$ and the other ${\x_m}$, similarly to the bridge sampling estimator. However, in the bridge sampling, the estimator is given by the ratio of two sums. It is possible here to construct an alternative estimator that more closely mirrors that structure. Indeed, {\it estimating}  also the constant value $N+M$ in Eq. \eqref{MIX-1}, i.e.,
\begin{align}\label{NMeq}
N+M \approx \sum_{i=1}^{N+M} \frac{q({\bf u})}{q_{\texttt{mix}}({\bf u})},
\end{align}
since we assume that $q$ is normalized (i.e., $\int_{\mathcal{Y}}q(\y) d\y=1$), by using the previous IS arguments,
$$
\frac{1}{N+M}  \sum_{i=1}^{N+M} \frac{q({\bf u})}{q_{\texttt{mix}}({\bf u})} \approx 1.
$$
Replacing \eqref{NMeq} into Eq. \eqref{MIX-1},
\begin{align}\label{MIX-2}
\widehat{Z}_{t+1}&=\frac{1}{\sum_{k=1}^{N+M} \frac{q({\bf u}_k)}{q_{\texttt{mix}}({\bf u}_k)}} \sum_{i=1}^{N+M} \frac{\phi({\bf u}_i)}{q_{\texttt{mix}}({\bf u}_i)},\\
  &=\frac{\sum_{i=1}^{N+M} \frac{\phi({\bf u}_i)}{q_{\texttt{mix}}({\bf u}_i)}}{\sum_{k=1}^{N+M} \frac{q({\bf u}_k)}{q_{\texttt{mix}}({\bf u}_k)}} ,
\end{align}
and replacing inside the expression of the mixture $q_{\texttt{mix}}(\y)= \alpha_1 \post(\y)+\alpha_2 q(\y)$,  we obtain the iterative procedure:\footnote{Note that the two $\widehat{Z}_t$ terms that should appear in the numerators cancel each other out, as in the bridge sampling expression.}
\begin{align}\label{MIX-3}
\fbox{$\widehat{Z}_{t+1}=\dfrac{\sum\limits_{i=1}^{N+M} \dfrac{ \phi({\bf u}_i)}{\alpha_1  {\phi}({\bf u}_i)+\alpha_2 \widehat{Z}_{t} q({\bf u}_i)}}{\sum\limits_{k=1}^{N+M} \dfrac{ q({\bf u}_k)}{\alpha_1  {\phi}({\bf u}_k)+\alpha_2 \widehat{Z}_{t} q({\bf u}_k)}}, \qquad \{{\bf u}_i\}=\left\{\y_n\right\} \cup  \left\{\x_m\right\}.$}
\end{align}
The expression above is very similar to Eq. \eqref{optimal_bridge_eq} with the difference that both summations consider all the data $\{{\bf u}_i\}_{i=1}^{N+M}$ in Eq, \eqref{EqU2}, instead of just $\y_n$ or $\x_m$ in Eq. \eqref{EqU0}.  We name this estimator as {\bf Self-IS-with-mix}.

{\Remark In  \cite{meng1996simulating}, the authors assert that both estimators in Eqs.~\eqref{MIX-1}Ð\eqref{MIX-3} converge to the solution given by optimal bridge sampling estimator, expressed as \eqref{optimal_bridge_eq}. As demonstrated in the simulation study in  Section \ref{NumSect}, however, the convergence rates of the corresponding iterative methods differ depending also on the starting point. }  

{\Remark Within the EBM framework, the observed data $\{\y_n\}_{n=1}^N$ are assumed to be generated directly by the model itself; consequently, the issue of sampling from a posterior distribution in Bayesian inference, that is central in standard bridge sampling applications, does not arise here, i.e., in the frequentist inference for  EBMs.  }
\newline
\newline
After looking \eqref{MIX-1} and \eqref{MIX-3}, symmetrically we can find an estimator only involving the denominator of Eq. \eqref{MIX-3}. 
\begin{align}
\E_{q_{\texttt{mix}}}\left[ \frac{q({\bf u})}{\alpha_1 {\bar {\phi}}({\bf u})+\alpha_2 q({\bf u})}\right]=\E_{q_{\texttt{mix}}}\left[ \frac{Zq({\bf u})}{\alpha_1 \phi({\bf u})+\alpha_2 Zq({\bf u})}\right]&=1, \\
Z\E_{q_{\texttt{mix}}}\left[ \frac{q({\bf u})}{\alpha_1 \phi({\bf u})+\alpha_2 Zq({\bf u})}\right]&=1, \\
\E_{q_{\texttt{mix}}}\left[ \frac{q({\bf u})}{\alpha_1 \phi({\bf u})+\alpha_2 Zq({\bf u})}\right]&=\frac{1}{Z}, 
\end{align}   
The last expression suggests the estimator:
\begin{align}
 \widehat{Z}&\approx \frac{1}{\frac{1}{N+M} \sum_{i=1}^{N+M} \frac{q({\bf u}_i)}{\alpha_1 \phi({\bf u}_i)+\alpha_2  \widehat{Z}q({\bf u}_i)}} \\
 &\approx\frac{N+M}{ \sum_{i=1}^{N+M} \frac{q({\bf u}_i)}{\alpha_1 \phi({\bf u}_i)+\alpha_2  \widehat{Z}q({\bf u}_i)}}, \quad {\bf u}_i \sim q_{\texttt{mix}}({\bf u}),
\end{align} 
Thus, considering to use a deterministic mixture sampling approach,  the final iterative expression is:
\begin{align}\label{MIX-4}
\fbox{$\widehat{Z}_{t+1}=\dfrac{N+M}{\sum\limits_{k=1}^{N+M} \dfrac{ q({\bf u}_k)}{\alpha_1  {\phi}({\bf u}_k)+\alpha_2 \widehat{Z}_{t} q({\bf u}_k)}}, \qquad \{{\bf u}_i\}=\left\{\y_n\right\} \cup  \left\{\x_m\right\}.$}
\end{align}

\subsection{Samples from one density}
Considering only $q(\y)$ or only $\post(\y)$, we have the standard IS estimator and the reverse IS estimator, respectively \cite{llorenteREV,Neal2008harmonic}. The first one is derived from the following equality,
\begin{align}
\E_q\left[\frac{\phi(\y)}{q(\y)}\right]=\int_{\mathcal{Y}} \frac{\phi(\y)}{q(\y)} q(\y) d\y=\int_{\mathcal{Y}} \phi(\y) d\y=Z.
\end{align}
and the {\bf standard IS estimator (Stand-IS)} has the form:
\begin{align}\label{standISest_eq}
\fbox{$\displaystyle\widehat{Z}=\frac{1}{M} \sum_{m=1}^{M} \frac{\phi({\bf x}_m)}{q({\bf x}_m)},  \qquad  \bx_m \sim q(\y).$}
\end{align}
The reverse IS estimator is based on the following equality,
\begin{align*}
\\E_{\bar{\phi}}\left[\frac{q(\y)}{\post(\y)}\right]= \int_{\mathcal{Y}}\frac{q(\y)}{\post(\y)}\post(\y)  d\y&=1, \\
 Z\int_{\mathcal{Y}}\frac{q(\y)}{\phi(\y)}\post(\y)  d\y&=1, \\
Z \E_{\bar{\phi}}\left[\frac{q({\bf y})}{\phi({\bf y})}\right]&=1, \\
\E_{\bar{\phi}}\left[\frac{q({\bf y})}{\phi({\bf y})}\right]&=\frac{1}{Z}.
\end{align*}
where we have used the fact that $q(\y)$ is normalized, i.e., $\int_{\mathcal{Y}}q(\y) d\y=1$. Therefore,  the {\bf reverse IS (RIS) estimator} has the form:
\begin{align}\label{RISestEq}
\fbox{$\displaystyle\widehat{Z}=\left(\frac{1}{N} \sum_{n=1}^{N} \frac{q({\bf y}_n)}{\phi({\bf y}_n)}\right)^{-1},  \quad \y_n \sim \bar{\phi}(\y)=\frac{1}{Z}\phi(\y)$}, 
\end{align}
Note that the quantity $\widehat{A}=\frac{1}{N} \sum_{n=1}^{N} \frac{q({\bf y}_n)}{\phi({\bf y}_n)}$ is an unbiased estimate of $1/Z$, i.e., $\E[\widehat{A}]=1/Z$. However, by Jensen's inequality, we have $\E\left[\frac{1}{\widehat{A}}\right]\geq \frac{1}{\E[\widehat{A}]}=Z$. Hence,
the RIS  estimator is positively biased, i.e., overestimates $Z$. 
\newline
Both estimators above do not require recursion. Finally, another related estimator is the so-called optimal umbrella estimator \cite{torrie1977nonphysical,chen1997monte,llorenteREV}. In this case, we draw samples  from a single density
\begin{align}
{\bar r}(\y)&\propto r(\y)=\left|\post(\y) - q(\y)\right|, \\
&= \frac{1}{c} \left|\frac{1}{Z}\phi(\y) - q(\y)\right|,
\end{align}
where $c=\int_{\mathcal{Y}} \left|\frac{1}{Z}\phi(\y) - q(\y)\right| d\y$ is generally unknown and intractable. Hence, drawing $\widetilde{{\bf x}}_1,...,\widetilde{{\bf x}}_{M+N}$ samples from ${\bar d}(\y)$, we have 
\begin{align}
Z &=\frac{c}{(M+N)} \sum_{i=1}^{M+N}\frac{\phi(\widetilde{{\bf x}}_i)}{|\frac{1}{Z}\phi( \widetilde{{\bf x}}_i) - q(\widetilde{{\bf x}}_i)|}, \label{SallyEq0}  
\end{align}
and
\begin{align}
1&=\frac{c}{(M+N)} \sum_{i=1}^{M+N}\frac{q(\widetilde{{\bf x}}_i)}{|\frac{1}{Z}\phi( \widetilde{{\bf x}}_i) - q(\widetilde{{\bf x}}_i)|}, \\
\frac{1}{c}&=\frac{1}{M+N} \sum_{i=1}^{M+N}\frac{q(\widetilde{{\bf x}}_i)}{|\frac{1}{Z}\phi( \widetilde{{\bf x}}_i) - q(\widetilde{{\bf x}}_i)|}, \\
c&=\left(\frac{1}{M+N} \sum_{i=1}^{M+N}\frac{q(\widetilde{{\bf x}}_i)}{|\frac{1}{Z}\phi( \widetilde{{\bf x}}_i) - q(\widetilde{{\bf x}}_i)|}\right)^{-1}  \label{SallyEq1} 
\end{align}
where we have used again  that $q(\y)$ is normalized, i.e., $\int_{\mathcal{Y}}q(\y) d\y=1$.  Replacing the expression of $c$ in Eq. \eqref{SallyEq1} into \eqref{SallyEq0},
we obtain (after some simple algebra) the final fixed point and consequently recursive equation,
\begin{align}
\fbox{$\widehat{Z}_{t+1} =\dfrac{ \sum\limits_{i=1}^{M+N}\dfrac{\phi(\widetilde{{\bf x}}_i)}{|\phi( \widetilde{{\bf x}}_i) - \widehat{Z}_t q(\widetilde{{\bf x}}_i)|}}{\sum\limits_{k=1}^{M+N}\dfrac{ q(\widetilde{{\bf x}}_k)}{|\phi( \widetilde{{\bf x}}_k) - \widehat{Z}_t q(\widetilde{{\bf x}}_k)| }}, \qquad \widetilde{{\bf x}}_i \sim \bar{r}(\y).$} \label{SallyEq2}   
\end{align}	
that is the the optimal umbrella sampling estimator ({\bf Opt-Umb}) \cite{torrie1977nonphysical,chen1997monte,llorenteREV}. However, we need another Monte Carlo method to draw samples from $\bar{r}(\y)\propto \left|\post(\y) - q(\y)\right|$ (it is not a straightforward task).  See Table \ref{TableFF} for a summary of the described estimators. { Some interesting comments regarding the connections between extended NCE schemes and umbrella sampling is given in \cite{Chehab2023OptimizingNoise}.}

\section{{Alternative classification cost functions for EBM inference}}\label{NovelSect}

{
In this section, we introduce additional cost functions designed for classification,  which can also be used to estimate both \(\btheta_{\texttt{tr}}\) and \(Z(\btheta_{\texttt{tr}})\). Since we remain within the standard classification framework, and the proposed objectives are still functions of the class probabilities, \(Z\) can be treated as a free parameter during minimization, following the same arguments as in NCE. These extended cost functions may lead to alternative estimation schemes, some of which could present some degree of novelty with respect to the existing literature.
}




\subsection{MIS arguments in NCE}
Building on observations from previous works \cite{OwenZhou2000,MIS2019},  { and  similarly to the approach in \cite{Matsuda2019Estimation}}, one can argue that treating ${\bf u}_i = {\y_n} \cup {\x_m}$ jointly as samples drawn from the mixture distribution $q_{\texttt{mix}}({\bf u})$ may lead to improved performance. Thus, one could design a cost function of type: 
\begin{align}
&J_{\texttt{MIS}}(\btheta,Z)= \nonumber \\
&-\alpha_1\sum_{k=1}^{M+N} \log  \frac{\phi({\bf u}_k|\btheta)}{\alpha_1 \phi({\bf u}_k|\btheta)+\alpha_2 Zq({\bf u}_k)}-\alpha_2\sum_{k=1}^{M+N} \log  \frac{Z q({\bf u}_k)}{\alpha_1 \phi({\bf u}_k|\btheta)+\alpha_2 Zq({\bf u}_k)}. \label{EqJmis}
\end{align}
 { Note that, as in Section \ref{NCEmixModelsSect} and in \cite{Matsuda2019Estimation}, within $J_{\texttt{MIS}}(\btheta,Z)$, we ignore the sample labels.}

{\Remark Fixing $\btheta$, differentiating the above expression with respect to $Z$ and setting the result equal to zero yields the self-IS-with-mixture estimator given in Eq.~\eqref{MIX-4}.
}
\newline
\newline
Indeed, the cost function  can be rewritten as:
\[
J_{\texttt{MIS}}(\btheta,Z) =
-\alpha_1
\sum_{k=1}^{M+N}
\log\!\left(
\frac{\phi({\bf u}_k|\btheta)}
{Zq_{\texttt{mix}}({\bf u}_k)}
\right)
-\alpha_2
\sum_{k=1}^{M+N}
\log\!\left(
\frac{Zq({\bf u}_k)}
{Zq_{\texttt{mix}}({\bf u}_k)}
\right).
\]

Expanding the logarithms, we obtain
\begin{align}
J_{\texttt{MIS}}(\btheta,Z)
&=
-\alpha_1\sum_{k=1}^{M+N}\log \phi({\bf u}_k|\btheta)
-\alpha_2\sum_{k=1}^{M+N}\log Z \nonumber\\
&\quad
-\alpha_2\sum_{k=1}^{M+N}\log q({\bf u}_k)
+(\alpha_1+\alpha_2)\sum_{k=1}^{M+N}\log (Zq_{\texttt{mix}}({\bf u}_k)).
\end{align}

Since $\alpha_1+\alpha_2=1$ and $Zq_{\texttt{mix}}({\bf u}_k)=\alpha_1 \phi({\bf u}_k|\btheta)+\alpha_2 Z q({\bf u}_k)$, we can replace above:
\begin{align} 
J_{\texttt{MIS}}(\btheta,Z)
&=
-\alpha_1\sum_{k=1}^{M+N}\log \phi({\bf u}_k|\btheta)
-\alpha_2\sum_{k=1}^{M+N}\log Z \nonumber\\
&\quad
-\alpha_2\sum_{k=1}^{M+N}\log q({\bf u}_k)
+\sum_{k=1}^{M+N}\log (\alpha_1 \phi({\bf u}_k|\btheta)+\alpha_2 Z q({\bf u}_k)).
\end{align}
Differentiating with respect to $Z$ and equal to zero gives
\begin{align}
\frac{\partial J_{\texttt{MIS}}(\btheta,Z)}{\partial Z}
&=
-\frac{\alpha_2(M+N)}{Z}+
\sum_{k=1}^{M+N}
\frac{\partial}{\partial Z}\log (\alpha_1 \phi({\bf u}_k|\btheta)+\alpha_2 Z q({\bf u}_k))=0,
\nonumber\\
&=
-\frac{\alpha_2(M+N)}{Z}
+
\alpha_2\sum_{k=1}^{M+N}
\frac{ q({\bf u}_k)}
{\alpha_1\phi({\bf u}_k|\btheta)+\alpha_2 Z q({\bf u}_k)}=0.
\end{align}
Assuming $\alpha_2>0$, we obtain
\begin{equation}
\frac{M+N}{Z}
=
\sum_{k=1}^{M+N}
\frac{q({\bf u}_k)}
{\alpha_1\phi({\bf u}_k|\btheta)+\alpha_2 Z q({\bf u}_k)}.
\end{equation}
With additional simple algebra, we arrive to
\begin{equation}
\fbox{$\displaystyle Z =
\frac{M+N}{\displaystyle \sum_{k=1}^{M+N} \frac{q({\bf u}_k)}
{\alpha_1\phi({\bf u}_k|\btheta)+\alpha_2 Z q({\bf u}_k)}}$,}
\end{equation}
that is the fix point equation related to Eq. \eqref{MIX-4}. 
\newline
\newline
 Given the results on prior MIS works (e.g., \cite{MIS2019}), one could expect that $J_{\texttt{MIS}}(\btheta,Z)$ and Eq.~\eqref{MIX-3} provide better results in the estimation of $Z$. For the other side, in terms of binary classification, $J_{\texttt{MIS}}(\btheta,Z)$ is expected to perform worse than $J_{\texttt{NCE}}(\btheta,Z)$, at least for estimating $\btheta$. Indeed, $J_{\texttt{NCE}}(\btheta,Z)$ leverages class label information, whereas $J_{\texttt{MIS}}(\btheta,Z)$ does not. The numerical simulations in Section \ref{NumSect} partially support this intuition: the performance minimizing $J_{\texttt{MIS}}$ in the $\btheta$-space  depends strongly on the choice of the proposal parameters. While, under certain ideal conditions, minimizing $J_{\texttt{MIS}}$ in the $Z$-space provides the best performance. 

\subsection{Multiple proposal densities in NCE and bridge sampling}

All the considerations and connections given in the previous sections allow us to extend the optimal bridge sampling using multiple proposal densities. { The use of different reference distributions in NCE schemes has been analyzed in different works \cite[Section 4.2]{Matsuda2019Estimation} and \cite{ceylan2018conditional,ChehabHyvarinenRisteski2023,Olmin2024}.}
 Let us consider $K$ proposal densities $\{q_k(\y)\}_{k=1}^K$, and we draw $M$ samples from each of them, i.e.,
$$
\x_{k,1},...,\x_{k,M} \sim q_k(\y).
$$
We also recall that we have $N$ observed data from the model, i.e., $\y_1,...,\y_N \sim \post({\bf y}_n|{\bm \theta})$. Thus, similarly as in Section \ref{NCEsect}, we can design a classification problem {\it with $K+1$ classes}, with cost function:
\begin{align}
J({\bm \theta},Z)&=-\sum_{n=1}^{N} \log\left[\frac{N\post({\bf y}_n|{\bm \theta})}{N\post({\bf y}_n|{\bm \theta})+M\sum_{j=1}^K q_j({\bf y}_n)}\right]+ \nonumber \\
&-\sum_{k=1}^K\sum_{m=1}^{M}\log \left[\frac{M q_k({\bf x}_{k,m})}{N\post({\bf x}_{k,m}|{\bm \theta})+M \sum_{j=1}^K q_j({\bf x}_{k,m}) }\right],
 \label{CLeqFinMIS}
\end{align}
Deriving the expression above with respect to $Z$ as in Section \ref{ArriveToBridge}, we obtain:
\begin{align}\label{optimal_bridge_eq2}
\fbox{$\displaystyle \widehat{Z}_{t+1}=\dfrac{\dfrac{1}{M}\sum\limits_{k=1}^{K} \sum\limits_{m=1}^{M}\dfrac{N\phi({\bf x}_{k,m}|{\bm \theta})}{N\phi({\bf x}_{k,m}|{\bm \theta})+\widehat{Z}_t M \sum_{j=1}^{K}  q_j({\bf x}_{k,m})}
 }{\dfrac{1}{N}\sum\limits_{k=1}^{K}\sum\limits_{n=1}^N\dfrac{M_kq_k(\y_n)}{N \phi(\y_n|{\bm \theta})+ \widehat{Z}_t M\sum_{j=1}^{K}  q_j(\y_n)}},$}
\end{align}
This iterative procedure could be easily integrated into the NCE optimization through an alternating optimization scheme with respect to $\btheta$ and $Z$. The use of multiple proposal densities is particularly interesting for designing adaptive schemes, as suggested in \cite{Cappe04,bugallo2017adaptive}. Furthermore, the use of different proposal densities can be combined with the idea of including tempered models in bridge sampling to help the exploration of the state-space. However, in this case we have one than more unknown normalizing constants to be estimated as in RLR.

\subsection{{Different scoring rules} for binary classifiers}
We can consider other loss in the binary classification problem described in Section \ref{NCEsect}. Let us consider a positive, decreasing, convex function {$V(\eta)$ defined in $\eta\in[0,1]$ (recall that $\eta({\bf u})=p(a=1|{\bf u})$); see Section \ref{NCEsect})}, that is also a {\it strictly proper scoring rule} \cite{gneiting2007strictly}. { See also Appendix \ref{Vscore} for more specific definitions and conditions. In particular, in order to be a proper score, $V(\eta)$ must satisfy the condition:
\begin{align}
\boxed{\displaystyle \eta V^{\prime}(\eta)=(1-\eta) V^{\prime}(1-\eta),}
\end{align}
where $V^{\prime}(\eta)=\frac{dV}{d\eta}$, as shown in App. \ref{Vscore}.
}
The NCE procedure described above is also valid considering the cost function:
\begin{align}\label{GenNCEeq}
\fbox{$\displaystyle J\big({\bm \theta},Z\big)=\sum_{n=1}^{N} V\left(\eta\left({\bf y}_n, {\bm \theta},Z\right)\right) +\sum_{m=1}^{M}V\left(1- \eta\left({\bf x}_m, {\bm \theta},Z\right)\right), $}
\end{align}
that can be {minimized} with respect to ${\bm \xi}=[{\bm \theta},Z]$ {to obtain} an estimators of ${\bm \theta}_{\texttt{tr}}$ and $Z({\bm \theta}_{\texttt{tr}})$, since this is a solution of a binary classification problem. 
{Hence, minimizing \eqref{GenNCEeq}, we can treat $Z$ as a free parameter as in standard NCE.} 
Repeating the procedure done in Section \ref{ArriveToBridge}, we can derive the cost function above $J\big({\bm \theta},Z\big)$ with respect to $Z$,
\begin{align}\label{55Eq_imp}
\fbox{$\displaystyle \frac{\partial J}{\partial Z}=\sum_{n=1}^{N} \frac{d V}{d\eta} \dot{\eta}(\y_n,{\bm \theta},Z)- \sum_{m=1}^{M} \left.\frac{d V}{d\eta}\right|_{1-\eta} \dot{\eta}(\x_m,{\bm \theta},Z), $}
\end{align}
where we have denoted $\dot{\eta}=\frac{d\eta}{d Z}$ we have used $\frac{d V(1-\eta)}{d\eta}=-\left.\frac{d V(\eta)}{d\eta}\right|_{1-\eta}$. Recalling 
\begin{align}\label{EtaHereAgain}
\eta\left({\bf u}, {\bm \theta},Z\right)=\frac{\post\left({\bf u}|  \boldsymbol{\theta}\right)}{ \post\left(\cdot |  \boldsymbol{\theta}\right)+ \nu q\left({\bf u}\right)}= \frac{\phi\left({\bf u} |  \boldsymbol{\theta}\right)}{ \phi\left({\bf u}|  \boldsymbol{\theta}\right)+ \nu Z q\left({\bf u}\right)},
\end{align}
hence {we can also compute its first derivative with respect to $Z$:}
\begin{align}\label{DerEta}
\fbox{$\displaystyle  \dot{\eta}({\bf u},{\bm \theta},Z)=-\frac{\nu \phi\left({\bf u} |  \boldsymbol{\theta}\right)  q\left({\bf u}\right)}{(\phi\left({\bf u} |  \boldsymbol{\theta}\right)+ \nu Z q\left({\bf u}\right))^2}.$}
\end{align}
{Hence, minimizing the generic cost function in Eq. \eqref{GenNCEeq} yields estimators of both $\btheta$ and $Z$; in particular, $Z$ can considered in  the binary classification problem as free parameter, as in standard NCE. Moreover, by fixing $\btheta$ and setting Eq. \eqref{55Eq_imp} equal to zero, one can derive alternative estimators of $Z$ and, if necessary, develop iterative procedures to obtain these estimates, as discussed above.}


\subsubsection{Example 1 with a proper scoring rule}
Let us consider a proper scoring rule, $V(\eta)=(1-\eta)^2$. In this case,  we have
\begin{align}\label{EqVnow1_0}
\frac{d V(\eta)}{d\eta}=-2(1-\eta)
= -  \frac{2\nu Z q\left({\bf u}\right)}{\phi\left({\bf u}|  \boldsymbol{\theta}\right)+ \nu Z q\left({\bf u}\right)},
\end{align}
\begin{align}\label{EqVnow2_0}
\frac{d V(1-\eta)}{d\eta}=- \left.\frac{d V(\eta)}{d\eta}\right|_{1-\eta}=2\eta
=\frac{2\phi\left({\bf u}|  \boldsymbol{\theta}\right)}{\phi\left({\bf u}|  \boldsymbol{\theta}\right)+ \nu Z q\left({\bf u}\right)}.
\end{align}
where we have also used the  definition $\eta$ recalled in Eq. \eqref{EtaHereAgain}.
Replacing \eqref{EqVnow1_0}-\eqref{EqVnow2_0} and \eqref{DerEta} into Eq. \eqref{55Eq_imp}, we obtain:
\begin{align*}
2 \nu^2 Z^2\sum_{n=1}^N  \frac{\phi\left({\bf y}_n |  \boldsymbol{\theta}\right) q\left({\bf y}_n\right)^2}{(\phi\left({\bf y}_n|  \boldsymbol{\theta}\right)+ \nu Z q\left({\bf y}_n\right))^3} -2  \nu Z \sum_{m=1}^M \frac{\phi\left({\bf x}_m |  \boldsymbol{\theta}\right)^2 q\left({\bf x}_m\right)}{(\phi\left({\bf x}_m|  \boldsymbol{\theta}\right)+ \nu Z q\left({\bf x}_m\right))^3}
&=0 \\
\nu Z\sum_{n=1}^N  \frac{\phi\left({\bf y}_n |  \boldsymbol{\theta}\right)  q\left({\bf y}_n\right)^2}{(\phi\left({\bf y}_n|  \boldsymbol{\theta}\right)+ \nu Z q\left({\bf y}_n\right))^3} -\sum_{m=1}^M \frac{\phi\left({\bf x}_m |  \boldsymbol{\theta}\right)^2 q\left({\bf x}_m\right)}{(\phi\left({\bf x}_m|  \boldsymbol{\theta}\right)+ \nu Z q\left({\bf x}_m\right))^3}
&=0.
\end{align*}
Isolating the first $Z$ in one side, we find a fixed point equation over $Z$ and can write the final iterative procedure:
\begin{align}
\fbox{$\displaystyle  \widehat{Z}_{t+1}=\dfrac{\dfrac{1}{M}\sum\limits_{m=1}^M \dfrac{\phi\left({\bf x}_m |  \boldsymbol{\theta}\right)^2 q\left({\bf x}_m\right)}{(\phi\left({\bf x}_m|  \boldsymbol{\theta}\right)+ \nu  \widehat{Z}_{t} q\left({\bf x}_m\right))^3}}{\dfrac{1}{N}\sum\limits_{n=1}^N  \dfrac{\phi\left({\bf y}_n |  \boldsymbol{\theta}\right)  q\left({\bf y}_n\right)^2}{(\phi\left({\bf y}_n|  \boldsymbol{\theta}\right)+ \nu  \widehat{Z}_{t} q\left({\bf y}_n\right))^3} }.$} 
\end{align}
we could also obtain the estimator above from Eq. \eqref{BridgeFirstId}, setting as bridge function: 
\begin{align}
b({\bf y})=\dfrac{\phi\left({\bf y} |  \boldsymbol{\theta}\right)  q\left({\bf y}\right)}{(\phi\left({\bf y}|  \boldsymbol{\theta}\right)+ \nu  Z q\left({\bf y}\right))^3}.
\end{align}
{\Remark From this result, we could speculate that there is a correspondence between proper scoring rules $V(\eta)$ and bridge functions 
 $b(\y)$ in Eq. \eqref{BridgeFirstId}.}

\subsubsection{Example 2 with a non-proper scoring rule}
Let us consider now a non-proper scoring rule, for instance,  assume $V(\eta)=1/\eta$. { The corresponding function cannot be directly employed for estimating $\btheta_{\texttt{tr}}$ without additional transformations. Indeed,} in this scenario, we could obtain highly-biased estimators that require some corrections. { See Appendix \ref{Vscore} for a proper transformation in this scenario}. However,  with $V(\eta)=1/\eta$,  we have:
\begin{align}\label{EqVnow1}
\frac{d V(\eta)}{d\eta}=-\frac{1}{\eta^{2}}
&= -\frac{\left[\phi\left(\mathbf{y}_n |  \boldsymbol{\theta}\right)+ \nu Z q\left(\mathbf{y}_n\right)\right]^{2} }{\phi\left(\mathbf{y}_n |  \boldsymbol{\theta}\right)^{2}},
\end{align}
\begin{align}\label{EqVnow2}
\frac{d V(1-\eta)}{d\eta}=- \left.\frac{d V(\eta)}{d\eta}\right|_{1-\eta}=\frac{1}{(1-\eta)^{2}}=\frac{\left[\phi\left({\bf x}_m |  \boldsymbol{\theta}\right)+ \nu Z q\left({\bf x}_m\right)\right]^{2} }{\left[\nu Z q\left({\bf x}_m\right)\right]^{2}},
\end{align}
where we have substituted  the  definition of $\eta$ in Eq. \eqref{EtaHereAgain}.
Replacing \eqref{EqVnow1}-\eqref{EqVnow2} and \eqref{DerEta} into Eq. \eqref{55Eq_imp}, we obtain
\begin{align*}
&\sum_{n=1}^N \left[-\frac{\left(\phi\left(\mathbf{y}_n |  \boldsymbol{\theta}\right)+ \nu Z q\left(\mathbf{y}_n\right)\right)^{2} }{\phi\left(\mathbf{y}_n |  \boldsymbol{\theta}\right)^{2}}\right] \left[-\frac{\nu \phi\left(\y_n |  \boldsymbol{\theta}\right)  q\left(\y_n\right)}{(\phi\left(\y_n |  \boldsymbol{\theta}\right)+ \nu Z q\left(\y_n\right))^2}\right]+\\
&\sum_{m=1}^M\left[\frac{\left(\phi\left({\bf x}_m |  \boldsymbol{\theta}\right)+ \nu Z q\left({\bf x}_m\right)\right)^{2} }{\left[\nu Z q\left({\bf x}_m\right)\right]^{2}}\right]\left[-\frac{\nu \phi\left({\bf x}_m |  \boldsymbol{\theta}\right)  q\left({\bf x}_m\right)}{(\phi\left({\bf x}_m |  \boldsymbol{\theta}\right)+ \nu Z q\left({\bf x}_m\right))^2}\right]=0,
\end{align*}
so that 
\begin{align*}
&\nu\sum_{n=1}^N  \frac{  q\left(\y_n\right)}{\phi\left(\y_n |  \boldsymbol{\theta}\right)}
-\frac{1}{\nu Z^{2}}\sum_{m=1}^M\frac{ \phi\left({\bf x}_m |  \boldsymbol{\theta}\right)  }{ q\left({\bf x}_m\right)}=0, \\
&\nu^2 Z^{2}\sum_{n=1}^N  \frac{  q\left(\y_n\right)}{\phi\left(\y_n |  \boldsymbol{\theta}\right)}
-\sum_{m=1}^M\frac{ \phi\left({\bf x}_m |  \boldsymbol{\theta}\right)  }{ q\left({\bf x}_m\right)}=0,
\end{align*}
and isolating $Z^2$ in one side, we get
\begin{align*}
 Z^{2}=\dfrac{1}{\nu^2}\dfrac{\sum\limits_{m=1}^M\dfrac{ \phi\left({\bf x}_m |  \boldsymbol{\theta}\right)  }{ q\left({\bf x}_m\right)}}{ \sum\limits_{n=1}^N  \dfrac{  q\left(\y_n\right)}{\phi\left(\y_n |  \boldsymbol{\theta}\right)}}=\dfrac{N}{M}\dfrac{\dfrac{1}{M}\sum\limits_{m=1}^M\dfrac{ \phi\left({\bf x}_m |  \boldsymbol{\theta}\right)  }{ q\left({\bf x}_m\right)}}{\dfrac{1}{N} \sum\limits_{n=1}^N  \dfrac{  q\left(\y_n\right)}{\phi\left(\y_n |  \boldsymbol{\theta}\right)}}.
\end{align*}
Finally, we obtain a ``bad'' estimator
\begin{align}
 \widehat{Z}_{\texttt{bad}}=\sqrt{\dfrac{N}{M}}\sqrt{\dfrac{\dfrac{1}{M}\sum\limits_{m=1}^M\dfrac{ \phi\left({\bf x}_m |  \boldsymbol{\theta}\right)  }{ q\left({\bf x}_m\right)}}{\dfrac{1}{N} \sum\limits_{n=1}^N  \dfrac{  q\left(\y_n\right)}{\phi\left(\y_n |  \boldsymbol{\theta}\right)}},}
\end{align}
that can have highly biased with $M\neq N$, for finite values $M$ and $N$. Indeed, note that the numerator is the  stand-IS estimator and the denominator is the RIS estimator, i.e.,
$$
\dfrac{1}{M}\sum\limits_{m=1}^M\dfrac{ \phi\left({\bf x}_m |  \boldsymbol{\theta}\right)  }{ q\left({\bf x}_m\right)} \approx Z, \qquad \left(\dfrac{1}{N} \sum\limits_{n=1}^N  \dfrac{q\left(\y_n\right)}{\phi\left(\y_n |  \boldsymbol{\theta}\right)}\right)^{-1}\approx Z,
$$ 
so that $\widehat{Z}_{\texttt{bad}} \approx \sqrt{\frac{N}{M}} Z$.  Therefore, we can easily improve this estimator defining a scaled version, i.e.,
\begin{align}\label{GeoEq}
\fbox{$\widehat{Z}_{\texttt{geo}}=\sqrt{\dfrac{M}{N}}  \widehat{Z}_{\texttt{bad}}=\sqrt{\dfrac{\dfrac{1}{M}\sum\limits_{m=1}^M\dfrac{ \phi\left({\bf x}_m |  \boldsymbol{\theta}\right)  }{ q\left({\bf x}_m\right)}}{\dfrac{1}{N} \sum\limits_{n=1}^N  \dfrac{ q\left(\y_n\right)}{\phi\left(\y_n |  \boldsymbol{\theta}\right)}}}$,}
\end{align}
that also represents the {\it geometric mean} between the stand-IS and the RIS estimators. {The corrected cost function corresponding to $\widehat{Z}_{\texttt{geo}}$ is given in Appendix \eqref{AppB}. Other information regarding the scoring rule $V(\eta)=\frac{1}{\eta}$ is given also in  Appendix \eqref{AppBantes}.} Table \ref{TableFF} summarizes the main described estimators.

{\Remark { Note that the estimator \(\widehat{Z}_{\texttt{geo}}\) {\bf cannot} be derived from the bridge identity; that is, there exists no bridge function \(b(\y)\) that yields \(\widehat{Z}_{\texttt{geo}}\).}}

\begin{table}[H]
		\centering
	\caption{\small Summary of the estimators of $Z$ using $q(\y)$ and/or $\post(\y)$. The last column shows if a recursive procedure is required. The first four rows correspond to estimators using samples from $\post(\y)$ and $q(\y)$. The last four rows correspond to estimators using samples from a single proposal density.  \label{TableFF}}
	\vspace{0.1cm}
	\small
	\begin{tabular}{|c|c|c|c|}
	\hline
{\bf Name}    & {\bf Estimator} & {\bf Samples} & {\bf Rec.} \\
	\hline
	\hline
	& & & \\
 Opt-Bridge    & $\widehat{Z}_{t+1}=\dfrac{\dfrac{1}{M}\sum\limits_{m=1}^M\dfrac{ \phi({\bf x}_m)}{ \alpha_1 \phi({\bf x}_m)+\alpha_2  \widehat{Z}_t q({\bf x}_m)} }{\dfrac{1}{N}\sum\limits_{n=1}^N\dfrac{q(\y_n)}{\alpha_1 \phi(\y_n)+\alpha_2 \widehat{Z}_t q(\y_n)}}$ &  $\y_n\sim\post(\y)$, $\quad$ $\x_m \sim q(\y)$ &  \CheckmarkBold  \\
       &  & &   \\
       &  &  & \\
  MIS    & $\widehat{Z}_{t+1}=\dfrac{1}{N+M} \sum\limits_{i=1}^{N+M} \dfrac{\widehat{Z}_t \phi({\bf u}_i)}{ \alpha_1 \phi({\bf u}_i)+\alpha_2 \widehat{Z}_t q({\bf u}_i)}$ &  $\{{\bf u}_i\}=\left\{\y_n\right\} \cup  \left\{\x_m\right\}$ & \CheckmarkBold \\
     & & &\\ 
     & & &\\
  Self-IS-with-mix    & $\widehat{Z}_{t+1}=\dfrac{\sum\limits_{i=1}^{N+M} \dfrac{ \phi({\bf u}_i)}{\alpha_1  {\phi}({\bf u}_i)+\alpha_2 \widehat{Z}_{t} q({\bf u}_i)}}{\sum\limits_{k=1}^{N+M} \dfrac{ q({\bf u}_k)}{\alpha_1  {\phi}({\bf u}_k)+\alpha_2 \widehat{Z}_{t} q({\bf u}_k)}}$  &  $\{{\bf u}_i\}=\left\{\y_n\right\} \cup  \left\{\x_m\right\}$ & \CheckmarkBold \\
   & & & \\
           & & & \\  
    Geo        & $\widehat{Z}_{\texttt{geo}}=\sqrt{\dfrac{\dfrac{1}{M}\sum\limits_{m=1}^M\dfrac{ \phi\left({\bf x}_m \right)  }{ q\left({\bf x}_m\right)}}{\dfrac{1}{N} \sum\limits_{n=1}^N  \dfrac{ q\left(\y_n\right)}{\phi\left(\y_n\right)}}}$ &  $\y_n\sim\post(\y)$, $\quad$ $\x_m \sim q(\y)$  & \XSolidBrush\\
           & & & \\
           \hline
           & & & \\  
    Stand-IS        & $\widehat{Z}_{\texttt{IS}}=\dfrac{1}{M} \sum\limits_{m=1}^{M} \dfrac{\phi({\bf x}_m)}{q({\bf x}_m)}$ &  $\x_m \sim q(\y)$ & \XSolidBrush\\
           & & & \\
           & && \\  
     RIS    & $ \widehat{Z}_{\texttt{RIS}}=\left(\dfrac{1}{N} \sum\limits_{n=1}^{N} \dfrac{q({\bf y}_n)}{\phi({\bf y}_n)}\right)^{-1}$ &   $\y_n\sim\post(\y)$ & \XSolidBrush \\     
           & & & \\  
 & & & \\             
   Opt-Umb          & $   \widehat{Z}_{t+1} =\dfrac{ \sum\limits_{i=1}^{N+M}\dfrac{\phi(\widetilde{{\bf x}}_i)}{|\phi( \widetilde{{\bf x}}_i) - \widehat{Z}_t q(\widetilde{{\bf x}}_i)|}}{\sum\limits_{k=1}^{N+M}\dfrac{ q(\widetilde{{\bf x}}_k)}{|\phi( \widetilde{{\bf x}}_k) - \widehat{Z}_t q(\widetilde{{\bf x}}_k)| }}$ & $\widetilde{{\bf x}}_i \sim \bar{r}(\y) \propto \left|\post(\y) - q(\y)\right|$ &  \CheckmarkBold\\
         & & & \\
\hline
	\end{tabular}
\end{table}

{
\section{Beyond standard classification: extended objective functions for EBM inference}\label{BClassSect}

In this section, we explore further extensions of the NCE objective function. The main ideas presented here have been introduced and analized in different works
\cite{Pihlaja2010,Chehab2023OptimizingNoise,ChehabHyvarinenRisteski2023,GutmannHirayama2011,UeharaMatsudaKomaki2018}. In particular, we investigate general conditions and classes of objective functions under which \(Z(\btheta)\) can be treated as a free parameter \(Z\) during optimization, as in standard NCE. Unlike in the previous section, we do not derive these objective functions from a classification framework; in general, they are not expressed in terms of class probabilities.\footnote{{ However, in \cite{Chehab2023OptimizingNoise,GutmannHirayama2011,UeharaMatsudaKomaki2018,Xing2022},  the authors establish a connection (regarding some material in this section) with alternative classification loss functions and use this connection to derive distance-based expressions for the mean squared error (MSE) of the estimator of \(Z\). Moreover, some additional theoretical connections with bridge sampling is remarked in \cite{Chehab2023OptimizingNoise}.}} Hence, here we use a more general approach with respect to the previous section. Moreover, unlike other works in the literature, we explicitly derive the theoretical conditions and provide illustrative examples, consistently using the same notation and maintaining a uniform level of theoretical detail throughout.
\newline
 First of all,  recall that we assume that  the true model,  $p_{\texttt{tr}}(\y)= \bar{\phi}(\y|\btheta_{\texttt{tr}})=\frac{\phi(\by|\btheta_{\texttt{tr}})}{Z(\btheta_{\texttt{tr}})}$ is contained in the EBM family that we are considering, the model is identifiable (for simplicity). Let us now  consider a generic theoretical cost function $J_\texttt{g}(\btheta,Z)=J_\texttt{g}(\btheta,Z(\btheta))$. Generally, we cannot treat $Z=Z(\btheta)$ as {\it free parameter} in the minimization of $J_\texttt{g}$, since it depends on $\btheta$. One possibility is to first estimate the partition function $Z(\btheta)$, obtaining $\widehat{Z}(\btheta)$ with some specific method, so that finally we have just to optimize with respect to $\btheta$, i.e., $J(\btheta)=J(\btheta,\widehat{Z}(\btheta))$. If we use an IS estimator for $\widehat{Z}(\btheta)$  this coincides with the Geyer's approach \cite{Geyer1994Convergence,MartinoScaffidi26}.
However, we can desire to treat $Z=Z(\btheta)$ as {\it free parameter} in the optimization problem to skip the estimation of $Z(\btheta)$, as in the NCE approach. In this scenario,  we would like that
\begin{align}
[\btheta^*, Z^*]=\arg \min J_\texttt{g}(\btheta,Z),
\end{align}
with 
\begin{gather}\label{SuperIMPwhatWewant}
\left\{
\begin{split}
\btheta^*& =\btheta_{\texttt{tr}}, \\
Z^*& =Z(\btheta_{\texttt{tr}}).
\end{split}
\right.
\end{gather}
A condition that ensures the equalities is 
\begin{align}
J_\texttt{g}(\btheta,Z) &\geq  J_\texttt{g}(\btheta^*,Z^*), \nonumber\\
J_\texttt{g}(\btheta,Z(\btheta)) &\geq  J_\texttt{g}(\btheta_{\texttt{tr}},Z(\btheta_{\texttt{tr}})).
\end{align}
It is important to remark that, from a practical point of view, we can only evaluate an empirical cost function $J_N(\btheta,Z)$ computed using samples, at least from $p_{\texttt{tr}}(\y)$. Minimizing $J_N$m we have  
$$
[\widehat{\btheta}, \widehat{Z}]=\arg \min J_N(\btheta,Z),
$$
and we would like, as $N$ grows, that:
\begin{gather}
\begin{split}
&J_N(\btheta,Z)  \rightarrow  J_\texttt{g}(\btheta,Z) \qquad \widehat{\btheta} \rightarrow \btheta^* =\btheta_{\texttt{tr}},  \qquad \widehat{Z}  \rightarrow Z^*=Z(\btheta_{\texttt{tr}}).
\end{split}
\end{gather}
To obtain Eq \eqref{SuperIMPwhatWewant}, then we need:
\begin{gather}\label{SuperIMPwhatWewant2}
\left\{
\begin{split}
&\nabla_{\btheta} J_\texttt{g} =0 
 \longrightarrow \btheta^* =\btheta_{\texttt{tr}}, \\
&\frac{\partial J_\texttt{g}}{\partial Z}=0 \longrightarrow Z^* =Z(\btheta_{\texttt{tr}}).
\end{split}
\right.
\end{gather}
We give some example where $Z$ can or cannot be treated as a free parameter in the next sections.


%
%
%

\subsection{An example in which $Z$  cannot be treated as a free-parameter:  the negative log-likelihood }

If we consider $Z(\btheta)$ known or estimated with $\widehat{Z}(\btheta)$, the negative log-likelihood of the EBMs can be minimized to find a consistent estimator for $\btheta$. This is the idea of the Geyer's approach \cite{Geyer1994Convergence,MartinoScaffidi26}. However, $Z$ cannot be used as a free parameter in the optimization problem. Indeed, the (ideal) negative log-likelihood function with infinite number of data ($N\rightarrow \infty$) can be expressed in term of expectation as a cross-entropy, 
\begin{align}
 J_\texttt{I-NNL}(\btheta)=J_\texttt{I-NNL}(\btheta,Z(\btheta))&=-\mathbb{E}_{p_{\texttt{tr}}}\left[\log \post(\y | \btheta)\right], \label{GeyerMLinfEq}Ê\\
 &= -\mathbb{E}_{p_{\texttt{tr}}}\left[\log \phi(\y | \btheta)\right]+\log Z(\btheta),  \nonumber \\
&= -\int \big[\log \phi(\y | \btheta)\big] \frac{\phi\left(\y | \btheta_{\mathrm{tr}}\right)}{Z\left(\btheta_{\mathrm{tr}}\right)} d\y +\log  \int \phi(\y | \btheta) d\y. \nonumber
\end{align} 
First of all,  note that
\begin{align*}
\nabla_{\boldsymbol{\theta}} J_\texttt{I-NNL}(\btheta)=-\mathbb{E}_{p_{\texttt{tr}}}\left[\nabla_{\boldsymbol{\theta}} \log \phi(\y | \boldsymbol{\theta})\right]+\nabla_{\boldsymbol{\theta}} \log Z(\boldsymbol{\theta}),
\end{align*}
with 
\begin{align*}
\nabla_{\boldsymbol{\theta}} \log Z(\boldsymbol{\theta})& =\frac{1}{Z(\boldsymbol{\theta})} \int \phi(\y | \boldsymbol{\theta}) \nabla_{\boldsymbol{\theta}} \log \phi(\y | \boldsymbol{\theta}) d \y,  =\mathbb{E}_{\post}\left[\nabla_{\boldsymbol{\theta}} \log \phi(\y | \boldsymbol{\theta})\right],
\end{align*}
then, replacing in the expression above, we finally get:
\begin{align*}
\nabla_{\boldsymbol{\theta}} J_\texttt{I-NNL}(\btheta)=-\mathbb{E}_{p_{\texttt{tr}}}\left[\nabla_{\boldsymbol{\theta}} \log \phi(\y | \boldsymbol{\theta})\right]+\mathbb{E}_{\post}\left[\nabla_{\boldsymbol{\theta}} \log \phi(\y | \boldsymbol{\theta})\right].
\end{align*}
At $\btheta=\btheta_{\texttt{tr}}$,  we have $\post(\y|\btheta_{\texttt{tr}})=p_{\texttt{tr}}(\y)$ and, as a consequence, the gradient of $J_\texttt{I-NNL}$ is null:
\begin{align*}
\left. \nabla_{\boldsymbol{\theta}} J_\texttt{I-NNL}\right|_{\btheta=\btheta_{\texttt{tr}}}=-\mathbb{E}_{p_{\texttt{tr}}}\left[\nabla_{\boldsymbol{\theta}} \log \phi(\y | \boldsymbol{\theta})\right]+\mathbb{E}_{p_{\texttt{tr}}}\left[\nabla_{\boldsymbol{\theta}} \log \phi(\y | \boldsymbol{\theta})\right]={\bf 0}.
\end{align*}
Hence $\btheta=\btheta_{\texttt{tr}}$ is a stationary point, as desired and expected. However, if we consider $Z$ as a free parameter, we have:
$$
\fbox{$\displaystyle J_\texttt{I-NNL}(\btheta,Z)=
  -\mathbb{E}_{p_{\texttt{tr}}}\left[\log \phi(\y | \btheta)\right]+\log Z.$}
$$
Computing the partial derivative with respect to $Z$, i.e.,
$$
\frac{\partial J_\texttt{I-NNL}}{\partial Z}=\frac{1}{Z}>0,
$$
we obtain a quantity that is always positive (never null at a finite value $Z^*$), hence we cannot consider $Z$ as a free parameter in the optimization problem.  Hence, in this scenario, we can only apply the Geyer's approach replacing  $Z(\btheta)$, in the cost function, with an estimation $\widehat{Z}(\btheta)$. One possibility is to use another expectation, $\widehat{Z}(\btheta) \approx Z(\btheta)=\mathbb{E}_q\left[\frac{\phi(\y|\btheta)}{q(\y)}\right]$ (that is the IS equality), to express $J_\texttt{I-NNL}$:
\begin{align}\label{INNL_doubleExpEq}
 J_\texttt{I-NNL}(\btheta)=
  -\mathbb{E}_{p_{\texttt{tr}}}\left[\log \phi(\y | \btheta)\right]+\log \left(\mathbb{E}_q\left[\frac{\phi(\y|\btheta)}{q(\y)}\right]\right).
\end{align} 
Drawing $N$ samples from $\y_n \sim p_{\texttt{tr}}(\y)=\post(\y|\btheta_{\texttt{tr}})$ and $M$ samples from the proposal $\x_m\sim q(\y)>0$, then we can write the empirical version of the ideal cost function above (this version is denoted with the subindex E-NNL), i.e., an approximation of the negative log-likelihood $J_\texttt{NNL}(\btheta) =  -\sum_{n=1}^N \log \phi(\y_n | \btheta)+\log Z(\btheta)$ (it is an approximation since $Z(\btheta)$ is estimated): 
\begin{align}
 J_\texttt{NNL}(\btheta) \approx J_\texttt{E-NNL}(\btheta)= -\frac{1}{N}\sum_{n=1}^N \log \phi(\y_n | \btheta)+\log\left[ \sum_{m=1}^M \frac{1}{M} \frac{\phi(\x_m|\btheta)}{q(\x_m)}\right], \\
 N \cdot J_\texttt{E-NNL}(\btheta)= -\sum_{n=1}^N \log \phi(\y_n | \btheta)+N \log\left[ \sum_{m=1}^M \frac{1}{M} \frac{\phi(\x_m|\btheta)}{q(\x_m)}\right],  \label{NNLGeyerEq}
\end{align} 
Fixing all the samples $\{\y_n\}_{n=1}^N$ and $\{\x_m\}_{m=1}^M$, $J_\texttt{E-NNL}(\btheta)$ is a given, deterministic function to minimize. See \cite{Geyer1994Convergence,MartinoScaffidi26,merda2025} for more details.


\subsection{Pihlaja's cost function in \cite{Pihlaja2010}}

The previous objective function $J_\texttt{I-NNL}$ can be ``corrected'' in order to allow the minimization in the extended space $(\btheta,Z)$ considering $Z$ as a free parameter. Indeed, let us consider the Pihlaja's cost function introduced in  \cite{Pihlaja2010}: 
\begin{align}
 J_\texttt{Ph}(\btheta,Z(\btheta))&=- \mathbb{E}_{p_{\texttt{tr}}}\left[\log \frac{\phi(\y | \btheta)}{Z(\btheta)}\right]+\int \frac{\phi(\y  | \btheta)}{Z(\btheta)} d\y, \\
 &=- \mathbb{E}_{p_{\texttt{tr}}}\left[\log \phi(\y | \btheta)\right]+ \mathbb{E}_{p_{\texttt{tr}}}\left[\log Z(\btheta)\right]+\int \frac{\phi(\y  | \btheta)}{Z(\btheta)} d\y, \\
 &=- \mathbb{E}_{p_{\texttt{tr}}}\left[\log \phi(\y | \btheta)\right]+ \mathbb{E}_{p_{\texttt{tr}}}\left[\log Z(\btheta)\right]+1, 
\end{align} 
where $\int \frac{\phi(\y  | \btheta)}{Z(\btheta)} d\y=1$ by definition.
Hence considering as function of $\btheta$, we have $J_\texttt{Ph}(\btheta)=J_\texttt{I-NNL}(\btheta)+1$ and the gradient with respect to $\btheta$ is the same,
$$
\nabla_{\btheta} J_\texttt{Ph}(\btheta,Z(\btheta))= \nabla_{\btheta}  J_\texttt{Ph}(\btheta)= \nabla_{\btheta}  J_\texttt{I-NNL}(\btheta).
$$ 
However, considering $Z$ as a free parameter (independent on $\btheta$), we have:
\begin{align}
 J_\texttt{Ph}(\btheta,Z)&=- \mathbb{E}_{p_{\texttt{tr}}}\left[\log \phi(\y | \btheta)\right]+ \mathbb{E}_{p_{\texttt{tr}}}\left[\log Z\right]+\int \frac{\phi(\y  | \btheta)}{Z} d\y,  \\
 & =- \mathbb{E}_{p_{\texttt{tr}}}\left[\log \phi(\y | \btheta)\right]+ \log Z+ \frac{\int\phi(\y  | \btheta) d\y}{Z} ,
\end{align}
where 
$$
\mathbb{E}_{p_{\mathrm{tr}}}[\log \phi(\y | \btheta)]=\int \big[\log \phi(\y | \btheta)\big] \frac{\phi\left(\y | \btheta_{\mathrm{tr}}\right)}{Z\left(\btheta_{\mathrm{tr}}\right)} d \y.
$$
Fixing $\btheta$, we set the partial derivative with respect to $Z$ equal to zero:
$$
\frac{\partial J_\texttt{Ph}}{\partial Z}=\frac{1}{Z}-\frac{\int \phi(\y  | \btheta) d\y}{Z^2}=0 \Longrightarrow Z^{\star}=\int \phi(\y  | \btheta) d\y=Z(\btheta), 
$$
where we have used the definition $Z(\btheta)=\int \phi(\y  | \btheta) d\y$.
Thus, if $\btheta^*=\btheta_{\texttt{tr}}$ (as is the case), minimizing $ J_\texttt{Ph}(\btheta,Z)$ treating $Z$ as a free variable, we have obtain the solution in Eq. \eqref{SuperIMPwhatWewant}, i.e., $\btheta^*=\btheta_{\texttt{tr}}$ and $Z^{\star}=Z(\btheta_{\texttt{tr}})$. Hence, we can consider the minimization of  $J_\texttt{Ph}(\btheta,Z)$ in the extended space $[\btheta,Z]$, i.e., treating $Z$ as a free parameter. However, we still need to evaluate/approximate the integral $\int \phi(\y | \btheta) d\y$, as remarked below. Note also that, the cost function above can be interpreted as the sum of the cross-entropy (that empirically is a negative log-likelihood) and a regularizer, i.e., 
\begin{align}\label{PijamaEqHere}
 J_\texttt{Ph}(\btheta,Z)&=\underbrace{- \mathbb{E}_{p_{\texttt{tr}}}\left[\log \post(\y | \btheta)\right]}_{\texttt{cross-entropy}}\underbrace{+\frac{1}{Z}\int \phi(\y | \btheta) d\y}_{\texttt{regularizer}}, \\
  J_\texttt{Ph}(\btheta,Z)&= J_\texttt{I-NNL}(\btheta,Z)+\frac{1}{Z}\int \phi(\y | \btheta) d\y.  \\
  &= \underbrace{- \mathbb{E}_{p_{\texttt{tr}}}\left[\log \phi(\y | \btheta)\right]+\log Z}_{J_\texttt{I-NNL}(\btheta,Z)}+\frac{1}{Z}\int \phi(\y | \btheta) d\y. 
\end{align} 
Above, we can also see the relationship with the cost function $J_\texttt{I-NNL}$ in Eq. \eqref{GeyerMLinfEq}, and we can observe that $\frac{1}{Z}\int \phi(\y | \btheta) d\y$ plays the role of an additional regularized term. However, the evaluation of $J_\texttt{Ph}(\btheta,Z)$ still depends on $\int \phi(\y | \btheta) d\y$.

{\Remark Treating $Z$ as a free parameter allows it to be optimized independently of $\btheta$, rather than imposing the constraint $Z=Z(\btheta)$ throughout the optimization. Nevertheless, the model-induced partition function $Z(\btheta)=\int \phi(\y | \btheta) d\y$ still appears in the cost function and must therefore be evaluated or approximated. Hence, considering $Z$ as a free parameter does not remove the need to approximate the integral defining  $Z(\btheta)=\int \phi(\y | \btheta) d\y$; it only decouples its optimization from that of  $\btheta$.}
\newline
\newline
We can solve this problem as done with $J_\texttt{I-NNL}(\btheta,Z)$ in Eq. \eqref{INNL_doubleExpEq}, i.e.,  using an additional expectation,
\begin{align}\label{PijamaEqDouble}
\fbox{$\displaystyle J_\texttt{Ph}(\btheta,Z)= - \mathbb{E}_{p_{\texttt{tr}}}\left[\log \phi(\y | \btheta)\right]+\log Z+\frac{1}{Z} \mathbb{E}_{q}\left[\frac{\phi(\y | \btheta)}{q(\y)}\right]$,}
\end{align} 
Given independent samples  $\y_n\sim p_{\texttt{tr}}(\y)$  and $\x_m \sim q(\y) >0$, the corresponding empirical objective function, such that $ J_\texttt{E-Ph}(\btheta,Z) \approx J_\texttt{Ph}(\btheta,Z)$, is:
\begin{align}\label{PijamaEqEmp}
 \boxed{J_\texttt{E-Ph}(\btheta,Z)= -\frac{1}{N}\sum_{n=1}^N \log \phi(\y_n | \btheta)+\log Z+\frac{1}{M Z} \sum_{m=1}^M \frac{\phi(\x_m | \btheta)}{q(\x_m)},} 
 \end{align} 
 or
\begin{align} 
 NÊ\cdot J_\texttt{E-Ph}(\btheta,Z) = -\sum_{n=1}^N \log \phi(\y_n | \btheta)+N\log Z+\frac{N}{M Z} \sum_{m=1}^M \frac{\phi(\x_m | \btheta)}{q(\x_m)}. 
\end{align} 
{\bf Deriving the stand-IS estimator.} Note that the expression above is different from the cost function in Eq. \eqref{NNLGeyerEq}. Differentiating with respect to $Z$ and setting the derivative equal to zero, 
\begin{align}\label{PijamaDer}
 \frac{\partial J_\texttt{E-Ph}(\btheta,Z)}{\partial Z}=\frac{1}{Z}-\frac{1}{M Z^2} \sum_{m=1}^M \frac{\phi(\x_m | \btheta)}{q(\x_m)} =0 \Longrightarrow \fbox{$\displaystyle\widehat{Z}=\frac{1}{M } \sum_{m=1}^M \frac{\phi(\x_m | \btheta)}{q(\x_m)}$,}
\end{align} 
we obtain the standard IS estimator in Eq. \eqref{standISest_eq}. Recall that, since minimizing  $J_\texttt{E-Ph}(\btheta,Z)$ in extended space we obtain the pair $\btheta^*=\btheta_{\texttt{tr}}$ and $Z^*=\widehat{Z}(\btheta^*)$, we actually obtain an estimation of only one point of the function $Z(\btheta)$, that is $\widehat{Z}(\btheta_{\texttt{tr}})$.

\subsection{Generalized Pihlaja's cost functions}\label{GenPihSect}

In \cite{Pihlaja2010,Chehab2023OptimizingNoise,GutmannHirayama2011}, the authors suggested a general family of cost functions based on two expectations:
\begin{align}\label{GenPijamaEqDouble}
\fbox{$\displaystyle J_\texttt{G-Ph}(\btheta,Z)=  -\mathbb{E}_{p_{\texttt{tr}}}\left[h_1\left(\frac{\post(\y | \btheta)}{q(\y)}\right)\right]+\mathbb{E}_{q}\left[h_2\left(\frac{\post(\y | \btheta)}{q(\y)}\right)\right].$}
\end{align} 
where  $h_1(\cdot)$ and $h_2(\cdot)$  two non-linear functions, defined in $\mathbb{R}^+$ and taking values in $\mathbb{R}$. Clearly, with $h_1(v)=\log(v)$ and $h_2(v)=v$, we come back to an objective function equivalent to the Pihlaja's cost function in Eq. \eqref{PijamaEqDouble}.\footnote{{In this scenario, we have $J_\texttt{G-Ph}(\btheta,Z)=  -\mathbb{E}_{p_{\texttt{tr}}}\left[\log\post(\y | \btheta)\right]+\mathbb{E}_{p_{\texttt{tr}}}\left[\log q(\y)\right]+\mathbb{E}_{q}\left[\frac{\post(\y | \btheta)}{q(\y)}\right]$. Since $\mathbb{E}_{p_{\texttt{tr}}}\left[\log q(\y)\right]$ does not depends on $\btheta$, then minimizing $J_\texttt{G-Ph}(\btheta,Z)$ is in this case equivalent to minimze $J_\texttt{Ph}(\btheta,Z)$. Note also that $\mathbb{E}_{q}\left[\frac{\post(\y | \btheta)}{q(\y)}\right]=1$, hence its empirical version is an estimation of $1$.}}
Given the $N$ observed samples  $\y_n\sim p_{\texttt{tr}}(\y)$ and the $M$ samples $\x_m \sim q(\y)>0$, the empirical version is 
\begin{align}\label{GenPijamaEqDoubleEmp}
\fbox{$\displaystyle J_\texttt{EG-Ph}(\btheta,Z)=  -\frac{1}{N}\sum_{n=1}^N h_1\left(\frac{\post(\y_n | \btheta)}{q(\y_n)}\right)+ \frac{1}{M}\sum_{m=1}^M h_2\left(\frac{\post(\x_m | \btheta)}{q(\x_m)}\right).$}
\end{align} 
In \cite{Pihlaja2010,GutmannHirayama2011}, the authors establish  {\it sufficient conditions} on $h_1$ and $h_2$ under which the pair \(\btheta^*=\btheta_{\texttt{tr}}\) and \(Z^*=Z(\btheta^*)\) is the unique stationary point of \(J_\texttt{G-Ph}(\btheta,Z)\), with \(Z\) treated as a free parameter during minimization:
\begin{itemize}
\item $h_1(v)$ and $h_2(v)$ are strictly increasing functions, i.e., $h_1'(v)>0$, $h_2'(v)>0$,
\item and satisfy:
\begin{align}\label{CondonHeq}
\fbox{$\displaystyle \frac{h_2'(v)}{h_1'(v)}=v,$}
\end{align}
with $h_1^{\prime}(v) \neq 0$.
\end{itemize}
 Note that the dependence on $Z$ is inside $\post(\y | \btheta)=\frac{\phi(\y | \btheta)}{Z(\btheta)}=\frac{\phi(\y | \btheta)}{Z}$.  To prove in a simpler way the second condition, it is convenient to include  directly $Z$ (or $\log Z$) in the optimization problem, defining $\widetilde{\btheta}=[\btheta, Z]^{\top}$ and $r(\y,\widetilde{\btheta})=\frac{\post(\y | \widetilde{\btheta})}{q(\y)}$. With this notation, we have:
\begin{align}
J_\texttt{G-Ph}(\widetilde{\btheta})&=- \int p_{\texttt{tr}}(\y) h_1\left(\frac{\post(\y | \widetilde{\btheta})}{q(\y)}\right) d\y +\int q(\y) h_2\left(\frac{\post(\y | \widetilde{\btheta})}{q(\y)}\right) d\y. \\
&=- \int p_{\texttt{tr}}(\y) h_1\left(r(\y,\widetilde{\btheta})\right) d\y +\int q(\y) h_2\left(r(\y,\widetilde{\btheta})\right) d\y.
\end{align}
Assume that $q(\y)>0$, that $h_1, h_2$ are differentiable, and that differentiation can be interchanged with integration.  By the chain rule, we can write:
$$
\nabla_{\widetilde{\btheta}} r(\y,\widetilde{\btheta})=\frac{1}{q(\y)} \nabla_{\widetilde{\btheta}} \post(\y | \widetilde{\btheta}).
$$
Differentiating $J_\texttt{G-Ph}(\widetilde{\btheta})$ gives
$$
\begin{aligned}
\nabla_{\widetilde{\btheta}} J_\texttt{G-Ph}(\widetilde{\btheta})= & \int p_{\texttt{tr}}(\y) h_1^{\prime}\left(r(\y,\widetilde{\btheta})\right) \frac{1}{q(\y)} \nabla_{\widetilde{\btheta}} \post(\y | \widetilde{\btheta})d\y \\
& -\int q(\y) h_2^{\prime}\left(r(\y,\widetilde{\btheta})\right) \frac{1}{q(\y)} \nabla_{\widetilde{\btheta}} \post(\y | \widetilde{\btheta})d\y .
\end{aligned}
$$
Canceling $q(\y)$ in the second integral, we obtain the compact expression:
\begin{align}\label{GengradientEq}
\nabla_{\widetilde{\btheta}} J_\texttt{G-Ph}(\widetilde{\btheta})=\int\left[\frac{p_{\texttt{tr}}(\y)}{q(\y)} h_1^{\prime}\left(r(\y,\widetilde{\btheta})\right)-h_2^{\prime}\left(r(\y,\widetilde{\btheta})\right)\right] \cdot \nabla_{\widetilde{\btheta}} \post(\y | \widetilde{\btheta}) d\y .
\end{align}
Recalling that   $p_{\texttt{tr}}(\y)= \bar{\phi}(\y|\btheta_{\texttt{tr}})=\frac{\phi(\by|\btheta_{\texttt{tr}})}{Z(\btheta_{\texttt{tr}})}$, to ensure that the gradient vanishes at the true density $p_{\texttt{tr}}(\y)$ (hence at $\btheta^*=\btheta_{\texttt{tr}}$), a sufficient condition is that the term in square brackets must be zero for every $\y$ such that $p_{\texttt{tr}}(\y)>0$, namely:
$$
\frac{p_{\texttt{tr}}(\y)}{q(\y)} h_1^{\prime}\left(\frac{p_{\texttt{tr}}(\y)}{q(\y)}\right)=h_2^{\prime}\left(\frac{p_{\texttt{tr}}(\y)}{q(\y)}\right).
$$
Setting $v=\frac{p_{\texttt{tr}}(\y)}{q(\y)}$, the condition above becomes:
$$
v h_1^{\prime}(v)=h_2^{\prime}(v)Ê\Longrightarrow \frac{h_2^{\prime}(v)}{h_1^{\prime}(v)}=v, \quad v>0, 
$$
that is exactly Eq. \eqref{CondonHeq}.This condition is sufficient, but not generally necessary, for the gradient to vanish at $\btheta^*=\btheta_{\texttt{tr}}$ and $Z^*=Z(\btheta_{\texttt{tr}})$. The gradient in Eq. \eqref{GengradientEq} could also vanish because of other cancellations within the integral. The pointwise condition, that we have applied, is a stronger requirement that guarantees stationarity independently of such alternative cancellations. Other conditions required for the consistency of the estimators (not just the stationarity) are given in \cite[Section 2.3]{Pihlaja2010}.
\newline
\newline
The objective $J_\texttt{G-Ph}(\btheta,Z)$ has not been derived following a  classification arguments and generally does not required the expressions of the probability classes \cite{UeharaMatsudaKomaki2018}. However, the connections of the cost  function $J_\texttt{G-Ph}(\btheta,Z)$ with different classification losses have been clarified in
\cite{Chehab2023OptimizingNoise,GutmannHirayama2011, UeharaMatsudaKomaki2018,Xing2022}. Moreover, in \cite{Chehab2023OptimizingNoise,ChehabHyvarinenRisteski2023}, the authors write $J_\texttt{G-Ph}(\btheta,Z)$ using a unique convex function $f(v)$:
\begin{align}\label{GenPijamaEqDouble_conf}
{\small
\fbox{$\displaystyle J_\texttt{G-Ph}(\btheta,Z)=  -\mathbb{E}_{p_{\texttt{tr}}}\left[f^{\prime}\left(\frac{\post(\y | \btheta)}{q(\y)}\right)\right]+\mathbb{E}_{q}\left[f^{\prime}\left(\frac{\post(\y | \btheta)}{q(\y)}\right)\frac{\post(\y | \btheta)}{q(\y)}-f\left(\frac{\post(\y | \btheta)}{q(\y)}\right)\right].$}}
\end{align} 
In this form, it is also called Bregman classification loss \cite{GutmannHirayama2011}.
   The convex function $f(v)$ is related with $h_1(v)$ and $h_2(v)$ in this way: 
\begin{align}
f(v)=h_2(v) v-h_1(v), 
\end{align}
and 
\begin{align}
h_2(v)=f^{\prime}(v),  \quad\quad  h_1(v)=-f(v)+f^{\prime}(v) v.
\end{align}
Different choices of $f(v)$ should be equivalent to different scoring rules in binary classifiers \cite{Chehab2023OptimizingNoise,GutmannHirayama2011, UeharaMatsudaKomaki2018,Xing2022}. Under mild conditions, the estimators of $Z$ obtained minimizing $J_\texttt{G-Ph}(\btheta,Z)$ can be derived by the bridge sampling identity with a specific bridge function $b(\y)$ that can be expressed as function of second derivative of $f(v)$ and the gradient of $\post$, as shown \cite[Eqs. (17) and (31)]{Chehab2023OptimizingNoise}. 
 

\subsubsection{Cost function related to the reverse IS  estimator}

Let set now $h_1(v)=-1/v$ and $h_2(v)=\log(v)$, then  the objective function $J_\texttt{EG-Ph}(\btheta,Z)$ in Eq. \eqref{GenPijamaEqDoubleEmp} becomes:
\begin{align}
J(\btheta,Z)&=\frac{Z}{N} \sum_{n=1}^{N} \frac{q(\y_n)}{\phi(\y_n|\btheta)}+\frac{1}{M} \sum_{m=1}^{M} \log{\frac{\phi(\x_m|\btheta)}{Z q(\x_m)}}, \label{EstaConAppB}Ê\\
&=\frac{Z}{N} \sum_{n=1}^{N} \frac{q(\y_n)}{\phi(\y_n|\btheta)}+\frac{1}{M} \sum_{m=1}^{M} \log\phi(\x_m|\btheta) - \log Z -\frac{1}{M} \sum_{m=1}^{M} \log q(\x_m). \nonumber
\end{align}
Hence, removing constants with $\btheta$ and $Z$, we have:
\begin{align}\label{RIScostfun}
\boxed{J(\btheta,Z)=\frac{Z}{N} \sum_{n=1}^{N} \frac{q(\y_n)}{\phi(\y_n|\btheta)}+\frac{1}{M} \sum_{m=1}^{M} \log\phi(\x_m|\btheta) - \log Z.}
\end{align}
See also Appendix \ref{AppB} that shows a similar objective function.
Assuming that $\btheta$, the samples  $\y_m$, $\x_m$, and $q$ are fixed with
respect to $Z$, making the derivative and imposing the stationary-point condition, we have:
\begin{equation}
\frac{dJ(Z)}{dZ}
=
\frac{1}{N}\sum_{n=1}^{N}
\frac{q(\y_n)}{\phi(\y_n|\btheta)}
-\frac{1}{Z} =0.
\end{equation}
Hence, in this scenario, the estimator is:
\begin{equation}
\boxed{
\widehat{Z}
=
\frac{N}{
\sum_{n=1}^{N}
\frac{q(\y_n)}{\phi(\y_n|\btheta)}
} \qquad \y_n \sim \post(\y|\btheta),
}
\end{equation}
that is the RIS estimator in Eq. \eqref{RISestEq}.

\subsubsection{Cost function with squared root transformations}

Let us set $h_1(v)=-v^{-1/2}$ and $h_2(v)=v^{1/2}$,  so that the objective function $J_\texttt{EG-Ph}(\btheta,Z)$ in Eq. \eqref{GenPijamaEqDoubleEmp} becomes: 
\begin{align}\label{SpecialPigiama2}
\boxed{J(\btheta,Z)=\frac{\sqrt{Z}}{N} \sum_{n=1}^{N} \sqrt{\frac{q(\y_n)}{\phi(\y_n|\btheta)}}+\frac{1}{ \sqrt{Z} M} \sum_{m=1}^{M} \sqrt{\frac{\phi(\x_m|\btheta)}{q(\x_m)}},Ê}
\end{align}
We differentiate with respect to $Z$, keeping again $\btheta$, the
samples, and the functions $q$ and $\phi$ fixed. Since $Z>0$,
we have:
\[
\frac{d}{dZ}\sqrt{Z}=\frac{1}{2\sqrt{Z}},
\qquad
\frac{d}{dZ}\frac{1}{\sqrt{Z}}
=-\frac{1}{2Z^{3/2}}.
\]
Therefore,
$$
\frac{\partial J(\btheta,Z)}{\partial Z}
= \frac{1}{2N\sqrt{Z}}
\sum_{n=1}^{N}
\sqrt{\frac{q(\y_n)}{\phi(\y_n|\btheta)}}-
\frac{1}{2MZ^{3/2}}
\sum_{m=1}^{M}
\sqrt{\frac{\phi(\x_m|\btheta)}{q(\x_m)}}=0.
$$
Multiplying by $Z^{3/2}$ and moving a factor to the other side, we obtain:
\[
Z\,
\frac{1}{N}
\sum_{n=1}^{N}
\sqrt{\frac{q(\y_n)}{\phi(\y_n|\btheta)}}
=
\frac{1}{M}
\sum_{m=1}^{M}
\sqrt{\frac{\phi(\x_m|\btheta)}{q(\x_m)}}.
\]
Hence, the stationary point and hence the corresponding final estimator  is:
\[
\boxed{
\widehat{Z}
=
\frac{
\displaystyle\frac{1}{M}
\sum_{m=1}^{M}
\sqrt{\frac{\phi(\x_m|\btheta)}{q(\x_m)}}
}{
\displaystyle\frac{1}{N}
\sum_{n=1}^{N}
\sqrt{\frac{q(\y_n)}{\phi(\y_n|\btheta)}}
}.
}
\]
We can check that is a consistent estimator of $Z(\btheta)$. Fixing $\btheta$, the numerator is an estimator of $A=\int \sqrt{\phi(\y|\btheta)} \sqrt{q(\y)}d\y$, whereas the denominator is an estimator of $B=\int\sqrt{\frac{q(\y)}{\phi(\y|\btheta)}}\frac{\phi(\y|\btheta)}{Z(\btheta)}d\y=\frac{1}{Z(\btheta)}\int \sqrt{q(\y)} \sqrt{\phi(\y|\btheta)}d\y$. Hence, as $M,N \rightarrow \infty$, $\widehat{Z} \rightarrow \frac{A}{B}=Z(\btheta)$. The estimator above can be also obtained from Eq. 
Eq. \eqref{MainBridgeIdEq}, setting $b(\y)=[\phi(\y|\btheta)q(\y)]^{-1/2}$.
Finally, note also that the estimator above can resemble $\widehat{Z}_{\texttt{geo}}$ in Eq. \eqref{GeoEq} but they are different.

\subsubsection{Two additional objective functions and corresponding estimators }

Assuming $h_1(v)=v$ and $h_2(v)=\frac{1}{2} v^2$, then  the objective function $J_\texttt{EG-Ph}(\btheta,Z)$ in Eq. \eqref{GenPijamaEqDoubleEmp} becomes:
$$
\boxed{J(\btheta,Z)=-\frac{1}{NZ}\sum_{n=1}^{N}\frac{\phi(\y_n|\btheta)}{ q(\y_n)}+ \frac{1}{2MZ^2}\sum_{m=1}^{M}\left(\frac{\phi(\x_m|\btheta)}{ q(\x_m)}\right)^2,}
$$
For simplicity, let define
\begin{align*}
A=\frac{1}{N}\sum_{n=1}^{N}
\frac{\phi(\y_n|\btheta)}{q(\y_n)}, \quad B=\frac{1}{M}\sum_{m=1}^{M}
\left(\frac{\phi(\x_m|\btheta)}{q(\x_m)}\right)^2,
\end{align*}
so that, considering only $Z$ for simplicity, we  can write $J(Z)=-\frac{A}{Z}+\frac{B}{2Z^2}$.  Computing the first derivative with respect to $Z$ and imposing the stationary-point condition,
\begin{equation}
\frac{dJ(Z)}{dZ}
=\frac{A}{Z^2}-\frac{B}{Z^3}=0,
\end{equation}
we obtain
\begin{equation}
-AZ+B=0 \Longrightarrow Z^*=\frac{B}{A}.
\end{equation}
Hence, the estimator of $Z$ is:
\begin{equation}
\boxed{
\widehat{Z}=
\frac{
\displaystyle \frac{1}{M} \sum_{m=1}^{M}
\left(\frac{\phi(\x_m|\btheta)}{q(\x_m)}
\right)^2
}{
\displaystyle \frac{1}{N} \sum_{n=1}^{N}
\frac{\phi(\y_n|\btheta)}{q(\y_n)}}, \qquad \y_n \sim \post(\y |\btheta),
\quad \x_m \sim q(\y).
}
\end{equation}
Note that, fixing $\btheta$, the numerator is an estimator of $A=\int \frac{\phi(\y|\btheta)^2}{q(\y)} d\y$, whereas the denominator is an estimator of $B=\int\frac{\phi(\y|\btheta)}{q(\y)} \frac{\phi(\y|\btheta)}{Z(\btheta)}d\y=\frac{1}{Z(\btheta)}\int\frac{\phi(\y|\btheta)^2}{q(\y)}d\y$. Hence, as $M,N \rightarrow \infty$, $\widehat{Z} \rightarrow \frac{A}{B}=Z(\btheta)$ (the estimator is  consistent, as expected). The estimator above can be also obtained from Eq. 
Eq. \eqref{MainBridgeIdEq}, setting $b(\y)=\frac{\phi(\y|\btheta)}{q(\y)^2}$.
\newline
\newline
Alternatively, if we set $h_1(v)=-\frac{1}{2 v^2}$ and $h_2(v)=-\frac{1}{v}$, then  the objective function $J_\texttt{EG-Ph}(\btheta,Z)$ in Eq. \eqref{GenPijamaEqDoubleEmp} becomes:
$$
\boxed{J(\btheta,Z)=\frac{Z^2}{2N}\sum_{n=1}^{N}\left(\frac{q(\y_n)}{ \phi(\y_n|\btheta)}\right)^2- \frac{Z}{M}\sum_{m=1}^{M}\frac{q(\x_m)}{\phi(\x_m|\btheta)}.}
$$
Assuming that $\btheta$, the observations, and the functions $q$ and
$\phi$ are fixed, we can again differentiate with respect to $Z$ and equal to zero:
\[
\frac{\partial J}{\partial Z}
=
\frac{Z}{N}\sum_{n=1}^{N}
\left(\frac{q(\y_n)}{\phi(\y_n|\btheta)}\right)^2
-\frac{1}{M}\sum_{m=1}^{M}
\frac{q(\x_m)}{\phi(\x_m|\btheta)}=0.
\]
Thus, the stationary point is then given by
\[
\boxed{
\widehat{Z}=
\frac{
\displaystyle\frac{1}{M}\sum_{m=1}^{M}
\frac{q(\x_m)}{\phi(\x_m|\btheta)}
}{
\displaystyle\frac{1}{N}\sum_{n=1}^{N}
\left(\frac{q(\y_n)}{\phi(\y_n|\btheta)}\right)^2
}, \qquad \y_n \sim \post(\y |\btheta),
\quad \x_m \sim q(\y).
}
\]
Again fixing $\btheta$, the numerator is an estimator of $A=\int \frac{q(\y)^2}{\phi(\y|\btheta)} d\y$, whereas the denominator is an estimator of $B=\int\frac{ q(\y)^2}{\phi(\y|\btheta)^2} \frac{\phi(\y|\btheta)}{Z(\btheta)}d\y=\frac{1}{Z(\btheta)}\int\frac{q(\y)^2}{\phi(\y|\btheta)}d\y$. Hence, as $M,N \rightarrow \infty$, $\widehat{Z} \rightarrow \frac{A}{B}=Z(\btheta)$. The estimator above can be also obtained from Eq. 
Eq. \eqref{MainBridgeIdEq}, setting $b(\y)=\frac{q(\y)}{\phi(\y|\btheta)^2}$.
Finally, Table \ref{TablePigiama} summarizes all the previous examples, showing the corresponding functions $h_1(v)$, $h_2(v)$ as well. Note that the condition $\frac{h_2'(v)}{h_1'(v)}=v$ is always satisfied.

\subsection{Reducing the dimension of the optimization problem}\label{SuperGenOfPreviousPaper}

A final important observation is the following. Let us assume a generic suitable cost function $J(\btheta,Z)$ that provides consistent estimator of $\btheta_{\texttt{tr}}$ and $Z_{\texttt{tr}}=Z(\btheta_{\texttt{tr}})$, treating  $Z$ as a free parameter. The analyses presented here and in previous works \cite{Pihlaja2010,Chehab2023OptimizingNoise,ChehabHyvarinenRisteski2023,GutmannHirayama2011} establish connections between the cost function $J(\btheta,Z)$ and the corresponding estimator $\widehat{Z}(\btheta)$, obtained by fixing $\btheta$ and deriving an {\it explicit} expression for the optimal value of $Z$. These results can be exploited to reduce the dimensionality of the optimization problem by analytically eliminating $Z$, thereby yielding an equivalent objective function that depends only on $\btheta$:
\begin{align}
	\boxed{J(\btheta,Z) \Longrightarrow \widetilde{J}(\btheta)=J(\btheta,\widehat{Z}(\btheta)).}
\end{align}
Note that $J(\btheta,Z)$ and $ \widetilde{J}(\btheta)$ are completely equivalent in terms of optimization problem with respect to $\btheta$. This is an extension of the philosophy in \cite{MartinoScaffidi26}. For instance, the RIS cost function.
$$
J(\btheta,Z)=\frac{Z}{N} \sum_{n=1}^{N} \frac{q(\y_n)}{\phi(\y_n|\btheta)}+\frac{1}{M} \sum_{m=1}^{M} \log\phi(\x_m|\btheta) - \log Z,
$$
becomes:
\begin{align}
	 \widetilde{J}(\btheta)&= \frac{\frac{N}{
\sum_{n=1}^{N}
\frac{q(\y_n)}{\phi(\y_n|\btheta)}
}}{N} \sum_{n=1}^{N} \frac{q(\y_n)}{\phi(\y_n|\btheta)}+\frac{1}{M} \sum_{m=1}^{M} \log\phi(\x_m|\btheta) - \log \frac{N}{
\sum_{n=1}^{N}
\frac{q(\y_n)}{\phi(\y_n|\btheta)}
},\nonumber \\
&=1+\frac{1}{M} \sum_{m=1}^{M} \log\phi(\x_m|\btheta)  + \log 
\left[\sum_{n=1}^{N}
\frac{q(\y_n)}{\phi(\y_n|\btheta)}\right] - \log N, \nonumber \\
&=-\sum_{n=1}^{N} \log \phi(\y_n|\btheta)+\frac{1}{M} \sum_{m=1}^{M} \log\phi(\x_m|\btheta)+ C, 
\end{align}
where $C=1-\log N +\sum_{n=1}^{N}\log q(\y)$.

\begin{table}[h!]
{
		\centering
	\caption{\small Summary of the generalized Pihajala's cost functions and the corresponding functions $h_1$, $h_2$ and estimators of $Z$.  \label{TablePigiama}}
	\vspace{0.1cm}
	\small
	\begin{tabular}{|c|c|c|c|}
	\hline
{\bf Objective function}  & $h_1(v)$ & $h_2(v)$  & {\bf Estimator} \\
	\hline
	\hline
	&   & &  \\
$\displaystyle J(\btheta,Z)= -\frac{1}{N}\sum_{n=1}^N \log \frac{\phi(\y_n | \btheta)}{Z}+\frac{1}{M Z} \sum_{m=1}^M \frac{\phi(\x_m | \btheta)}{q(\x_m)}$ & $\displaystyle  \log v$  & $\displaystyle v$ & $\displaystyle  \widehat{Z}_{\texttt{IS}}=\dfrac{1}{M} \sum\limits_{m=1}^{M} \dfrac{\phi({\bf x}_m|\btheta)}{q({\bf x}_m)}$  \\
	&   & &  \\
	& & &  \\
	\hline       
	& & &  \\
     $\displaystyle J(\btheta,Z)=\frac{Z}{N} \sum_{n=1}^{N} \frac{q(\y_n)}{\phi(\y_n|\btheta)}+\frac{1}{M } \sum_{m=1}^{M} \log{\frac{\phi(\x_m|\btheta)}{Zq(\x_m)}} $  &  $\displaystyle  -1/v$ & $\displaystyle  \log v$ &  $\widehat{Z}_{\texttt{RIS}}
=\frac{\displaystyle N}{
\displaystyle  \sum_{n=1}^{N}
\frac{q(\y_n)}{\phi(\y_n|\btheta)}
}$ \\
       &   & &  \\
\hline
 &   & &  \\
$\displaystyle J(\btheta,Z)=\frac{\sqrt{Z}}{N} \sum_{n=1}^{N} \sqrt{\frac{q(\y_n)}{\phi(\y_n|\btheta)}}+\frac{1}{ \sqrt{Z} M} \sum_{m=1}^{M} \sqrt{\frac{\phi(\x_m|\btheta)}{q(\x_m)}}$ & $\displaystyle -v^{-1/2}$ & $\displaystyle  v^{1/2}$
& 
$\widehat{Z}
=
\frac{
\displaystyle\frac{1}{M}
\sum_{m=1}^{M}
\sqrt{\frac{\phi(\x_m|\btheta)}{q(\x_m)}}
}{
\displaystyle\frac{1}{N}
\sum_{n=1}^{N}
\sqrt{\frac{q(\y_n)}{\phi(\y_n|\btheta)}}
}$\\
 &   & &  \\
 \hline
       &   & &  \\
$\displaystyle J(\btheta,Z)=-\frac{1}{NZ}\sum_{n=1}^{N}\frac{\phi(\y_n|\btheta)}{ q(\y_n)}+ \frac{1}{2MZ^2}\sum_{m=1}^{M}\left(\frac{\phi(\x_m|\btheta)}{ q(\x_m)}\right)^2$ & $\displaystyle v$ & $\displaystyle  \frac{1}{2} v^2$ &  $\widehat{Z}=
\frac{
\displaystyle \frac{1}{M} \sum_{m=1}^{M}
\left(\frac{\phi(\x_m|\btheta)}{q(\x_m)}
\right)^2
}{
 \displaystyle \frac{1}{N} \sum_{n=1}^{N}
\frac{\phi(\y_n|\btheta)}{q(\y_n)}}$  \\
       &   & &  \\
\hline 
       &   & &  \\
$\displaystyle J(\btheta,Z)=\frac{Z^2}{2N}\sum_{n=1}^{N}\left(\frac{q(\y_n)}{ \phi(\y_n|\btheta)}\right)^2- \frac{Z}{M}\sum_{m=1}^{M}\frac{q(\x_m)}{\phi(\x_m|\btheta)}$ & $\displaystyle -\frac{1}{2v^2}$  & $\displaystyle -\frac{1}{v}$ 
& $\widehat{Z}=
\frac{\displaystyle \frac{1}{M}\sum_{m=1}^{M}
\frac{q(\x_m)}{\phi(\x_m|\btheta)}
}{
\displaystyle \frac{1}{N}\sum_{n=1}^{N}
\left(\frac{q(\y_n)}{\phi(\y_n|\btheta)}\right)^2
},$ \\
       &   & &  \\
		\hline 
	\end{tabular}}
\end{table}

%



}

\newpage

\section{Numerical Simulations}\label{NumSect}

{
\subsection{Experiments with an analytically known $Z(\btheta)$}
}
In this section, we provide some numerical results comparing different estimators of $Z$ and $\btheta$, {considering analytically known partition function $Z(\btheta)$ (i.e., a completely known  ground-truth)}. We assume finite values of $N$ and $M$, instead of asymptotical performance as in other studies \cite{RiouDurand2019NCE}. The purpose of this section is not to show performance on a complex model, but rather to illustrate the behavior of the estimators computing the  mean square error (MSE), under controlled scenarios, helping the reproducibility as well.\footnote{The code used is publicly available at \url{http://www.lucamartino.altervista.org/PUBLIC_CODE_NCE_BRIDGE.zip}.} For this reason, we consider a univariate Gaussian  target distribution as model,
\begin{align}\label{ModelHere}
\post\left(y | \theta\right) & =\frac{1}{\sqrt{2 \pi \theta^2}} \exp \left(-\frac{y^2}{2 \theta^2}\right), \quad \mbox{ hence } \quad \phi(y | \theta)  =\exp \left(-\frac{y^2}{2 \theta^2}\right), \\
&\qquad\qquad\qquad\qquad\qquad\qquad\mbox{ and }\qquad Z(\theta)=\sqrt{2 \pi \theta^2},
\end{align}
so that we also know {analytically} the ground-truth $Z(\theta)=\sqrt{2 \pi \theta^2}$.
Thus, given $\theta_{\mathrm{tr}}=1$,we also observe the data $y_1, \ldots, y_N$ are generated from the model above, i.e.,
$$
y_n \sim \post(y) =  \post\left(y | \theta_{\mathrm{tr}}\right)=\frac{1}{Z\left(\theta_{\mathrm{tr}}\right)} \exp \left(-\frac{y^2}{2 \theta_{\mathrm{tr}}^2}\right), \quad Z_{\mathrm{tr}}=Z\left(\theta_{\mathrm{tr}}\right)=\sqrt{2 \pi \theta_{\mathrm{tr}}^2},
$$
with $n=1, \ldots, N$. We also consider a Gaussian proposal/reference density,
\begin{align}\label{PropHere}
 q(y)=\frac{1}{\sqrt{2 \pi \sigma_p^2}} \exp \left(-\frac{\left(y-\mu_p\right)^2}{2 \sigma_p^2}\right), 
\end{align}
where we set $\mu_p=0$ and vary the value of $\sigma_p$. 

\subsubsection{Estimation of the normalizing constant $Z\left(\theta_{\mathrm{tr}}\right)$}

\begin{figure}[htbp]
   \centering
   \centerline{
   \includegraphics[width=7cm]{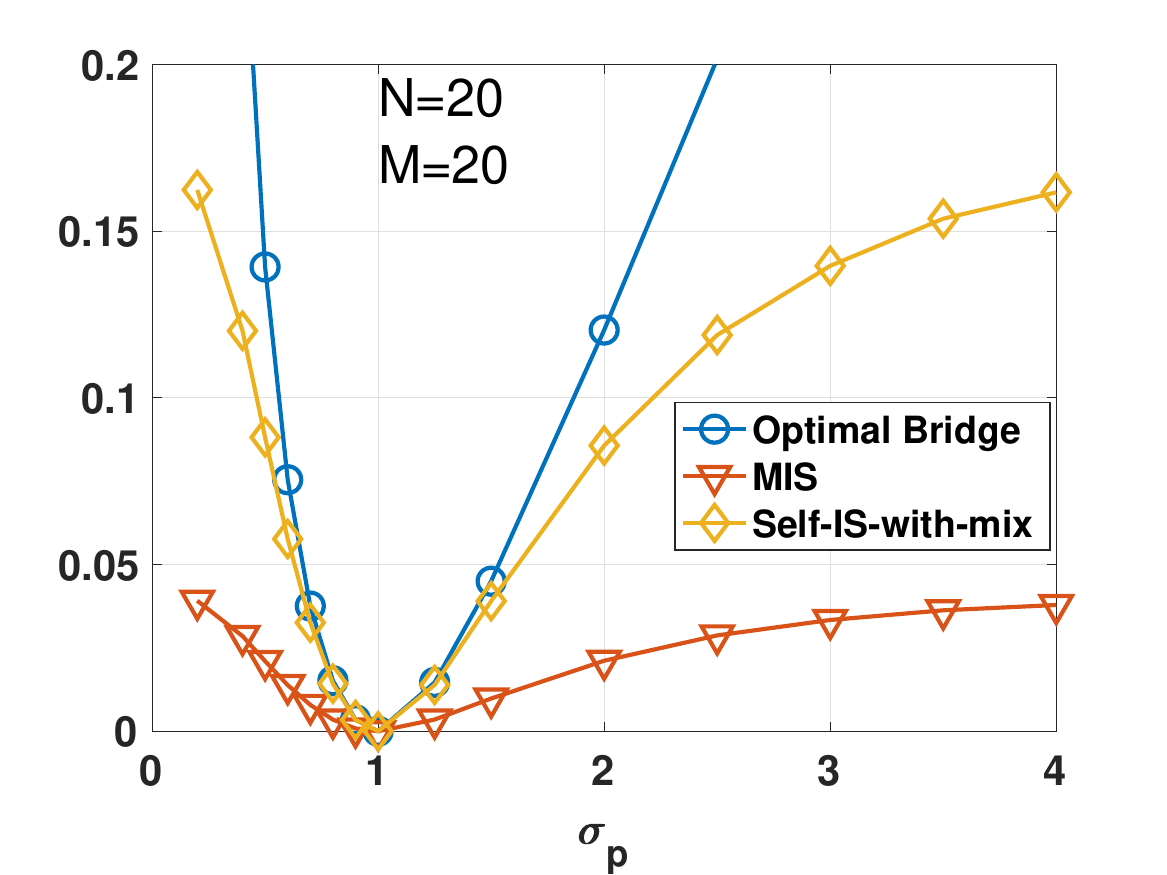} 
    \includegraphics[width=7cm]{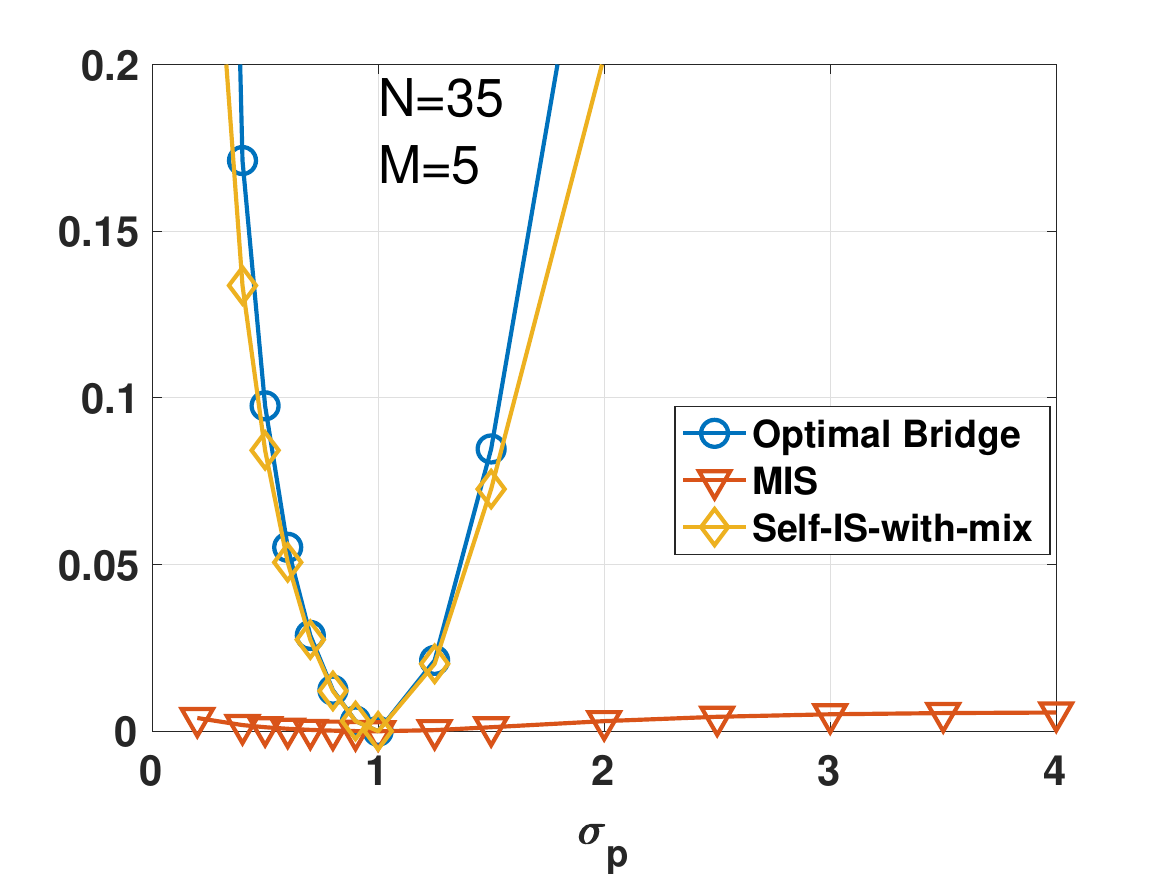} 
   \includegraphics[width=7cm]{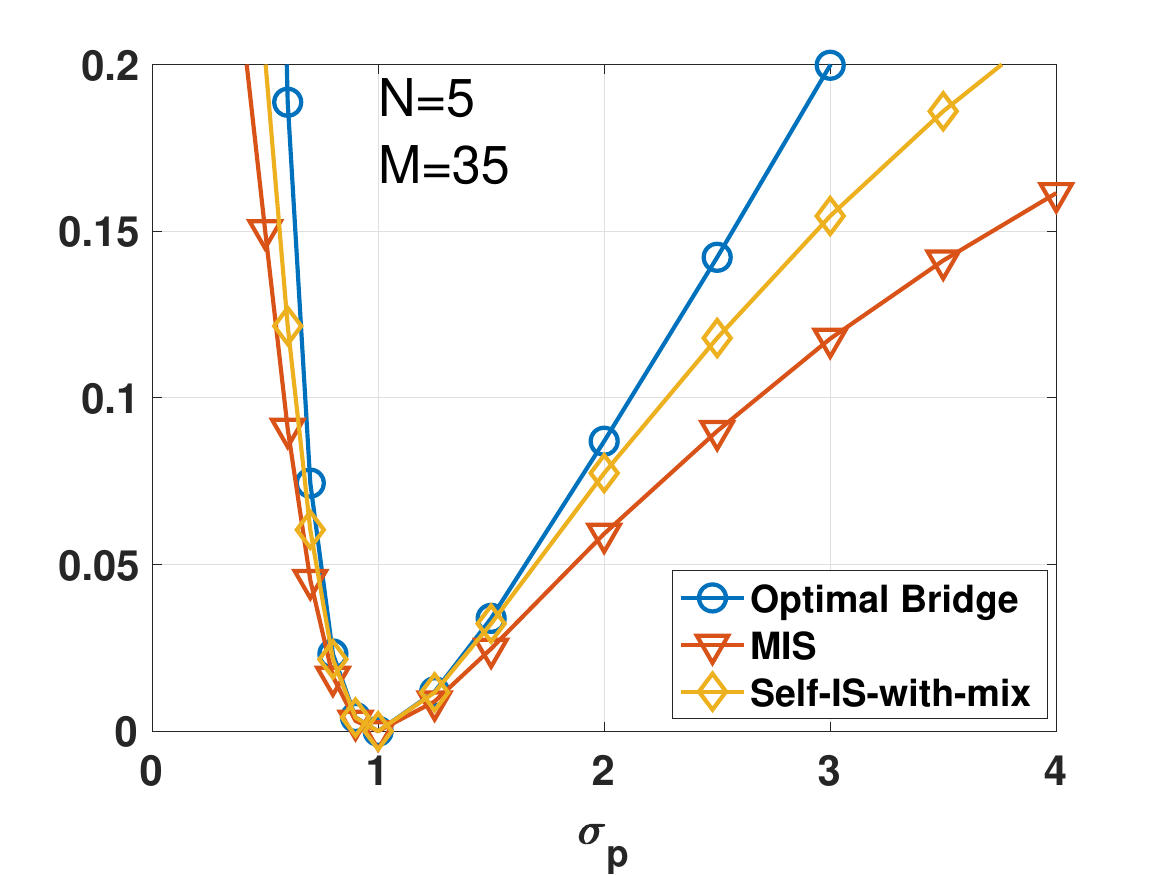} 

    }
   \caption{(Ideal scenario) MSE in the estimation of $Z_{\texttt{tr}}$ versus $\sigma_p$. We set $Z=Z_{\texttt{tr}}$ on the right side of Eqs. \eqref{optimal_bridge_eq}, \eqref{MIX-1}, and \eqref{MIX-3}, so that the resulting estimators do not require recursion. It can be interpreted as  $Z_{0}=Z_{\mathrm{tr}}$ and $T=1$. The figures differ for the numbers of $N\in\{5,20,35\}$ and $M\in\{5,20,35\}$ such that $N+M=40$. Surprisingly, the optimal bridge estimator provides the highest MSE values. }
   \label{fig:example1}
\end{figure}

\begin{figure}[htbp]
   \centering
   \centerline{
   \includegraphics[width=7cm]{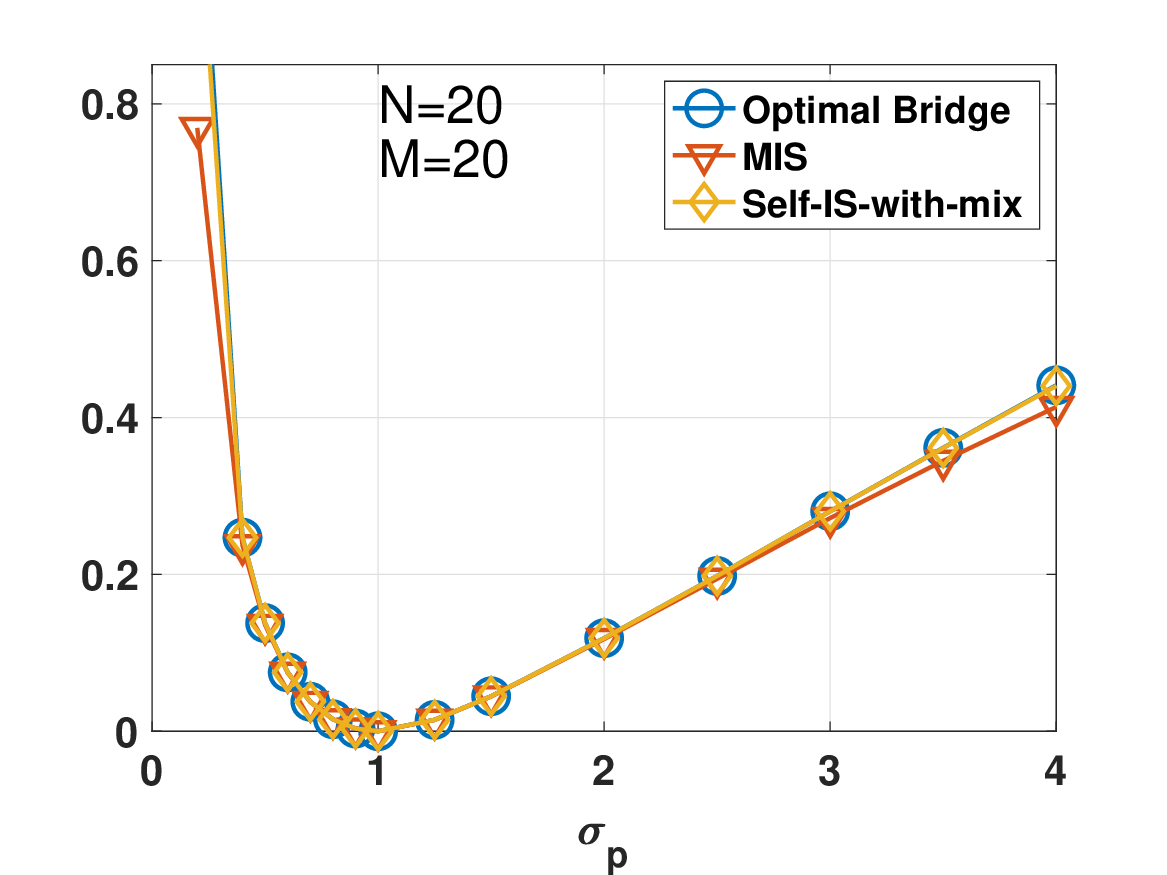} 
    \includegraphics[width=7cm]{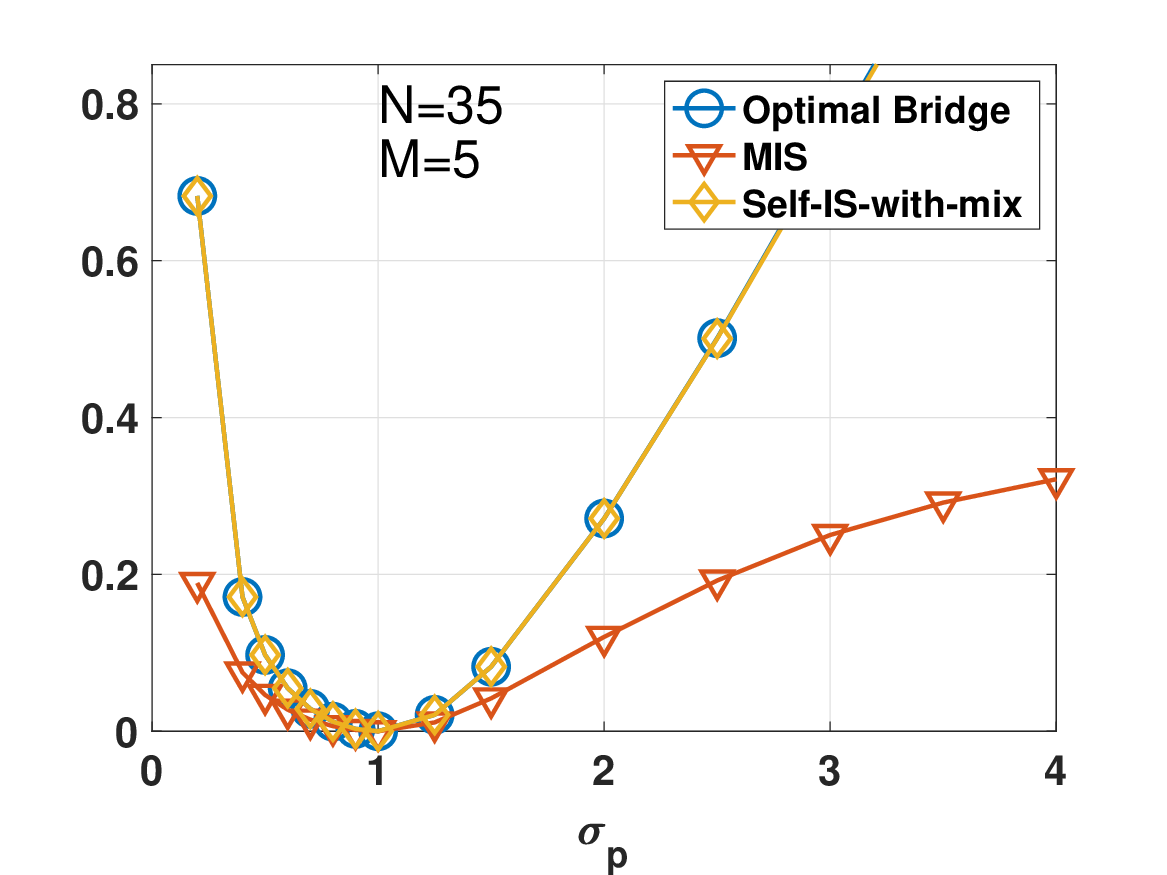} 
   \includegraphics[width=7cm]{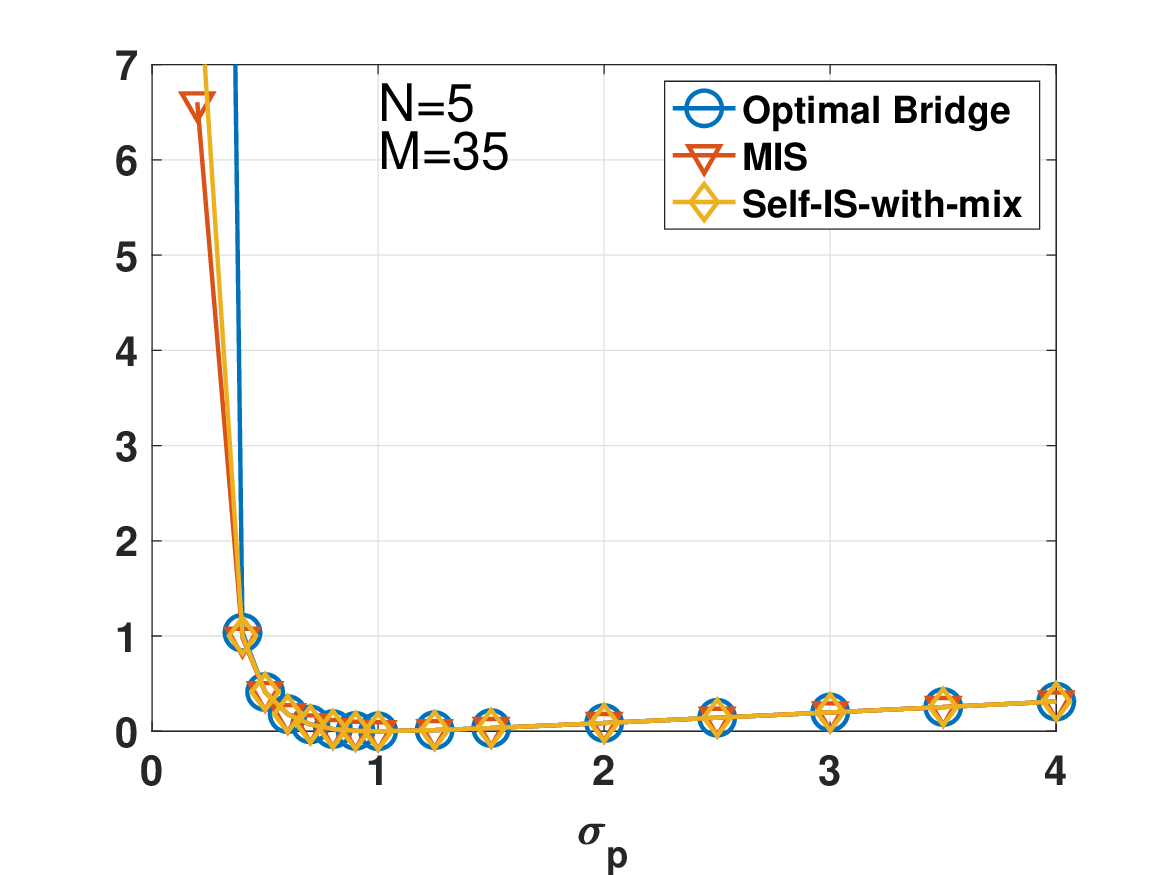} 
    }
   \caption{ (Almost-ideal scenario)  MSE in the estimation of $Z_{\texttt{tr}}$ versus $\sigma_p$. In this figure, we use $Z_{0}\approx Z_{\mathrm{tr}}$ and $T=10$. The figures differ for the numbers of $N\in\{5,20,35\}$ and $M\in\{5,20,35\}$ such that $N+M=40$.}
   \label{fig:example15}
\end{figure}

\begin{figure}[htbp]
   \centering
   \centerline{
   \includegraphics[width=7cm]{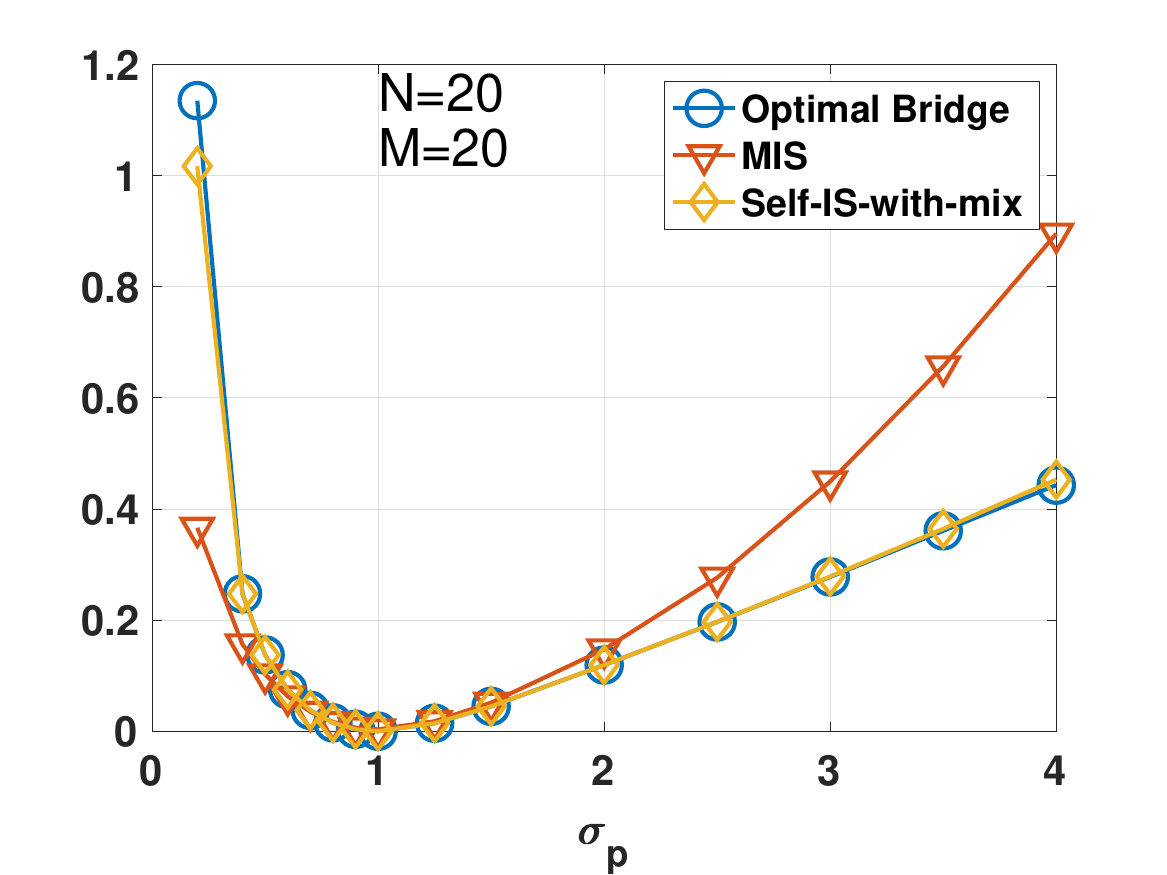} 
    \includegraphics[width=7cm]{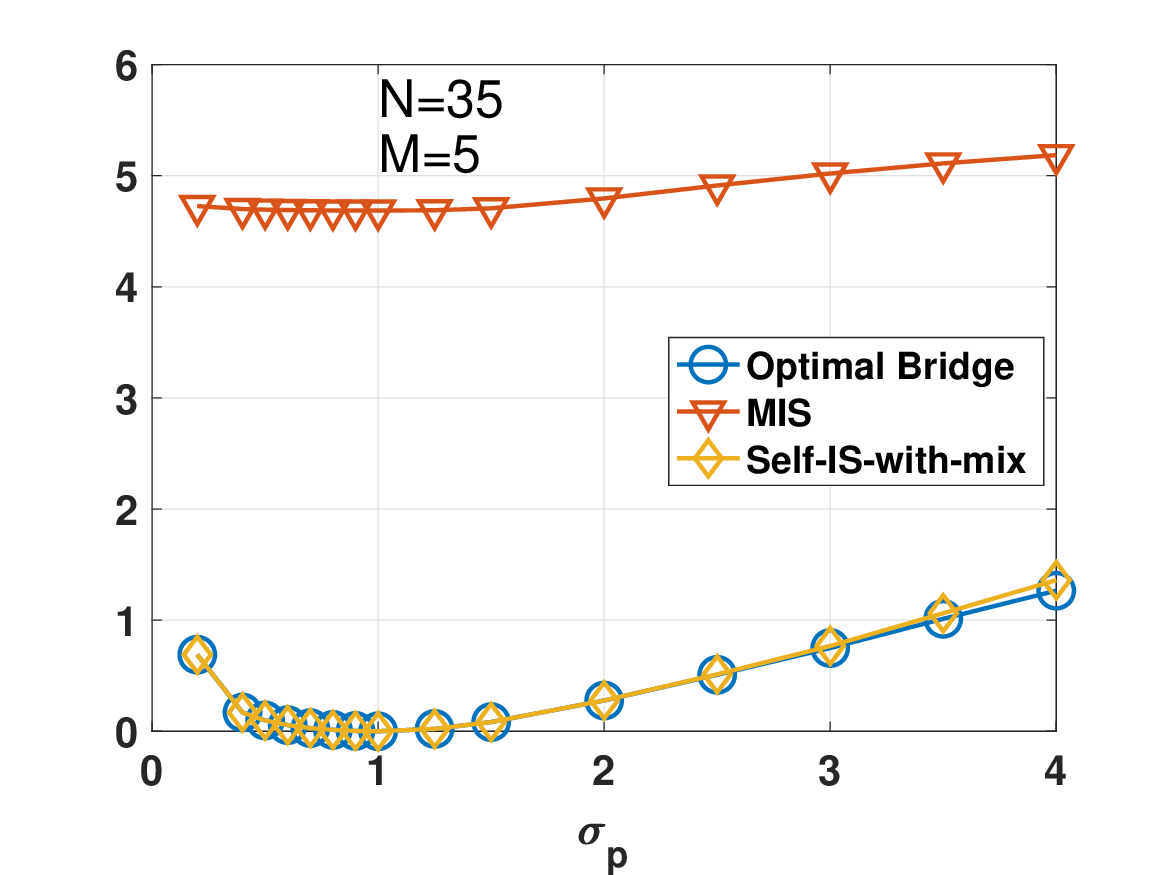} 
   \includegraphics[width=7cm]{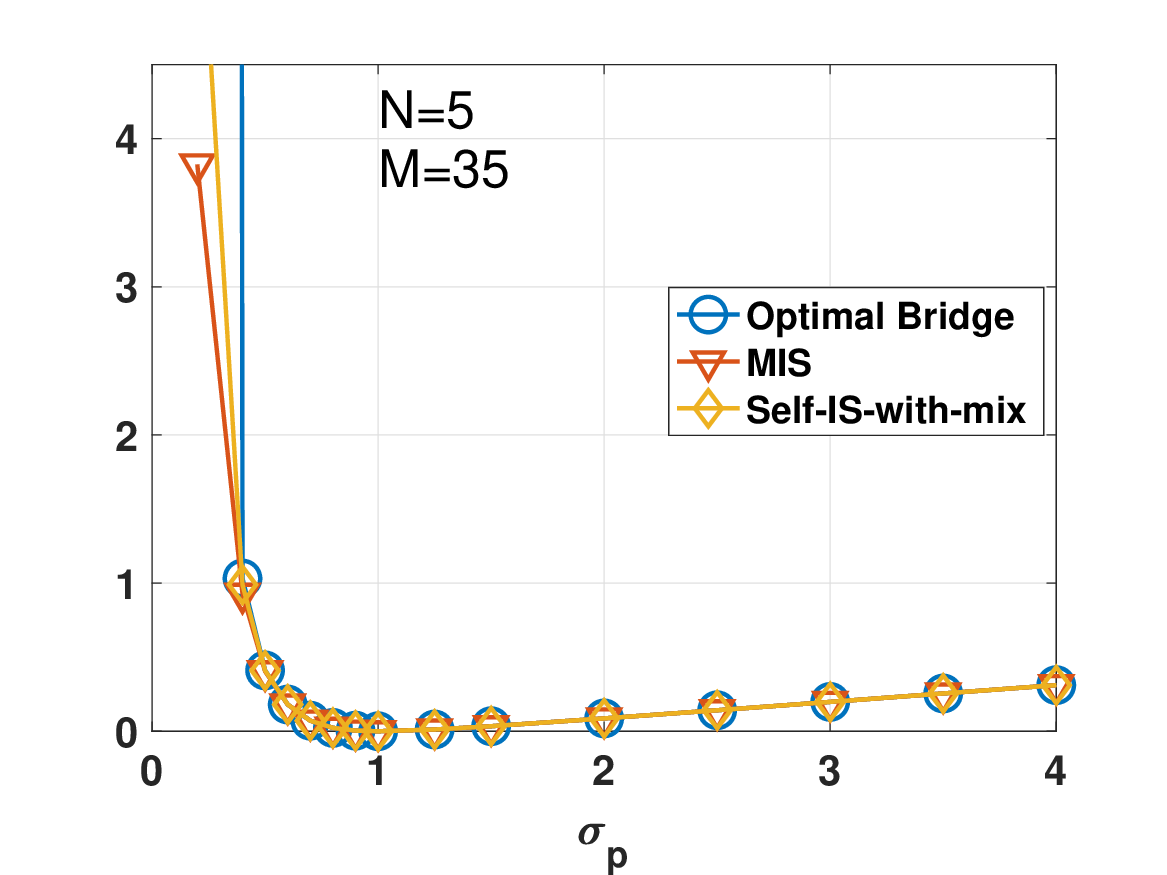} 
    }
   \caption{(Realistic scenario 1) MSE in the estimation of $Z_{\texttt{tr}}$ versus $\sigma_p$. In this figure, we use $Z_{0}=0.1$ and $T=10$. The figures differ for the numbers of $N\in\{5,20,35\}$ and $M\in\{5,20,35\}$ such that $N+M=40$.}
   \label{fig:example2}
\end{figure}

\begin{figure}[htbp]
   \centering
   \centerline{
   \includegraphics[width=7cm]{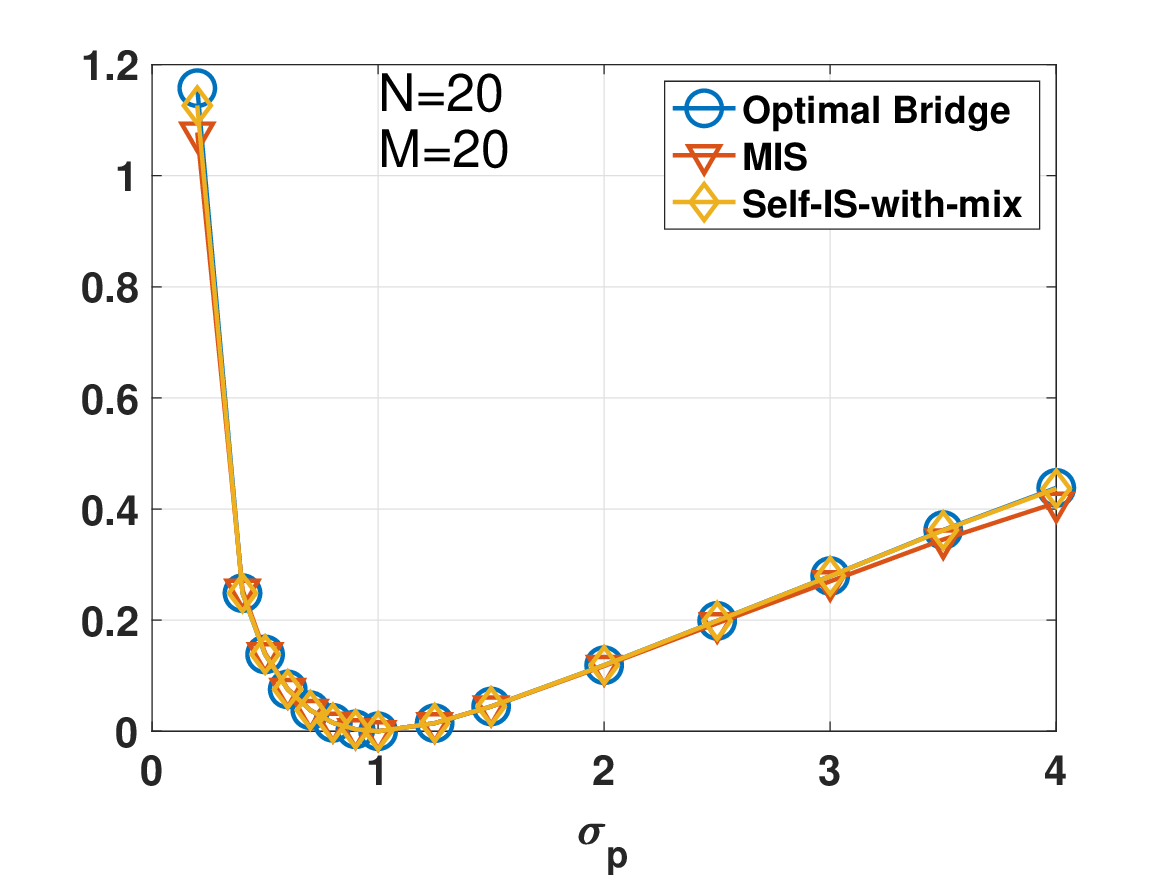} 
    \includegraphics[width=7cm]{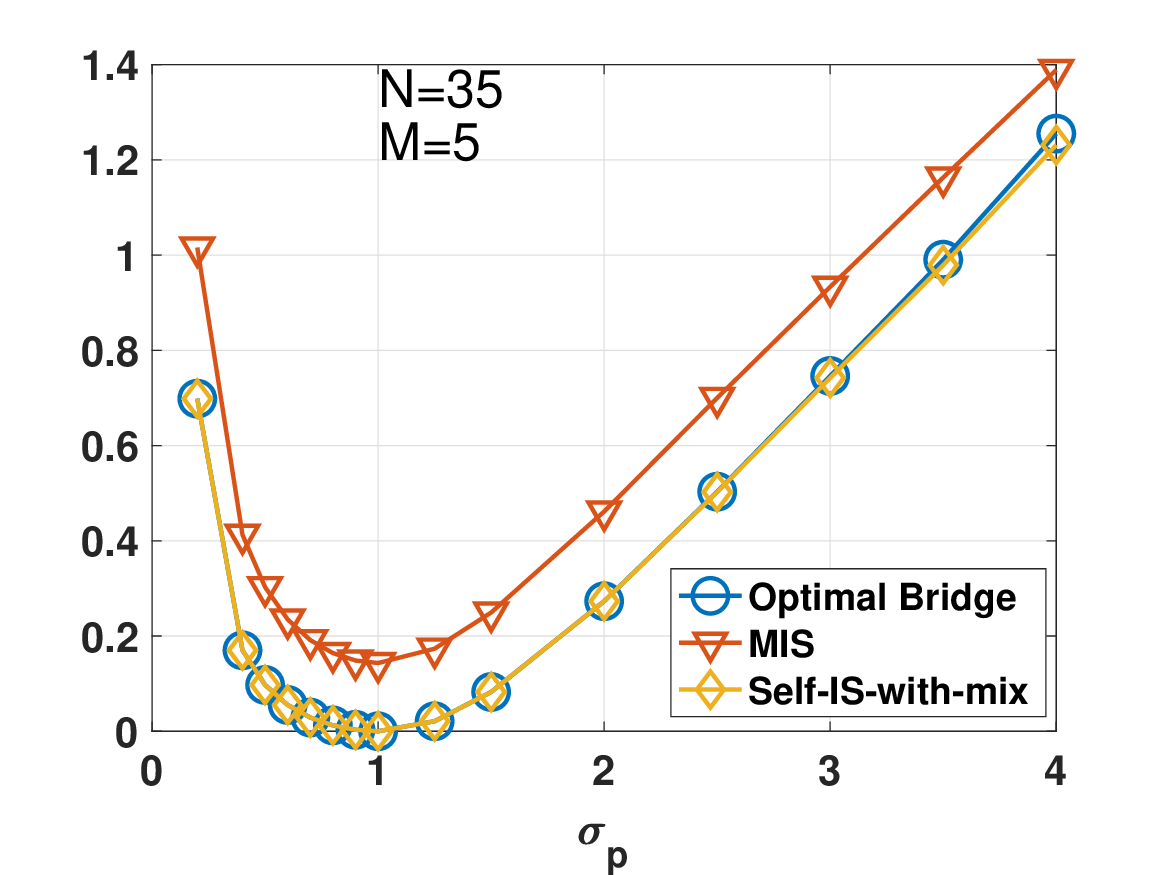} 
   \includegraphics[width=7cm]{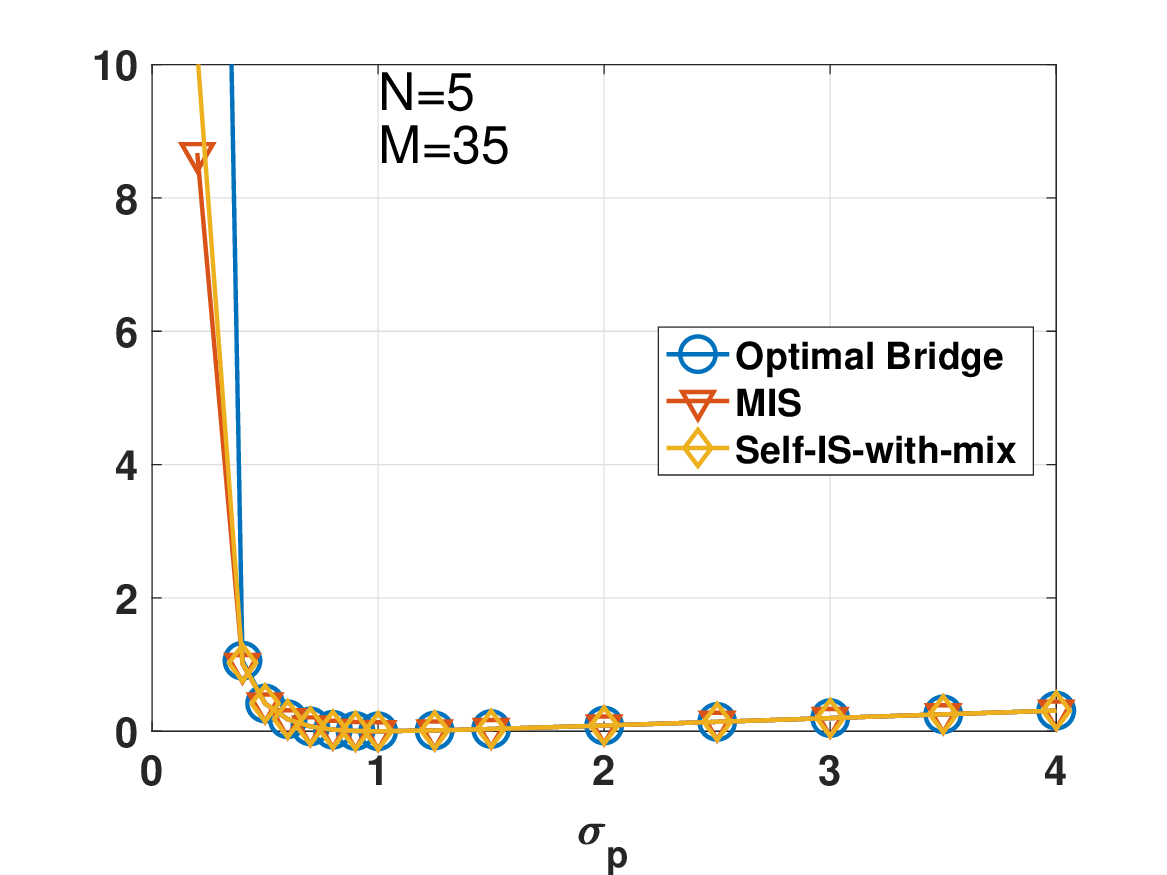} 
    }
   \caption{(Realistic scenario 2) MSE in the estimation of $Z_{\texttt{tr}}$ versus $\sigma_p$. In this figure, we use $Z_{0}=5$ and $T=10$. The figures differ for the numbers of $N\in\{5,20,35\}$ and $M\in\{5,20,35\}$ such that $N+M=40$.}
   \label{fig:example3}
\end{figure}

Setting $\theta_{\texttt{tr}}=1$ and generating $\{y_n\}_{n=1}^N$ by the model above, the goal is to estimate $ Z_{\mathrm{tr}}=Z\left(\theta_{\mathrm{tr}}\right)$ employing three estimators that use sets of samples from both densities, $x_m \sim q(y)$ and $y_n \sim \post(y)$, and require recursion. They are (a) the optimal bridge sampling, (b) the MIS and (c) the Self-IS-with-mix estimators, which are summarized in Table \ref{TableFF}. The comparison is done in terms of mean square error (MSE) versus different values of $\sigma_p$. The results are averaged over $10^6$ independent runs.
We set $M+N=40$, considering the three cases (a) $M=20$, $N=20$, (b) $M=5$, $N=35$ and (c) $M=35$, $N=5$. Furthermore,
  we consider four scenarios, one ideal and three more realistic scenarios, corresponding to whether we can evaluate $\post(y)=\frac{1}{Z}\phi(y)$ in the right side of the estimators (ideal and impossible scenario) or we can only evaluate $\phi(y)$ (realistic scenarios):
\begin{itemize}
\item {\bf Ideal scenario.} We replace $Z=Z_{\texttt{tr}}$ on the right side of Eqs. \eqref{optimal_bridge_eq}, \eqref{MIX-1}, and \eqref{MIX-3}, so that the resulting estimators do not require recursion. This setting can also be interpreted as initializing the iterative procedure at the true value, $Z_{0}=Z_{\mathrm{tr}}$ (i.e., a very good initialization), and performing a single iteration step, i.e., $T=1$. The first scenario is  for illustration purposes. The results are given in Figure \ref{fig:example1}. 
\item {\bf Almost-ideal scenario.} This is a realistic scenario since we apply the recursion using with  $T=10$ iterative steps. However, we start $Z_{0}\approx Z_{\texttt{tr}}$ very close to the true value. The corresponding results are given in Figure \ref{fig:example15}. 
\item {\bf Realistic scenario 1.} We set again $T=10$, but the initializing point is $Z_{0}= 0.1$. The corresponding  results are given in Figure \ref{fig:example2}. 
\item {\bf Realistic scenario 2.} We set again $T=10$, but the initializing point is  $Z_{0}= 5$. The corresponding results are given in Figure \ref{fig:example3}. 
\end{itemize}   
{\bf Results in ideal scenario.} As shown in Figure \ref{fig:example1},  the optimal bridge estimator provides the worst results in terms of MSE, whereas the MIS estimator provides the best results in line with the studies \cite{OwenZhou2000,MIS2019} that consider estimators where the proposal density (hence the denominators of the weights) can be completely evaluated. However, this is not a realistic case in our framework.
\newline
{\bf Results in the rest of scenarios.}  As shown in Figures \ref{fig:example15}, \ref{fig:example2}, and \ref{fig:example3}, the optimal bridge sampling gives the best results in the realistic scenarios, but the results of  the  Self-IS-with-mix  estimator \eqref{MIX-3} are very close and tends to be better for small values of $\sigma_p$ (smaller than $\theta_{\mathrm{tr}}=1$ that is the true standard deviation of the model). The MIS estimator provides the worst results except in Figure \ref{fig:example15} where we use a very good initialization, where provides the best results.




\subsubsection{Different cost functions for estimating ${\bf \theta}_{\texttt{tr}}$}

\begin{figure}[htbp]
   \centering
   \centerline{
   \includegraphics[width=10cm]{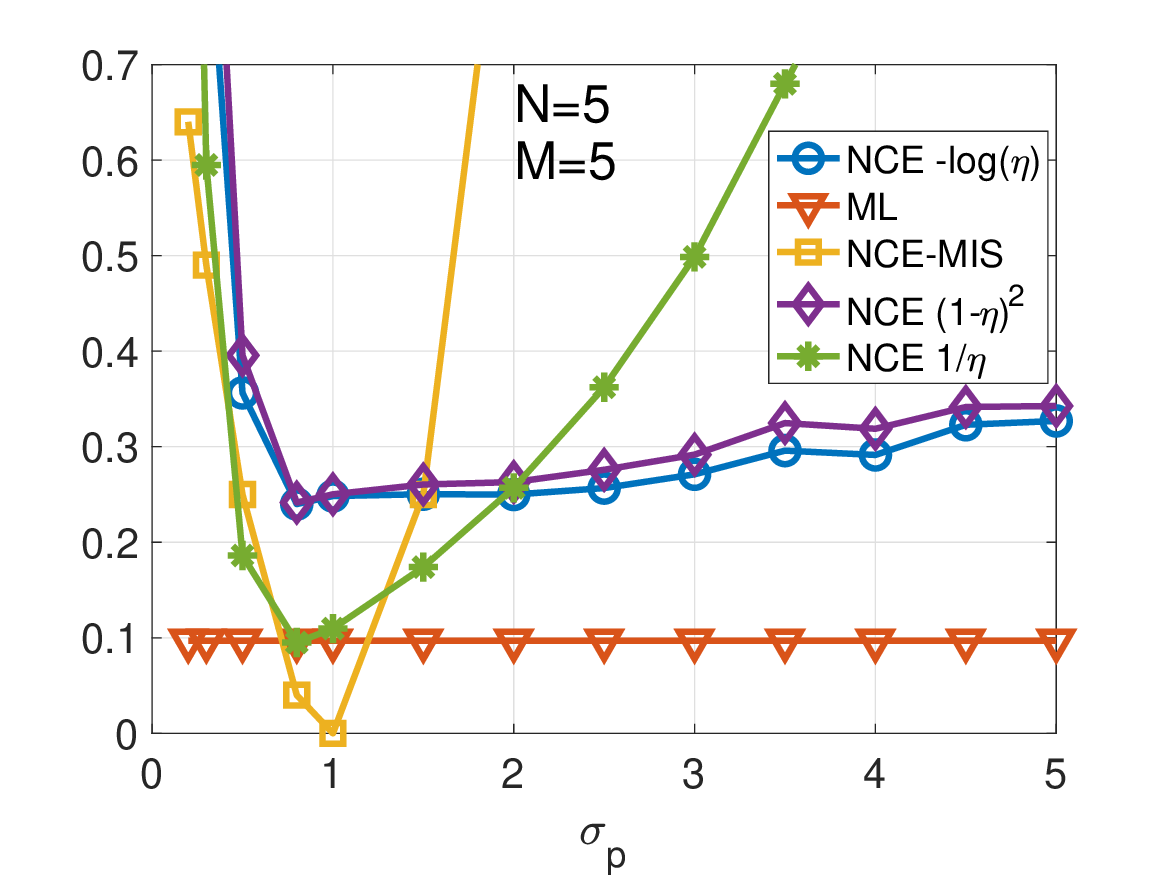} 
    \includegraphics[width=10cm]{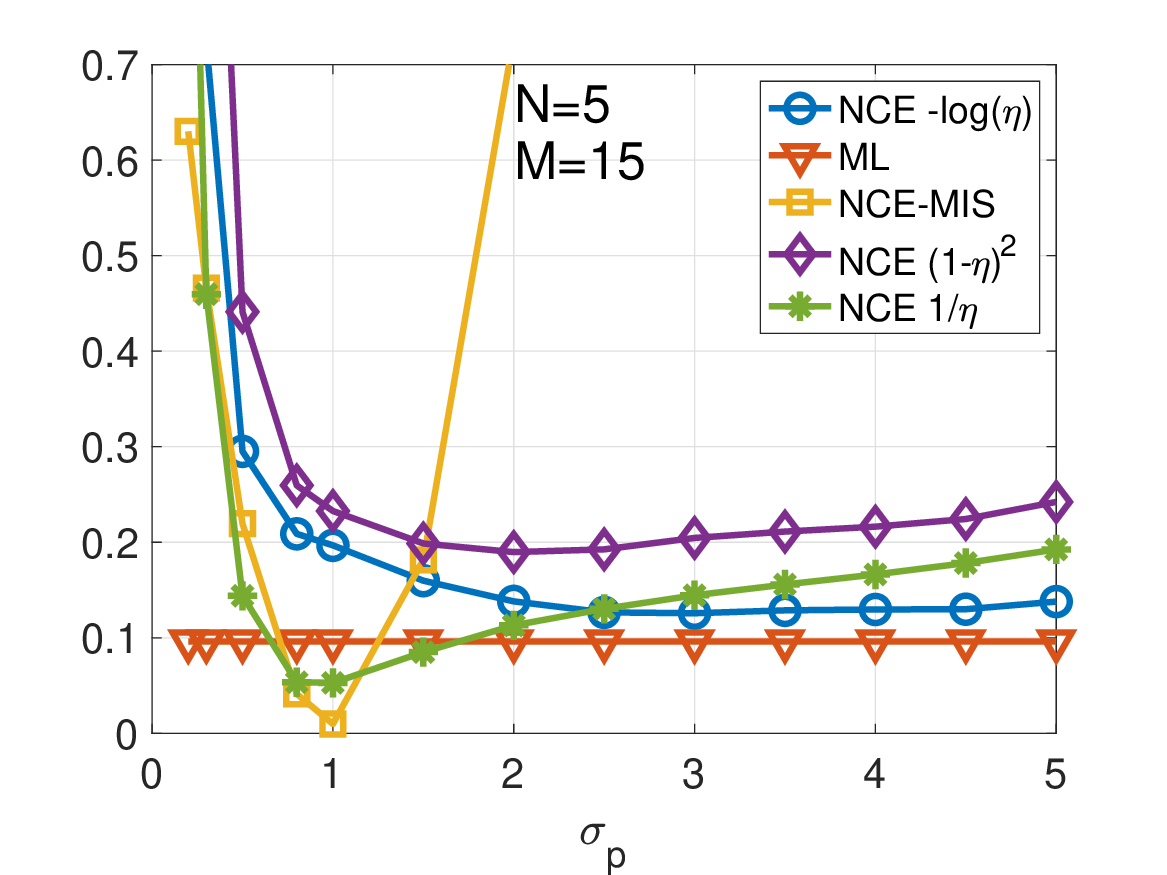} 
    }
      \centerline{
   \includegraphics[width=10cm]{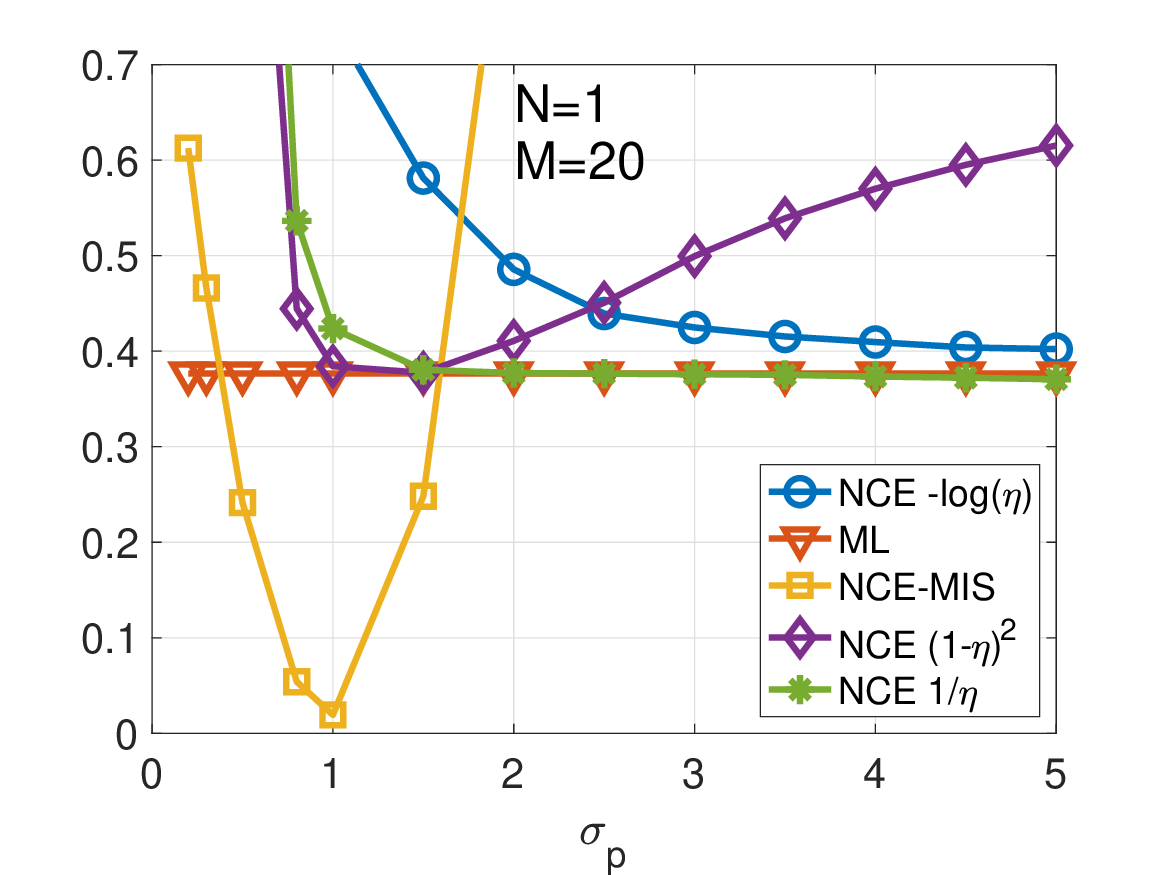} 
    \includegraphics[width=10cm]{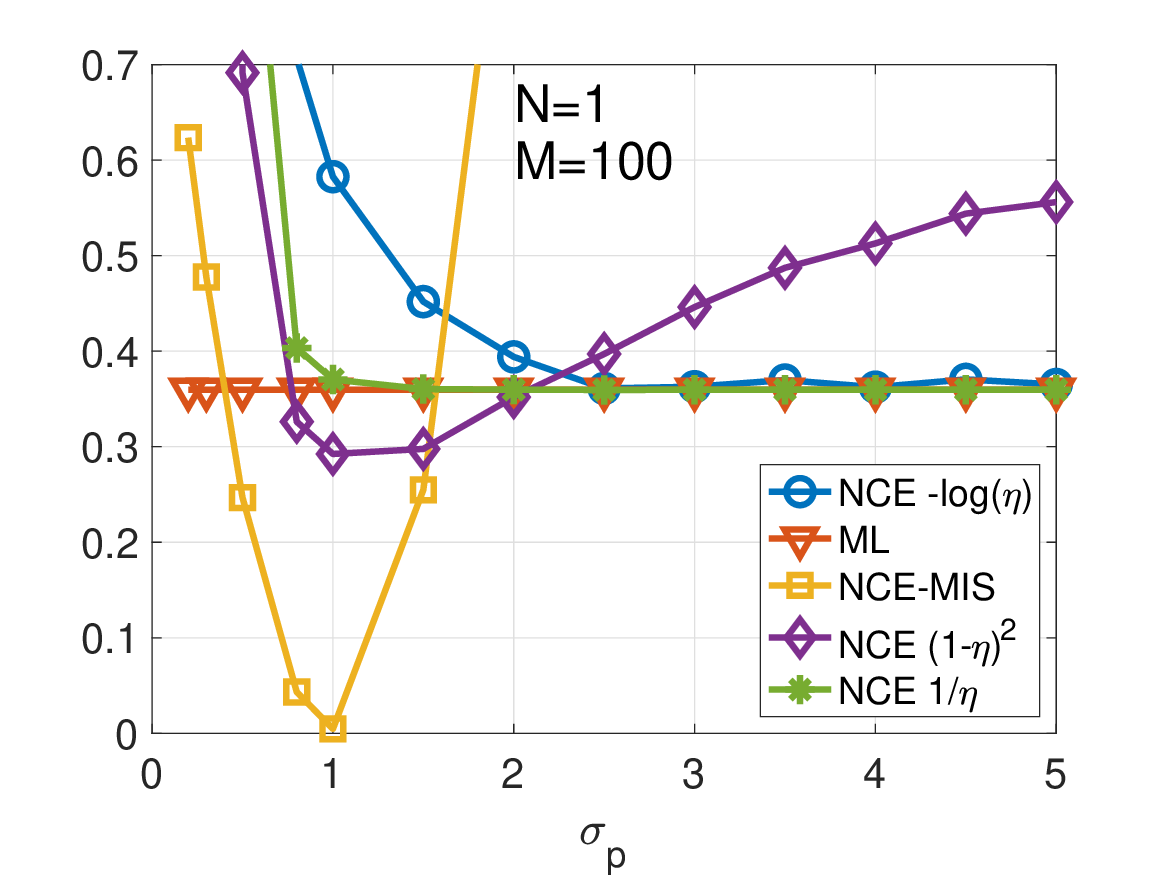} 
    }
   \caption{ MSE in the estimation of ${\bf \theta}_{\texttt{tr}}=1$ versus $\sigma_p$ (standard deviation of the proposal/reference density), for different values of $N$ and $M$. }
   \label{fig:example4}
\end{figure}

In this sub-section, we focus on the estimation of $ \theta_{\texttt{tr}}=1$ in EBM, fixing the true normalizing constant $Z_{\texttt{tr}}=Z(\theta_{\texttt{tr}})=\sqrt{2 \pi \theta_{\mathrm{tr}}^2}$ in the cost functions to minimize.  For the sake of simplicity, we assume again the model in Eq. \eqref{ModelHere} and the same proposal density in Eq. \eqref{PropHere}. 
\newline
We test different cost functions. We consider the cost function $J(\theta)=J(\theta, Z_{\texttt{tr}})$ in Eq. \eqref{GenNCEeq} with different choices of $V(\eta)$, more specifically:
\begin{itemize}
 \item  $V(\eta)=-\log(\eta)$ as in Eq. \eqref{CLeqFin0}, 
 \item $V(\eta)=(1-\eta)^2$, 
 \item   $V(\eta)=1/\eta$ and
 \item   $J_{\texttt{MIS}}(\theta)=J_{\texttt{MIS}}(\theta, Z_{\texttt{tr}})$ in Eq. \eqref{EqJmis}. 
 \end{itemize}
 Moreover, since $Z_{\texttt{tr}}$ is assumed to be known, we can also compare with the maximum likelihood (ML) estimator \cite{Geyer1994Convergence,Geyer1999LikelihoodInference}, which relies solely on $\left\{y_n\right\}_{n=1}^N$ and does not depend on the proposal density or on $\left\{x_m\right\}_{m=1}^M$.
 \newline
  We compute the MSE in estimation of  ${\bf \theta}_{\texttt{tr}}=1$ averaged over $5000$ independent runs. We vary the standard deviation $\sigma_p$ of the proposal density.  Since the ML solution does not depend on the proposal density, its MSE remains constant with respect to variations in $\sigma_p$. We also consider different pairs of $N$ and $M$ values, $\{N=5,M=5\}$, $\{N=5,M=15\}$, $\{N=1,M=20\}$  and  $\{N=1,M=100\}$. 
\newline
\newline
{\bf Results.}  The curves MSE versus $\sigma_p$ as depicted in Figure \ref{fig:example4}. Each figure corresponds to a pair of values of $N$ and $M$.  We can observe that the classical NCE with  $V(\eta)=-\log(\eta)$  generally yields good performance, particularly for larger values of $\sigma_p$, where its MSE approaches that of the ML solution. However, for certain values of $\sigma_p$, other cost functions seem to perform better specially for values of  $\sigma_p$ around the true value $\theta_{\texttt{tr}}$ (that is the standard deviation of the model).  Moreover, as $M$ grows and the classes are more unbalanced (having less true data $N$ and more artificial data $M$), other options of $V(\eta)$ seem to work better than $V(\eta)=-\log \eta$. 
 The cost function $J_{\texttt{MIS}}$ depends strongly on the choice of $\sigma_p$.
 Generally, the choice of the proposal is also a  relevant topic. 
The optimal proposal seems to be different for each cost functions \cite{Chehab2022OptimalNoise,Chehab2023OptimizingNoise,Llorente2025OptimalIS}. The analysis of these results suggests that, for $J_{\texttt{MIS}}$, the optimal proposal may be $q_{\texttt{opt}}(y)=\post(y|\theta_{\texttt{tr}})$.

{

\subsubsection{Joint estimation of ${\bf \theta}_{\texttt{tr}}$ and $Z_{\texttt{tr}}$}\label{SectjointEst}
Now, assuming again the same model and proposal, we focus on the joint estimation in the space $(\btheta,Z)$ of $ \theta_{\texttt{tr}}=1$ and the true normalizing constant $Z_{\texttt{tr}}=Z(\theta_{\texttt{tr}})=\sqrt{2 \pi \theta_{\mathrm{tr}}^2}$, considering $Z$ as free parameter in the optimization problem. We estimate both $\btheta_{\texttt{tr}}$ $Z_{\texttt{tr}}$. However, to ensure consistency with the existing literature and make our results directly comparable with previous contributions, we focus on the estimation error associated with $\btheta_{\texttt{tr}}$. Indeed, in frequentist inference for EBMs, the primary quantity of interest is typically the model parameter $\btheta_{\texttt{tr}}$, whereas estimating $Z_{\texttt{tr}}$ is often an intermediate step required to perform likelihood-based inference.
For the other hand, recall that estimating the partition function $Z(\btheta)$ over the parameter space is particularly useful for Bayesian inference and, more generally, for model selection. This also includes the estimation of $Z_{\texttt{tr}}$, as discussed in more detail in Section \ref{IMP_Z_sect}.
\newline
We test again different cost functions $J(\btheta, Z)$. For instance,
we consider {\bf (a)} $V(\eta)=-\log(\eta)$ (i.e., standard NCE) 
 {\bf (b)}$V(\eta)=(1-\eta)^2$,  and {\bf (c)}   $V(\eta)=1/\eta$  in Eq. \eqref{GenNCEeq}. Moreover, we test also  {\bf (d)} the Pihala's  objective function  in Eq. \eqref{PijamaEqEmp}, {\bf (e)} the special case in Eq. \eqref{RIScostfun} denoted as J-RIS and {\bf (f)}  the other special case in Eq. \eqref{SpecialPigiama2} denoted as J-sq. The results (averaged over $5000$ runs), in terms of MSE in log-scale versus $\sigma_p$ of the proposal, are given in Figure \ref{fig:example5}.  The standard NCE scheme appears to perform best as $N\to\infty$, consistently with several theoretical and practical findings reported in the literature. However, for finite values of $N$ and $M$, alternative objective functions may yield better results as shown in the figure. Furthermore, the performance of all methods is affected by the choice of $\sigma_p$. As commonly observed also in standard IS schemes, all the methods are particularly sensitive to proposal distributions that are too narrow, with values of $\sigma_p$ substantially smaller than the standard deviation of the target distribution leading to a marked deterioration in performance. Finally, we note that the J-sq cost function achieves results comparable to those of standard NCE.

\begin{figure}[htbp]
   \centering
   \centerline{
   \includegraphics[width=9cm]{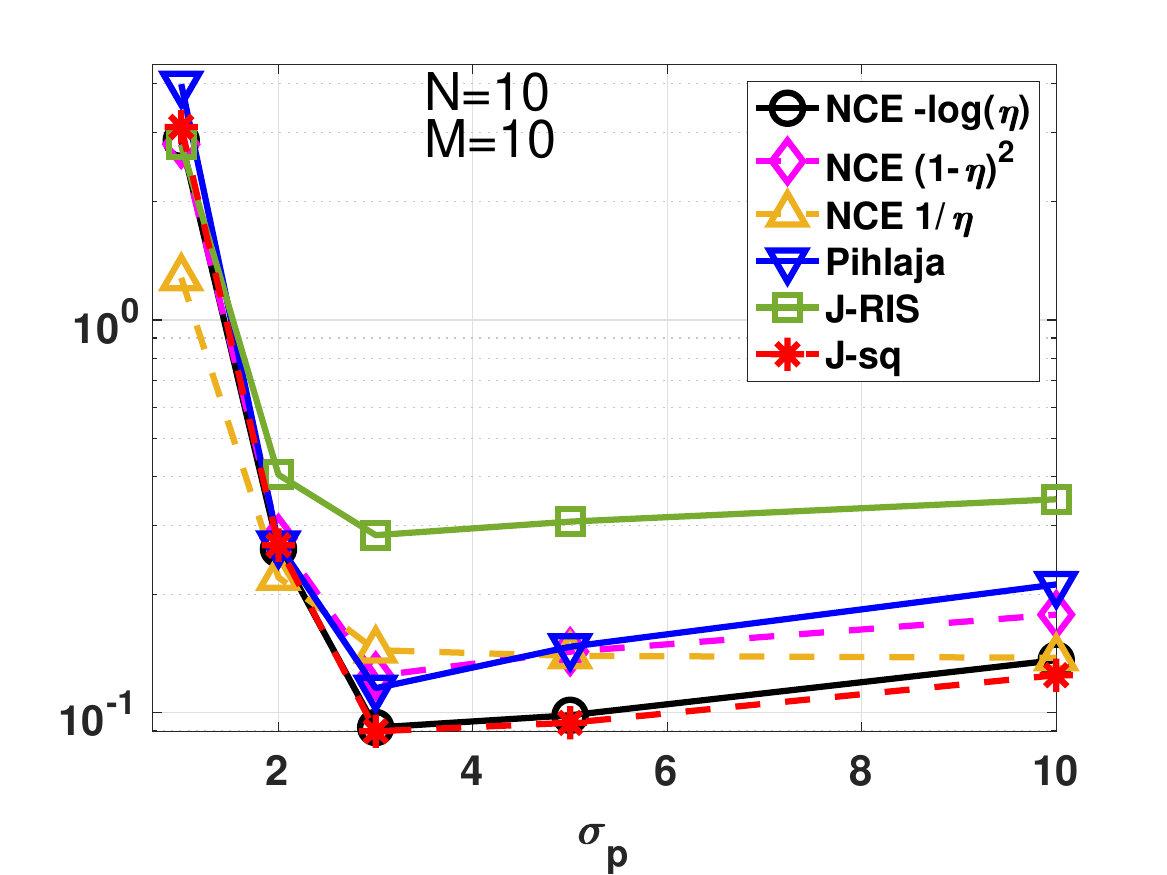} 
    \includegraphics[width=9cm]{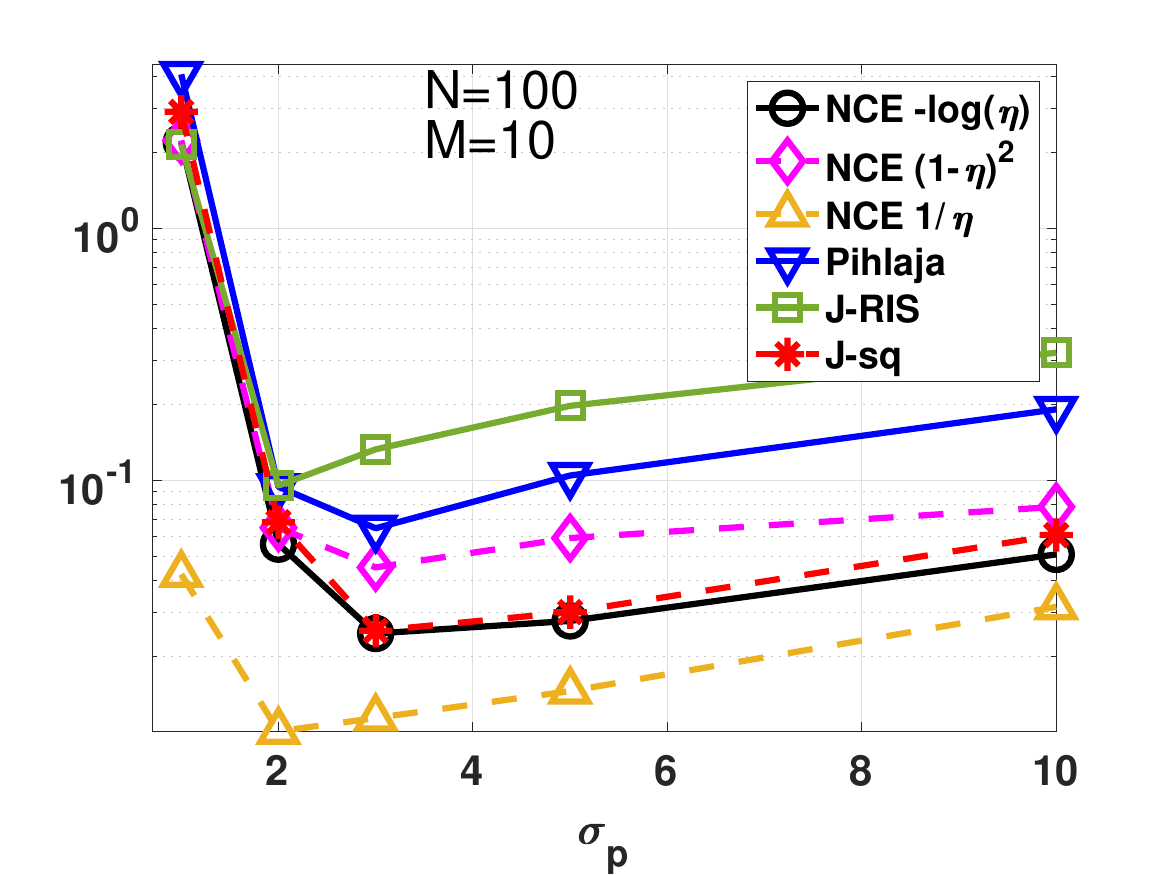} 
    }
       \centerline{
   \includegraphics[width=9cm]{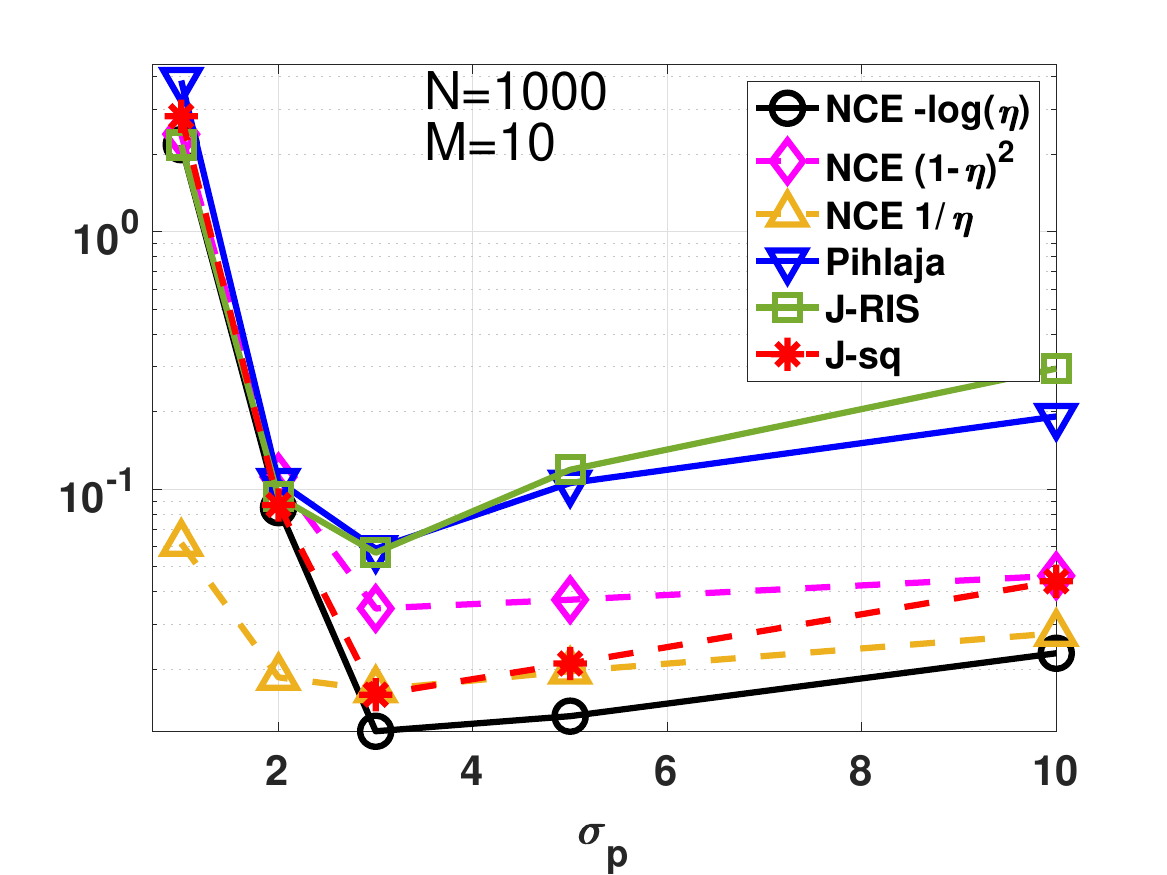} 
    \includegraphics[width=9cm]{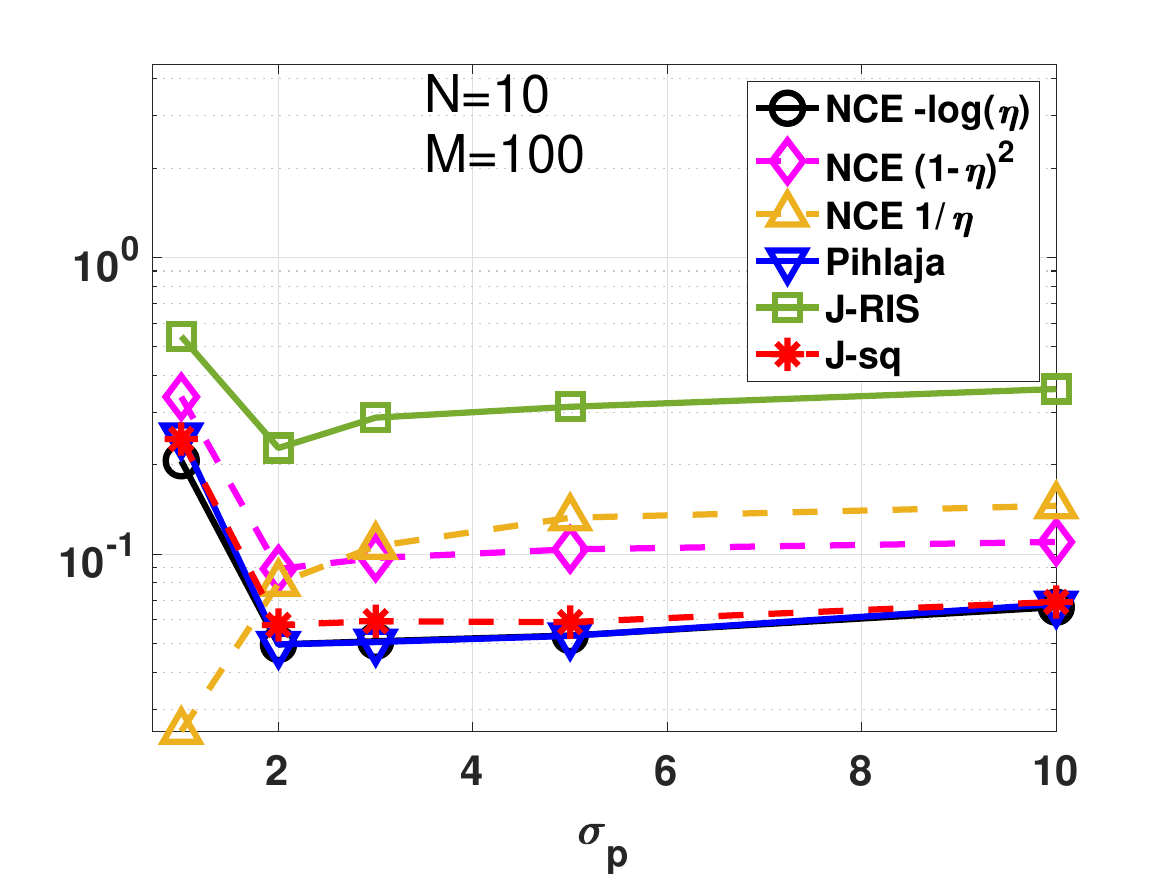} 
    }
   \caption{{ Results minimizing the cost functions in the joint space $(\btheta,Z)$ (Section \ref{SectjointEst}), treating $Z$ as free parameter, and estimating ${\bf \theta}_{\texttt{tr}}$ and $Z_{\texttt{tr}}=Z({\bf \theta}_{\texttt{tr}})$. Figures depict the MSE in the estimation of ${\bf \theta}_{\texttt{tr}}$   versus $\sigma_p$ (standard deviation of the proposal/reference density), for different values of $N$ and $M$.} }
   \label{fig:example5}
\end{figure}

\subsection{Experiments with an EBM}\label{SectEBMsimu}

{
In this section, we consider a nonlinear, non-Gaussian EBM in which the observed variable $y\in\mathbb{R}$ is characterized by an energy function parameterized by a scalar parameter $\theta\in\mathbb{R}$. We deliberately adopt this simple one-dimensional setting to make ground-truth results available, facilitate the interpretation of the underlying concepts, and keep the numerical experiments easy to implement and reproduce. Specifically, we consider the following nonlinear quartic energy function:
\begin{align}
E(y|\theta)=y^4- \theta y^2+16,
\end{align}
so that the corresponding unnormalized density is $ \phi(y|\theta)=\exp\left[-E(y|\theta)\right]$. Recall that  the partition function is defined as $Z_{\texttt{tr}}(\theta)=
\int_{\mathbb{R}}\exp\left[-E(y\mid\theta)\right]\,dy$.
Despite the apparent simplicity of this one-dimensional model, the quartic form of the energy function makes the resulting probability distribution non-Gaussian and generally prevents the partition function from being expressed in terms of elementary closed-form functions. Depending on the parameter $\theta$, the energy function can exhibit either a single global minimum or two symmetric global minima, giving rise to a unimodal or bimodal probability distribution, respectively. At the same time, the positive quartic term ensures that the energy grows sufficiently rapidly as \(|y|\) tends to infinity, so that the integral defining \(Z_{\texttt{tr}}(\theta)\) is finite for every finite value of \(\theta\). Thus, this model provides a simple yet informative example of a continuous EBM with a tractable unnormalized density but a nontrivial partition function.
\newline
At each run, we generate $N$ observations, $y_1,\ldots,y_N$, from the true model with $\theta_{\texttt{tr}}=2$, using Monte Carlo methods to obtain the samples. As in the previous experiment, we consider a Gaussian proposal (or reference) density,
$q(x)=\frac{1}{\sqrt{2\pi\sigma_p^2}}
\exp\left[-\frac{(x-\mu_p)^2}{2\sigma_p^2}\right]$,
with $\mu_p=0$. We also approximate $Z(\btheta)$ using Monte Carlo quadrature rules, for each value of $\btheta$, replacing this value in the different cost functions, we focus on the estimation of $\btheta_{\texttt{tr}}$.
 We assess the performance of the different methods under several experimental settings, varying the number of observations $N$, the number of proposal samples $M$, and the proposal standard deviation $\sigma_p$. Note that, in contrast to the first experiment, the proposal distribution $q$ considered here belongs to a different parametric family from that of the target model $\post$.
\newline
The results, averaged over $5000$ runs, are given in Figure \ref{fig:example6}. We test different cost functions and the maximum likelihood (ML) Geyer's approach.  The standard NCE scheme again appears to provide the best performance, yielding estimates closest to those obtained through  ML estimation. Other cost functions, such as J-sq and/or the Pihlaja's objective function, achieve very similar results, consistently with the findings reported in previous experiments. Once again, all methods exhibit considerable sensitivity to overly narrow proposal distributions: choosing values of $\sigma_p$ that are too small leads to a marked deterioration in performance. }

\begin{figure}[htbp]
   \centering
   \centerline{
   \includegraphics[width=9cm]{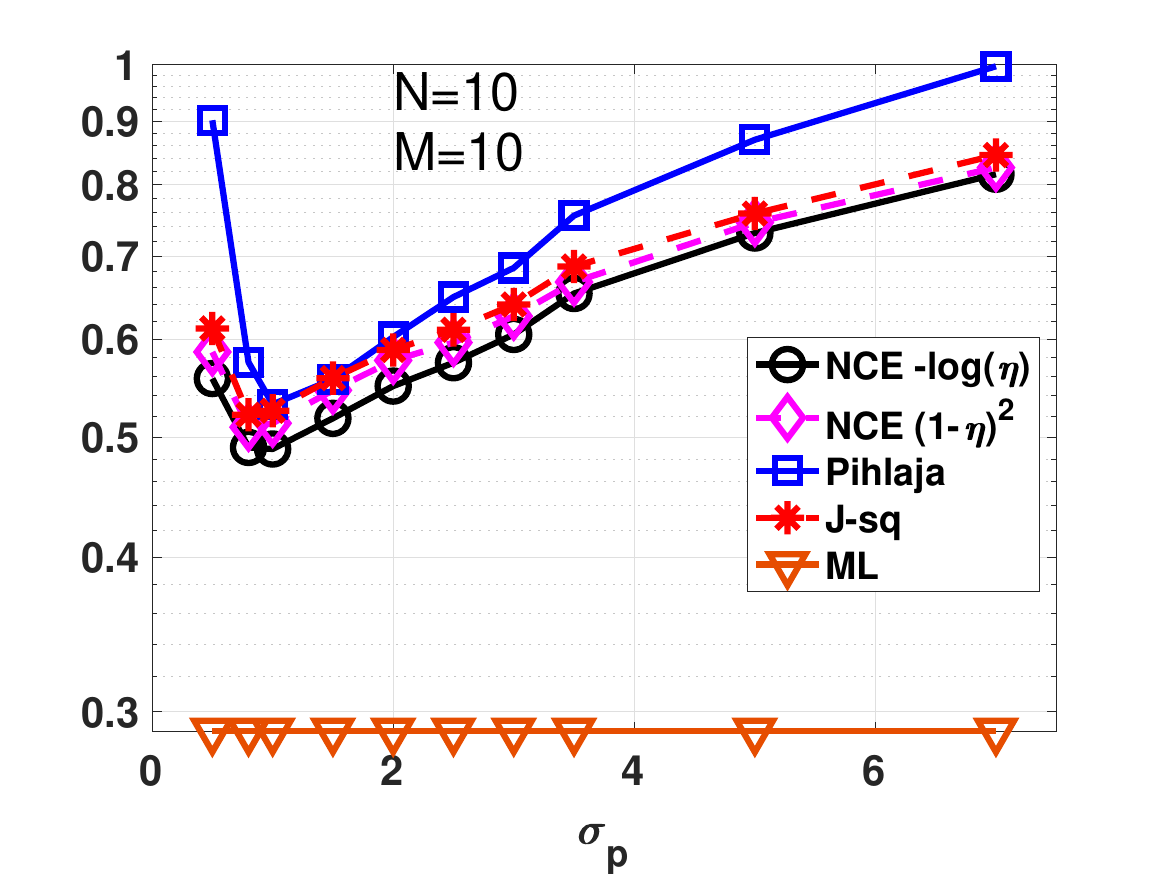} 
    \includegraphics[width=9cm]{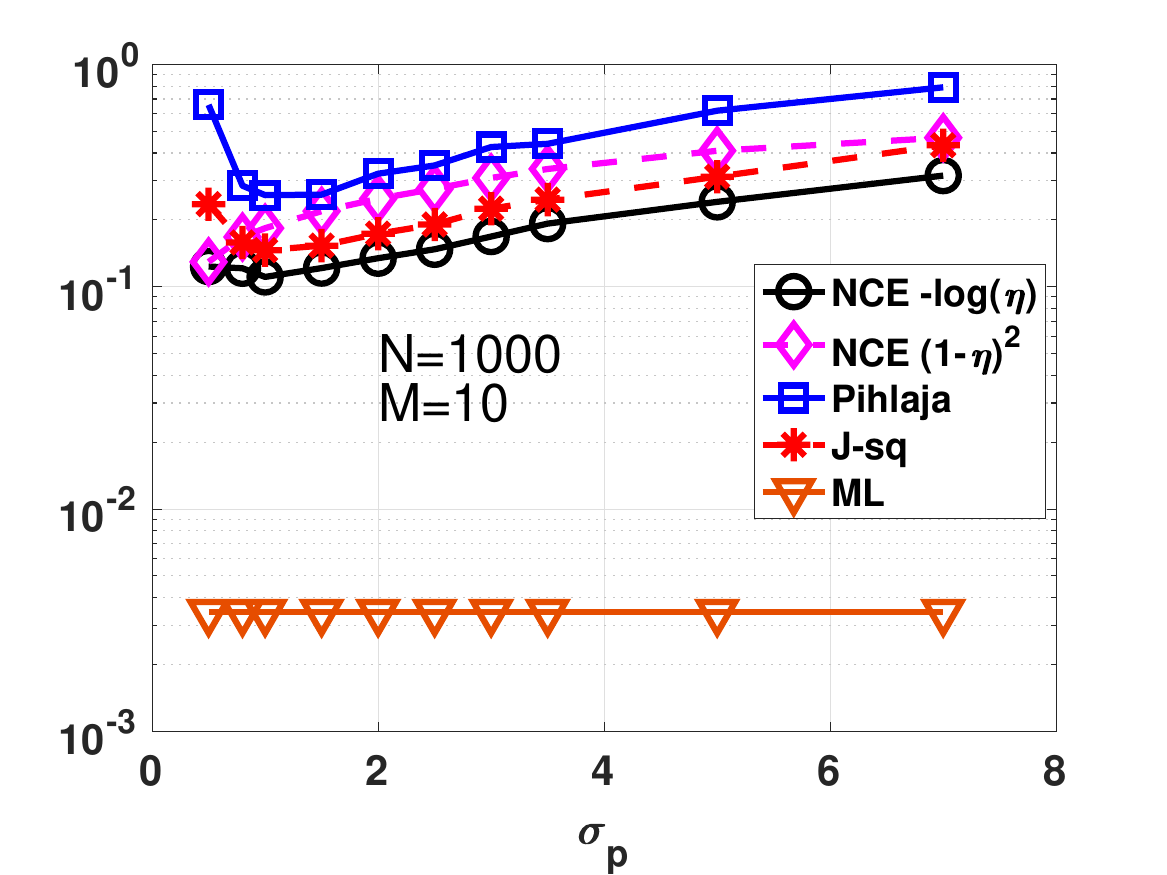} 
    }
       \centerline{
   \includegraphics[width=9cm]{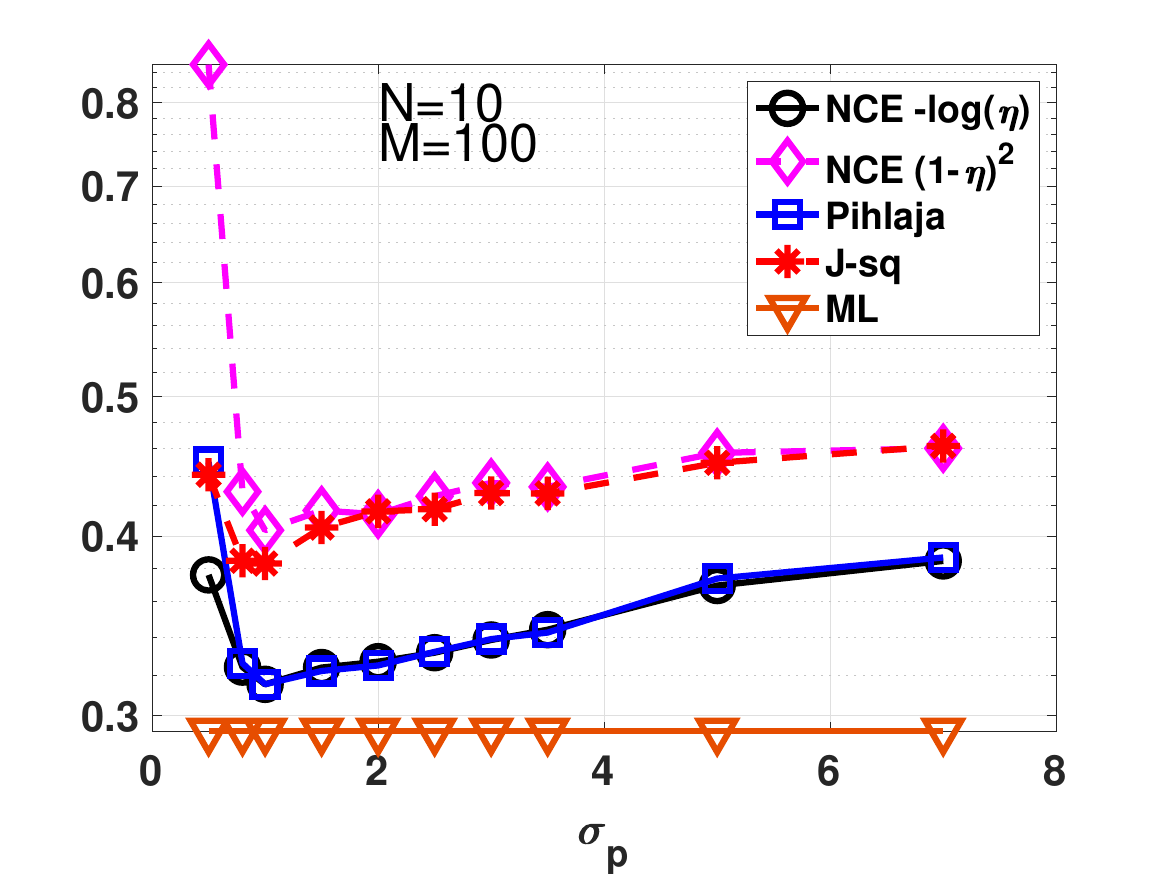} 
    \includegraphics[width=9cm]{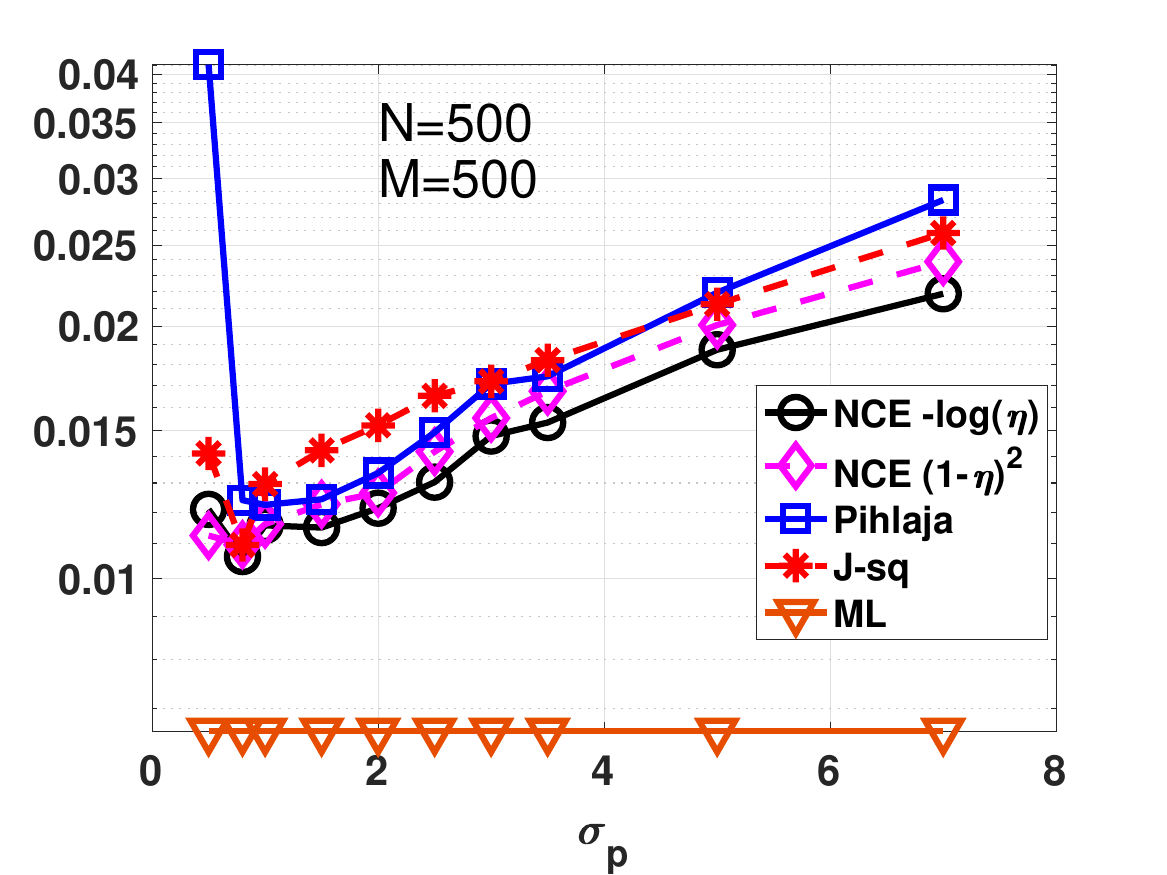} 
    }
   \caption{{ Results of the experiment in Section \ref{SectEBMsimu}. Figures depict the MSE in the estimation of ${\bf \theta}_{\texttt{tr}}$   versus $\sigma_p$ (standard deviation of the proposal/reference density), for different values of $N$ and $M$.} }
   \label{fig:example6}
\end{figure}
}

\section{Conclusions}
{ In this review paper, we provide a comprehensive overview and synthesis of contributions to noise-contrastive estimation (NCE) and related methods that have emerged across different research fields, highlighting their main ideas, methodological developments, and theoretical connections. Our aim is to offer a self-contained and accessible treatment, supported by clear and straightforward derivations, that brings these contributions together within a common framework and makes the main results, as well as the ideas underlying them, accessible to a broader readership.}  We have provided a unified perspective on several techniques that have been developed independently across the literature and different fields. We show the relationships among existing methods as NCE, multiple importance sampling, reverse logistic regression (RLR), and bridge sampling. This unified framework not only elucidates the relationships among existing methods, but also enables the principled design of novel estimators with potentially superior statistical and computational performance.
\newline
Contrastive learning, and in particular the NCE method \cite{Gutmann2010NCE,Gutmann2022Contrastive}, has become a widely adopted and highly successful approach, often regarded as a benchmark method. NCE is asymptotically equivalent to maximum likelihood estimation in the $\boldsymbol{\theta}$-space, as demonstrated in \cite{RiouDurand2019NCE,MartinoScaffidi26}, and it is also equivalent to the optimal bridge sampling solution in the $Z$-space \cite{Pihlaja2010,Chehab2023OptimizingNoise,GutmannHirayama2011}. This equivalence explains NCEÕs ability to estimate the normalizing constant and its success in the literature for inference in EBMs. Accordingly, NCE serves as a standard benchmark for frequentist inference in energy-based models. However, as shown in this work,  for specific choices of the proposal (or reference) density and for finite values of $N$ and $M$, alternative estimation schemes for $\boldsymbol{\theta}$ and $Z$ may yield improved performance. The related code has been made freely available to support reproducibility.
   This effect has been also highlighted in \cite{MartinoScaffidi26} regarding the inference in the $\boldsymbol{\theta}$-space. 
\newline
Recursive procedures commonly used for estimating normalizing constants $Z$ (as for the optimal bridge sampling) can also be incorporated into NCE optimization frameworks. Moreover, the joint selection of a specific scoring rule $V(\eta)$ and a proposal density $q(\mathbf{y})$ represents a promising direction for future research. { Other more sophisticated cost functions have been investigated \cite{Pihlaja2010,Chehab2023OptimizingNoise,GutmannHirayama2011,UeharaMatsudaKomaki2018}. The reduction of the optimization space,  based on the results presented here and in previous works, has been discussed in Section \ref{SuperGenOfPreviousPaper}. }
In addition, the use of multiple proposal distributions, for instance, those constructed through tempered versions of the EBM, represents a promising direction that deserves further investigation \cite{ChehabHyvarinenRisteski2023}. {This is particularly relevant for Bayesian inference, where such proposals may also be exploited to develop more efficient computational strategies \cite{Sonobe2026Contrastive}.}




{\small
\section*{{\small Acknowledgements}}
 L. Martino acknowledges  financial support by the  PIACERI Starting Grant BA-GRAPH (UPB 28722052144) of the University of Catania. 
  A preprint version of this work can be found at \url{https://arxiv.org/abs/2604.08116}. 
}

\bibliographystyle{IEEEtran}
\bibliography{biblio}

\appendix
 { 
\section{On scoring rules for binary classification}\label{Vscore}
Let $ \eta_{\mathrm{tr}}({\bm u})=P(a=1 | {\bm u})$ be the true conditional probability, and let $\eta({\bm u})$ be the probability predicted by a classifier. For a scoring rule $V(\eta)$, we can define the conditional risk
$$
R\left(\eta; \eta_{\mathrm{tr}}\right)=\eta_{\mathrm{tr}} V(\eta)+\left(1-\eta_{\mathrm{tr}}\right) V(1-\eta).
$$
The scoring rule can be:
\begin{itemize}
\item[-] {\bf Proper}, if $R\left(\eta ; \eta_{\mathrm{tr}}\right)$ is minimized at $\eta=\eta_{\mathrm{tr}}$.
\item[-] {\bf Strictly proper}, if this minimum is unique.
\item[-] {\bf Improper,} if the minimum is not necessarily attained at the true probability.
\end{itemize}
Therefore, strict properness is a sufficient condition for identifying the true conditional probability through population risk minimization.

\subsection{General conditions for proper scoring rules}
We can derive a general necessary and sufficient condition for propriety of your symmetric binary loss. You define the conditional risk as
$$
R\left(\eta ; \eta_{\text {tr }}\right)=\eta_{\text {tr }} V(\eta)+\left(1-\eta_{\text {tr }}\right) V(1-\eta),
$$
where $\eta$ is the predicted probability of class $a=1$ and $\eta_{\mathrm{tr}}$ is its true conditional probability.
A first-order condition is obtained by differentiating with respect to $\eta$:
$$
\frac{\partial R}{\partial \eta}=\eta_{\mathrm{tr}} V^{\prime}(\eta)-\left(1-\eta_{\mathrm{tr}}\right) V^{\prime}(1-\eta).
$$
For the scoring rule to be proper, the minimum must occur at $
\eta=\eta_{\mathrm{tr}}$. Substituting this condition into the derivative gives:
$\eta V^{\prime}(\eta)-(1-\eta) V^{\prime}(1-\eta)=0$.
Therefore, the general condition is
$$
\boxed{\eta V^{\prime}(\eta)=(1-\eta) V^{\prime}(1-\eta), \qquad 0<\eta<1.}
$$
{\bf Second-order condition.} Differentiating again,

$$
\frac{\partial^2 R}{\partial \eta^2}=\eta_{\mathrm{tr}} V^{\prime \prime}(\eta)+\left(1-\eta_{\mathrm{tr}}\right) V^{\prime \prime}(1-\eta) .
$$
At the true probability, $\eta=\eta_{\text {tr}}$, we have:
$$
R^{\prime \prime}(\eta ; \eta)=\eta V^{\prime \prime}(\eta)+(1-\eta) V^{\prime \prime}(1-\eta).
$$
A sufficient condition for strict propriety is that this quantity is strictly positive for every $0<\eta<1$, together with the first-order condition and appropriate boundary behaviour.

%
%
%
%
%

\subsubsection{Example: Brier score}

The Brier score for binary classification is $V(\eta)=(1-\eta)^2$. 
Let $\eta_{\mathrm{tr}}$ denote the true probability of class $a=1$. The expected loss is
\begin{align*}
R\left(\eta ; \eta_{\mathrm{tr}}\right)&=\eta_{\mathrm{tr}}(1-\eta)^2+\left(1-\eta_{\mathrm{tr}}\right) \eta^2=\eta_{\mathrm{tr}}-2 \eta_{\mathrm{tr}} \eta+\eta^2 .
\end{align*}
Computing the first derivative and Setting it equal to zero, we have:
$$
\frac{\partial R}{\partial \eta}=-2 \eta_{\mathrm{tr}}+2 \eta =0, \Longrightarrow \eta^*=\eta_{\mathrm{tr}}.
$$
Moreover, the second derivative is $\frac{\partial^2 R}{\partial \eta^2}=2>0$.
Therefore, this expected loss has a unique global minimum at the true probability. Hence,  $V(\eta)=(1-\eta)^2$ is a strictly proper scoring rule.
}

  { 
\subsection{Improper scoring rules}

With an  improper scoring rule, the risk is not minimized exactly at $\eta_{\mathrm{tr}}({\bf u})$. Suppose the population risk is minimized at
$$
\eta^*({\bf u})=T\left(\eta_{\mathrm{tr}}({\bf u})\right),
$$
where $T$ is a transformation different from the identity.
Then, even with infinitely many observations, the classifier converges to $\eta^*({\bf u})$, rather than to the true conditional probability  $\eta_{\mathrm{tr}}({\bf u})$. This is a form of inconsistency for probability estimation. However, there is an important qualification: if the transformation $T$ is known and invertible, it may be possible to recover the true probability through calibration. Thus, an improper scoring rule does not necessarily prevent consistent estimation of the underlying parameter.

\subsubsection{Example with $V(\eta)=1 / \eta$ } \label{AppBantes}
Consider first the population conditional risk, where the true probability of class 1 is $\eta_{\mathrm{tr}}$, i.e., $
R\left(\eta ; \eta_{\mathrm{tr}}\right)=\eta_{\mathrm{tr}} V(\eta)+\left(1-\eta_{\mathrm{tr}}\right) V(1-\eta)$, and
For $V(\eta)=1 / \eta$, we have:
$$
R\left(\eta ; \eta_{\mathrm{tr}}\right)=\frac{\eta_{\mathrm{tr}}}{\eta}+\frac{1-\eta_{\mathrm{tr}}}{1-\eta}.
$$
Differentiating with respect to $\eta$ and setting the derivative equal to zero, we obtain the following stationary condition:
$$
\frac{\partial R}{\partial \eta}=-\frac{\eta_{\mathrm{tr}}}{\eta^2}+\frac{1-\eta_{\mathrm{tr}}}{(1-\eta)^2}=0, \Longrightarrow \frac{\eta_{\mathrm{tr}}}{\eta^2}=\frac{1-\eta_{\mathrm{tr}}}{(1-\eta)^2} .
$$
By taking the positive square root of both sides, we obtain,
$$
\frac{\sqrt{\eta_{\mathrm{tr}}}}{\eta}=\frac{\sqrt{1-\eta_{\mathrm{tr}}}}{1-\eta} .
$$
Hence, with some algebra, we can solve with respect to $\eta$:
\begin{align*}
\sqrt{\eta_{\mathrm{tr}}}(1-\eta)&=\sqrt{1-\eta_{\mathrm{tr}}} \eta, \\
\left(\sqrt{\eta_{\mathrm{tr}}}+\sqrt{1-\eta_{\mathrm{tr}}}\right) \eta&=\sqrt{\eta_{\mathrm{tr}}},
\end{align*}
which finally gives
$$
\eta^*=T\left(\eta_{\mathrm{tr}}\right)=\frac{\sqrt{\eta_{\mathrm{tr}}}}{\sqrt{\eta_{\mathrm{tr}}}+\sqrt{1-\eta_{\mathrm{tr}}}}.
$$
Therefore, $V(\eta)=1 / \eta$ is not proper (as expected),
 \(\eta^*\neq\eta_{\mathrm{tr}}\), i.e., because the optimal prediction is not generally $\eta_{\mathrm{tr}}$. The solution is a transformed version of the true probability. However, It is  strictly consistent for the transformed probability $T\left(\eta_{\mathrm{tr}}\right)$. Namely, since $T$ is strictly increasing and invertible, we can compute $\eta_{\mathrm{tr}}=T^{-1}(\eta^*)$. Moreover, we can construct a proper scoring rule.
Indeed, since
$$
T^{-1}(u)=\frac{u^2}{u^2+(1-u)^2},
$$
we can define the transformed scoring rule as:
$$
\widetilde{V}(u)=V\left(\frac{u^2}{u^2+(1-u)^2}\right),
$$
where $V(\eta)=1 / \eta$. Replacing $V(\eta)=1 / \eta$ above gives
$$
\boxed{\widetilde{V}(u)
=1+\frac{(1-u)^2}{u^2}.}
$$
This is a proper scoring rule when used for both classes in the symmetric form, i.e., the following risk:
$$
R\left(u ; \eta_{\mathrm{tr}}\right)=\eta_{\mathrm{tr}} \widetilde{V}(u)+\left(1-\eta_{\mathrm{tr}}\right) \widetilde{V}(1-u),
$$
has a unique optimum exactly at $u=\eta_{\mathrm{tr}}$.
}

 { 
\subsection{Additional observations on $V(\eta)=\frac{1}{\eta}$ and variants} \label{AppB}

Let write the cost function $J(\btheta,Z)$ corresponding to $V(\eta)=\frac{1}{\eta}$, that is:
\begin{align}
J(\btheta,Z)&=\sum_{n=1}^N\frac{\phi\left({\bf y}_n|  \boldsymbol{\theta}\right)+ \nu Z q\left({\bf y}_n\right) }{\phi\left({\bf y}_n |  \boldsymbol{\theta}\right)}+ \sum_{m=1}^M\frac{\phi\left({\bf x}_m|  \boldsymbol{\theta}\right)+ \nu Z q\left({\bf x}_m\right) }{\nu Zq({\bf x}_m)},  \nonumber\\
&=\nu Z \sum_{n=1}^N\frac{q\left({\bf y}_n\right) }{\phi\left({\bf y}_n |  \boldsymbol{\theta}\right)}+ \frac{1}{\nu Z}  \sum_{m=1}^M\frac{\phi\left({\bf x}_m| \boldsymbol{\theta}\right)}{q({\bf x}_m)} +N+M, \nonumber \\
&=\frac{M}{N} Z \sum_{n=1}^N\frac{q\left({\bf y}_n\right) }{\phi\left({\bf y}_n |  \boldsymbol{\theta}\right)}+ \frac{N}{M Z}  \sum_{m=1}^M\frac{\phi\left({\bf x}_m| \boldsymbol{\theta}\right)}{q({\bf x}_m)} +C.
\end{align}
Note that this is quite similar in Eq. \eqref{EstaConAppB}.
If we remove the factors \(M\) and \(N\) multiplying the two summations (similarly as in the correction introduced to obtain \(\widehat{Z}_{\texttt{geo}}\)), we can further adjust the expression above achieving an even closer formula:
\begin{align}
\boxed{J_{\texttt{geo}}(\btheta,Z)=\frac{1}{N} Z \sum_{n=1}^N\frac{q\left({\bf y}_n\right) }{\phi\left({\bf y}_n |  \boldsymbol{\theta}\right)}+ \frac{1}{M Z}  \sum_{m=1}^M\frac{\phi\left({\bf x}_m| \boldsymbol{\theta}\right)}{q({\bf x}_m)}.}
\end{align}
The difference with respect to Eq.  \eqref{EstaConAppB} is that, in the present expression, the second summation does not contain the logarithmic transformation.  Moreover, it corresponds the to cost function of the estimator $\widehat{Z}_{\texttt{geo}}(\boldsymbol{\theta})$ in Eq. \eqref{GeoEq}.
Indeed, treating $\boldsymbol{\theta}$ as fixed, differentiation with respect to $Z$, and setting the derivative equal to zero, gives:
\begin{align*}
\frac{\partial J_{\text {geo}}}{\partial Z}&=\frac{1}{N} \sum_{n=1}^N \frac{q\left(\mathbf{y}_n\right)}{\phi\left(\mathbf{y}_n \mid \boldsymbol{\theta}\right)}-\frac{1}{M Z^2} \sum_{m=1}^M \frac{\phi\left(\mathbf{x}_m \mid \boldsymbol{\theta}\right)}{q\left(\mathbf{x}_m\right)}=0,  \\
&=A_N(\y_{1:N},\btheta)-\frac{1}{Z^2} B_M(\x_{1:M},\btheta)=0,  \Longrightarrow  \widehat{Z}_{\texttt{geo}}(\btheta)=\sqrt{\dfrac{B_M(\x_{1:M},\btheta)}{A_N(\y_{1:N},\btheta)}}.
\end{align*}
Therefore, we obtain again the geometric mean estimator of $Z$ in Eq. \eqref{GeoEq}:
$$
\boxed{\widehat{Z}_{\texttt{geo}}(\boldsymbol{\theta})=\sqrt{\frac{\frac{1}{M} \sum_{m=1}^M \frac{\phi\left(\mathbf{x}_m \mid \boldsymbol{\theta}\right)}{q\left(\mathbf{x}_m\right)}}{\frac{1}{N} \sum_{n=1}^N \frac{q\left(\mathbf{y}_n\right)}{\phi\left(\mathbf{y}_n \mid \boldsymbol{\theta}\right)}}}.}
$$
 Note that the estimator \(\widehat{Z}_{\texttt{geo}}\)  cannot be derived from the bridge identity and the cost function $J_{\texttt{geo}}(\btheta,Z)$ does not belong to the generalized Pihajala' s family. 
However, replacing $\widehat{Z}_{\texttt{geo}}(\boldsymbol{\theta})$ into $J_{\mathrm{geo}}$ we obtain:
\begin{align*}
 J_{\mathrm{geo}}(\boldsymbol{\theta}, \widehat{Z}_{\texttt{geo}}(\boldsymbol{\theta}))&= \sqrt{\dfrac{B_M(\x_{1:M},\btheta)}{A_N(\y_{1:N},\btheta)}} A_N(\y_{1:N},\boldsymbol{\theta})+ \sqrt{\dfrac{A_N(\y_{1:N},\btheta)}{B_M(\y_{1:M},\btheta)}} B_M(\y_{1:M},\btheta), \\
 &=2 \sqrt{A_N(\y_{1:N},\boldsymbol{\theta}) B_M(\x_{1:M},\boldsymbol{\theta})}.
 \end{align*}
 Thus, as $N, M \to \infty$, we have $A_N(\y_{1:N},\boldsymbol{\theta})\to \frac{1}{Z_{\texttt{tr}}}$ and $B_M(\y_{1:N},\boldsymbol{\theta})\to Z_{\texttt{tr}}$, and
 \begin{align*}
 J_{\mathrm{geo}}(\boldsymbol{\theta}, \widehat{Z}_{\texttt{geo}}(\boldsymbol{\theta})) \longrightarrow 2, \qquad N, M \to \infty.
 \end{align*}
 Hence, we cannot use  $ J_{\mathrm{geo}}$ for estimating $\btheta_{\texttt{tr}}$.
} 

\end{document}